\documentclass[acmlarge]{acmart}
\AtBeginDocument{%
  \providecommand\BibTeX{{%
    \normalfont B\kern-0.5em{\scshape i\kern-0.25em b}\kern-0.8em\TeX}}}

\setcopyright{cc}
\setcctype{by}
\acmJournal{IMWUT}
\acmYear{2026} 
\acmVolume{10} 
\acmNumber{3} 
\acmArticle{197}
\acmMonth{9} 
\acmDOI{10.1145/3832008}

\usepackage{algorithm}
\usepackage{algorithmicx}
\usepackage{algpseudocode}
\usepackage{multirow}
\usepackage{subfig}
\usepackage{appendix}
\usepackage{xcolor, xspace}
\usepackage{siunitx}

\usepackage{pifont}
\newcommand{\cmark}{\ding{51}} 
\newcommand{\xmark}{\ding{55}} 

\newcommand{\ie}{\emph{i.e.},\xspace}

\newcommand{\eg}{\emph{e.g.},\xspace}

\newcommand\figref[1]{Fig.~\ref{#1}}

\newcommand\tabref[1]{Tab.~\ref{#1}}

\newcommand\secref[1]{Sec.~\ref{#1}}

\newcommand{\fakeparagraph}[1]{\vspace{1mm}\noindent\textbf{#1.}}

\newcommand{\sysname}{{\sf UIAnchor}\xspace}
\newcommand{\sysnameposs}{{\sf UIAnchor's}\xspace}

\ifodd 1

\else

\fi

\begin{document}

\title{UIAnchor: Anchoring UI Perception and Action Execution for Reliable Service-Composed Mobile Task Automation with GUI Agents}


\author{Wentao Zhou}
\orcid{0009-0002-3165-4288}
\affiliation{%
  \institution{Northwestern Polytechnical University}
  \city{Xi'an}
  \country{China}
}
\email{never_gone.oldcat@mail.nwpu.edu.cn}

\author{Sicong Liu}
\orcid{0000-0003-4402-1260}
\authornote{Corresponding author.}
\affiliation{%
  \institution{Northwestern Polytechnical University}
  \city{Xi'an}
  \country{China}
}
\email{scliu@nwpu.edu.cn}

\author{Zimu Zhou}
\orcid{0000-0002-5457-6967}
\affiliation{%
  \department{School of Data Science}
  \institution{City University of Hong Kong}
  \city{Hong Kong}
  \country{China}
}
\email{zimuzhou@cityu.edu.hk}

\author{Yimeng Duan}
\orcid{0009-0002-6288-3403}
\affiliation{%
  \institution{Northwestern Polytechnical University}
  \city{Xi'an}
  \country{China}
}
\email{2935752563@mail.nwpu.edu.cn}

\author{Yongyan Cai}
\orcid{0009-0006-4610-4608}
\affiliation{%
  \institution{Northwestern Polytechnical University}
  \city{Xi'an}
  \country{China}
}
\email{yongyanc@mail.nwpu.edu.cn}

\author{Weiye Wu}
\orcid{0009-0002-8973-685X}
\affiliation{%
  \institution{Northwestern Polytechnical University}
  \city{Xi'an}
  \country{China}
}
\email{maxwellwwy@mail.nwpu.edu.cn}

\author{Teng Li}
\orcid{0009-0007-9033-1944}
\affiliation{%
  \institution{Northwestern Polytechnical University}
  \city{Xi'an}
  \country{China}
}
\email{liteng4735@mail.nwpu.edu.cn}

\author{Daqing Zhang}
\orcid{0000-0002-6608-1267}
\affiliation{%
  \department{Telecom SudParis}
  \institution{Institut Polytechnique de Paris}
  \city{Paris}
  \country{France}
}
\email{daqing.zhang@telecom-sudparis.eu}

\author{Zhiwen Yu}
\orcid{0000-0002-9905-3238}
\affiliation{%
  \institution{Harbin Engineering University}
  \city{Harbin}
  \country{China}
}
\email{zhiwenyu@nwpu.edu.cn}

\renewcommand{\shortauthors}{Zhou et al.}
\renewcommand{\shorttitle}{UIAnchor: Anchoring UI Perception and Action Execution}

\authorsaddresses{%
Authors' Contact Information:~
\href{https://orcid.org/0009-0002-3165-4288}{Wentao Zhou},
Northwestern Polytechnical University, Xi'an, China,
never\_gone.oldcat@mail.nwpu.edu.cn;
\href{https://orcid.org/0000-0003-4402-1260}{Sicong Liu} 
(corresponding author), Northwestern Polytechnical University, Xi'an, China,
scliu@nwpu.edu.cn;
\href{https://orcid.org/0000-0002-5457-6967}{Zimu Zhou},
School of Data Science, City University of Hong Kong, Hong Kong, China and City
University of Hong Kong Shenzhen Research Institute, Shenzhen, China,
zimuzhou@cityu.edu.hk;
\href{https://orcid.org/0009-0002-6288-3403}{Yimeng Duan},
Northwestern Polytechnical University, Xi'an, China,
2935752563@mail.nwpu.edu.cn;
\href{https://orcid.org/0009-0006-4610-4608}{Yongyan Cai},
Northwestern Polytechnical University, Xi'an, China,
yongyanc@mail.nwpu.edu.cn;
\href{https://orcid.org/0009-0002-8973-685X}{Weiye Wu},
Northwestern Polytechnical University, Xi'an, China,
maxwellwwy@mail.nwpu.edu.cn;
\href{https://orcid.org/0009-0007-9033-1944}{Teng Li},
Northwestern Polytechnical University, Xi'an, China,
liteng4735@mail.nwpu.edu.cn;
\href{https://orcid.org/0000-0002-6608-1267}{Daqing Zhang},
Telecom SudParis, Institut Polytechnique de Paris, Paris, France and
Peking University, Beijing, China,
daqing.zhang@telecom-sudparis.eu;
\href{https://orcid.org/0000-0002-9905-3238}{Zhiwen Yu},
Harbin Engineering University, Harbin, China and Northwestern Polytechnical University,
Xi’an, China,
zhiwenyu@nwpu.edu.cn.%
}

\begin{abstract}
Mobile task automation aims to streamline \emph{multi-step, cross-app} interactions on smartphones and in-vehicle systems.
Recent LLM-based GUI agents have advanced rapidly, yet reliability remains limited because modern mobile workflows are increasingly \emph{service-composed} (app switching, forms, pop-ups/permissions, copy-paste), requiring \emph{fine-grained} operations on dense, dynamic, and heterogeneous UIs, often in distraction-sensitive contexts (\eg hands-busy or attention-limited use).
Through an in-the-wild failure analysis of deployable GUI agents, we identify two dominant bottlenecks: agents often miss or misread actionable UI elements, and they execute actions without verifying target correctness or outcomes.
We present \sysname, a \textit{modular} multi-agent system that improves mobile GUI automation by \textit{anchoring} both perception and execution.
\sysname combines a two-stage UI parser for high-recall element capture and context-aware semantics with a meta-controller that performs pre-action verification, post-action outcome perception, per-step state tracking, and targeted recovery.
We further introduce an L1-L5 task taxonomy based on step length, cross-app scope, and UI granularity, showing that L5 remains beyond today’s practical frontier.
On the hardest practical tier, L4 tasks (20--30 steps, multi-app, targets $<100\times100$ px, $\sim$5\,mm on typical phones), \sysname improves success by 31.5\% over GPT-4o and 16.6\% over Mobile-Agent-v3.
With edge/cloud assistance, \sysname runs at $\sim$2\,s per step, comparable to human operation, while reducing per-step latency by up to 75.5\% and energy by 52.4\%.
It also generalizes to unseen apps and cross-platform GUIs.
\end{abstract}



\maketitle

\section{INTRODUCTION}
\label{sec:intro}
\begin{figure}[t]
  \centering
  \subfloat[Cycling]{
    \includegraphics[width=0.178\textwidth]{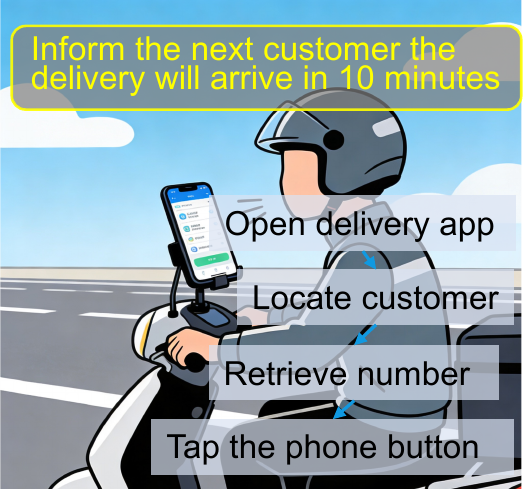}
    \label{fig:case1}
  }
  \subfloat[Driving]{
    \includegraphics[width=0.175\textwidth]{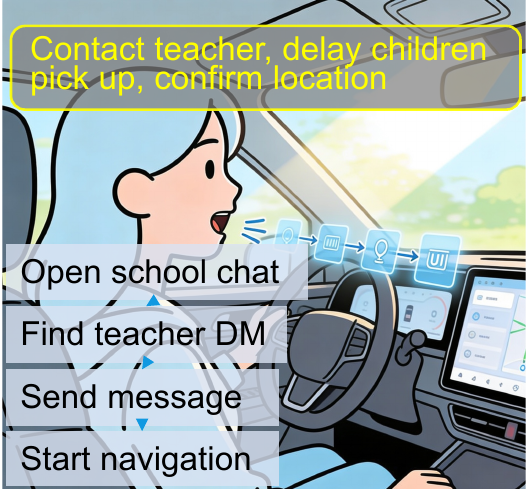}
    \label{fig:case2}
  }
  \subfloat[In motion]{
    \includegraphics[width=0.18\textwidth]{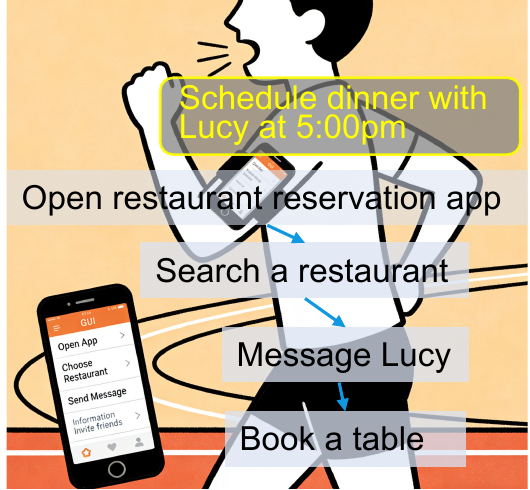}
    \label{fig:case4}
  }
  \subfloat[Inaccessible (blind)]{
    \includegraphics[width=0.18\textwidth]{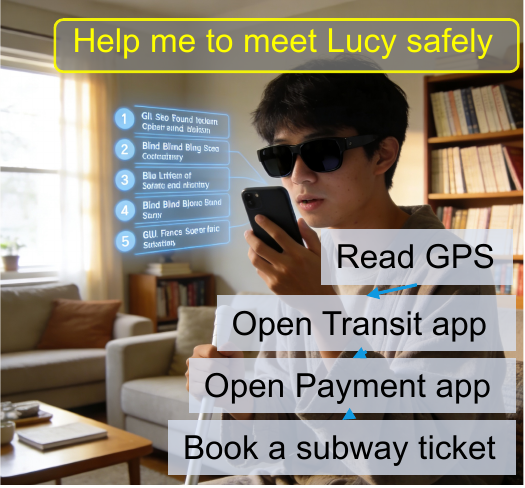}
    \label{fig:case5}
  }
    \subfloat[Science agent]{
    \includegraphics[width=0.18\textwidth]{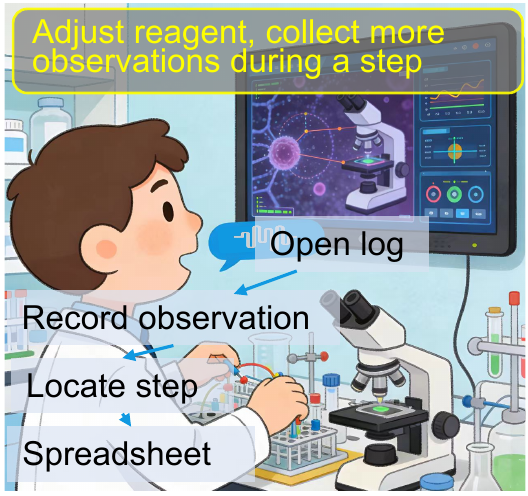}
    \label{fig:case3}
  }
  \vspace{-3mm}
    \caption{Illustration of \textbf{hands-busy} and \textbf{attention-limited} scenarios where automated GUI agents are especially valuable: manually performing dense, \textbf{multi-step}, \textbf{cross-app}, \textbf{mm-scale} interactions are inconvenient, unsafe, or inaccessible.}
  \vspace{-3mm}
  \label{fig:case}
\end{figure}

Human-computer interaction is undergoing a major \textit{shift}: from users \emph{operating interfaces} to \emph{delegating intent} and letting agents automatically execute \textbf{multi-step}, \textbf{cross-app}, \textbf{mm-scale} UI actions.
This shift is amplified by today’s \emph{service-composed} mobile workflows, where routine tasks increasingly involve permission pop-ups, copy--paste, form filling, and multi-app handoffs.
It is particularly valuable on mobile and in-vehicle devices, where interaction often occurs in \textbf{hands-busy} or \textbf{attention-limited} contexts (\eg driving, on-the-move, or for users with visual impairments; see~\figref{fig:case}), making \textit{dense} \textit{cross-app} tapping \textit{inconvenient}, \textit{unsafe}, or \textit{inaccessible}.

Motivated by this trend, \textit{mobile task automation} aims to autonomously carry out \textit{multi-step, cross-app} tasks, such as reordering a frequent meal (app 1) and inviting a friend (app 2), or messaging a contact (app 1) and launching navigation (app 2) while driving.
While existing platforms offer shortcuts~\cite{Apple2022ShortcutsGuide} and scripted routines (\eg Bixby Routines~\cite{Samsung2024BixbyRoutines}), they typically require \textit{manual} configuration and mainly support pre-defined, single-app, low-granularity operations, or rely on brittle UI assumptions. Thus, they remain fragmented and struggle to generalize across diverse \textit{third-party} apps and \textit{evolving} interfaces.

\textit{GUI agents} \cite{Agent-S2, xie2025gui_explorer, lai2025androidgen, gou2024UGround, wen2024autodroid, li2025mobileuse, dai2025advancing, wen2024autodroidv2, ye2025mobileagentv3, lee2024mobilegpt, wang2025mobile_e, cheng2024seeclick} provide a more general path to automation by operating \textit{through the interface} the same way humans do.
Given a natural-language instruction, a GUI agent observes the current screen, decides the next step, and executes touch actions such as tapping, typing, and scrolling.
This paradigm avoids privileged APIs and can generalize across system apps and third-party apps.
With recent advances in vision-language models (VLMs) \cite{wang2024cogvlm, chu2024mobilevlm,qwen3technicalreport,Qwen2.5-VL}, GUI agents have become increasingly capable at interpreting mobile screenshots and following a \textit{UI parse-action plan-action execution} loop for task execution \cite{lee2024mobilegpt, li2025mobileuse, ye2025mobileagentv3, wang2025mobile_e}.

However, even strong VLM-based GUI agents often fail on hard tasks, such as permission pop-ups, copy-paste, form filling, and multi-app handoffs.
More critically, these errors \textit{compound} in multi-step or cross-app workflows: a single mis-click can lead the agent to an unintended page, from which it continues with an incorrect understanding of the UI, spiraling into long failure chains. This propagation of errors is a major barrier to deploying GUI agents as reliable assistants, particularly in high-stakes or accessibility-critical scenarios.

Despite substantial advances in GUI grounding, deployment failures remain common in state-of-the-art VLM-driven GUI agents~\cite{lee2024mobilegpt}. 
Our analysis shows that these failures arise from the interaction between perception and execution, where residual perception errors propagate through long-horizon mobile workflows.
In real apps, small front-end errors easily cascade in \textit{service-composed} worklfows (\ie multi-step, cross-app, mm-scale), turning minor slips into long failure chains.

\noindent$\bullet$ \textbf{Fragile UI anchors: ambiguous perception in dense mobile UIs.}
Mobile UIs are dense and gesture-driven, where actionable targets are often mm-scale, visually similar, and text-sparse (\eg icon-only toolbars). 
Although recent GUI grounding models have substantially improved perception, existing single-stage parsers~\cite{li2025mobileuse, lee2024mobilegpt, ye2025mobileagentv3, wang2025mobile_e, hong2024cogagent, cheng2024seeclick, zhang2025appagent} still encounter two challenges: \textit{coverage failures}, where small or transient elements are missed, and \textit{semantic failures}, where detected elements remain ambiguous without sufficient context. Such errors may propagate downstream and lead to mis-grounded actions.

\noindent$\bullet$ \textbf{Limited runtime execution verification: open-loop actions without per-step state verification.}
Many agents~\cite{hong2024cogagent, cheng2024seeclick, dai2025advancing, zhang2025appagent} act in an open-loop manner, with limited checks that an action is valid or succeeded. 
Yet mobile apps hinge on transient or partially non-visual state changes (\eg toasts, short loaders, keyboard shifts, activity transitions, login/session changes). Without per-action verification, agents act on stale states, fail silently, and drift from the true UI, leading to brittle recovery and long failure chains.

To address these, we present \sysname, a mobile GUI agent that improves reliability by \textit{jointly anchoring both What to act on and Whether actions succeed}.
We retain the standard parse-plan-act paradigm while strengthening UI grounding and redesign the runtime per-action coordination to mitigate the failure modes above.

\noindent$\bullet$ \textbf{First, two-stage UI parsing for high coverage and context-aware semantics.}
We design a two-stage parser to \emph{anchor} GUI automation with both high coverage and context-grounded semantics, instead of forcing a single model to "detect+understand" in one shot.
In the \textit{capture} stage, we use segmentation-style extraction to maximize recall, preserving mm-scale, low-contrast, and icon-only candidates with optional OCR.
In the \textit{interpretation} stage, we infer each candidate’s functional intent from \emph{joint} evidence (appearance, OCR, and surrounding UI context), resolving icon ambiguity across pages.
We output a structured UI state with confidence scores, enabling robust grounding and allowing downstream agents to rank, abstain, or re-ground before errors cascade.

\noindent$\bullet$ \textbf{Second, meta-controller for per-action state perception and targeted recovery.}
We design a lightweight meta-controller that anchors each action with state evidence and prevents error cascades in long mobile workflows.
It performs \textit{pre-action verification} to ensure the target is confidently groundable, and \textit{post-action outcome perception} to verify success and detect state transitions.
A runtime \textit{meta-state memory} tracks transient signals (\eg toasts, activity transitions, overlay/keyboard shifts), while state differencing distinguishes real progress from silent failure.
Based on this, the meta-controller diagnoses failures and applies targeted recovery (\eg re-ground, retry, or navigation correction) instead of re-planning from scratch.

We implement UIAnchor as a \textit{modular} multi-agent system with four components: a UI Parser, Planner, Executor, and Meta-Controller. 
We build the \textit{UI Parser} as a lightweight pipeline for high-recall, structured UI extraction. 
For the remaining agents (Planner/Executor/Meta-Controller), we adopt a shared LVLM backbone over screenshots/UI states and specialize each role via \textit{Multi-LoRA}, without maintaining multiple full models.

We prototype \sysname on smartphones, tablets, and in-vehicle systems, and evaluate it on AndroidWorld~\cite{rawles2024androidworld}, MobileWorld~\cite{kong2025mobileworld}, ScreenSpot-v2~\cite{wu2024atlas}, MMBench-GUI~\cite{wang2025mmbenchgui}, and our meta-state-aware MobileTask episodes.
We stratify GUI tasks into five levels (L1--L5) by step length, cross-app scope, and UI granularity (\secref{sec_task_level}).
L5 approaches the practical ceiling of current frontier V/LLMs, while L4 captures the hardest attainable real-world workflows, including permission pop-ups, copy-paste, form filling, and multi-app handoffs.
On L4 tasks, \sysname improves accuracy by 20\% over \textit{GPT-4o} and 18\% over Mobile-Agent-v3.
The gain comes from a self-verifying \textit{meta-controller} that performs pre-action checks, post-action outcome perception, and targeted recovery.
\sysname also explicitly tracks transient meta-states, such as permission dialogs, toasts, activity transitions, and login/session changes, enabling more reliable execution and recovery.
Despite these gains, \sysname remains compact: it uses a 4B backbone with 83M LoRA parameters and runs at $\sim$2\,s/step, comparable to human operation.
It reduces per-step latency by up to 75.5\% and energy by 52.4\% over baselines, while generalizing to unseen apps and evolving cross-platform GUIs.
On resource-limited servers such as Jetson-class devices, where model size dominates latency, our 4B model avoids VRAM/bandwidth bottlenecks and cuts end-to-end latency by 65\% compared with 7B/8B GUI agents.
Our main contributions are summarized as follows.
\begin{itemize}
    \item 
    We revisit two complementary sources of unreliability in deployable mobile GUI agents \ie incomplete/ambiguous UI perception and the lack of per-action closed-loop coordination, and show how they lead to compounding execution errors.
    \item 
    We integrate structured UI grounding with runtime execution verification for reliable mobile GUI execution. 
    A two-stage parser produces confidence-aware UI states, and a lightweight meta-controller enables closed-loop execution through per-action verification and targeted recovery.
    \item 
    Extensive experiments show that \sysname achieves strong accuracy on the hardest practical tier L4 (cross-app, mm-scale, multi-step tasks) with only a 4B backbone, while running at human-like speed.
\end{itemize}
\section{PRELIMINARIES}
\label{sec:pre}

\subsection{Mobile Task Automation via GUI Agents}
\label{sec:pre:mobile}

Mobile task automation enables devices to autonomously execute multi-step workflows on behalf of users, which is particularly valuable when manual interaction is inconvenient, unsafe, or inaccessible (\eg while driving, during
emergencies, or for users with motor or visual impairments). 
Among various technical approaches, \textit{GUI-based} automation \cite{li2025mobileuse,ye2025mobileagentv3,wen2024autodroid,wen2024autodroidv2,qin2025uitar} prevails because it can operate across diverse apps and versions without relying on developer-provided APIs, making it broadly compatible with rapidly evolving mobile ecosystems.

A \textit{GUI agent} aims to translate a high-level natural-language instruction (\eg ``text Alice I’m running late'') into a sequence of low-level UI interactions (\eg \texttt{tap}, \texttt{swipe}, and \texttt{type}) that manipulate on-screen elements to complete the task.
This process is commonly modeled as a \textit{sequential decision-making} loop. 
At each step $t$, the agent observes the current screen state $S_t$ and produces an executable action $A_t$, which transitions the device to a new state $S_{t+1}$. 
The loop repeats until the task terminates (successfully or not):
\begin{equation}
\langle S_t, I\rangle \xrightarrow{\ \text{Agent}\ } A_t \xrightarrow{\ \text{Execute}\ } S_{t+1}.
\end{equation}

Most GUI agents implement this loop with three conceptual stages.
\begin{enumerate}
    \item \textbf{UI Parsing:} Extract actionable UI elements and their attributes from the current screen (\eg element type, text, icon semantics, and location).
    \item \textbf{Action Planning:} Decide the next action or interaction needed to advance the instruction under the current state.
    \item \textbf{Action Execution:} Emit a low-level operation (\eg \texttt{click(x,y)}, \texttt{swipe}, \texttt{input}) and apply it to the device to obtain the next state.
\end{enumerate}

\subsection{Limitations of Current GUI Agents}
\label{sec:pre:limitation}

\begin{figure}[t]
  \centering
  \subfloat[LLM-based]{
    \includegraphics[width=0.48\textwidth]{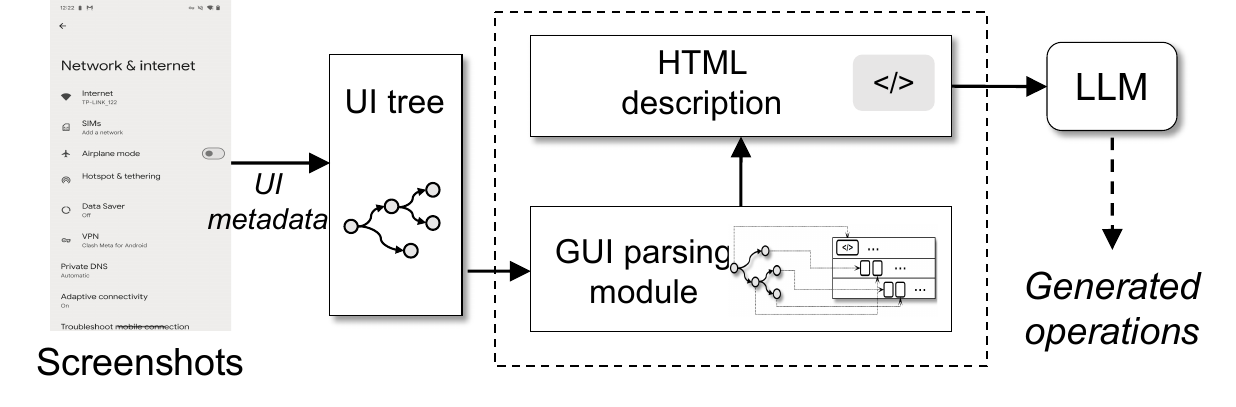}
    \label{fig:LLM-Based(a)}
  }\hfill
  \subfloat[VLM-based]{
    \includegraphics[width=0.48\textwidth]{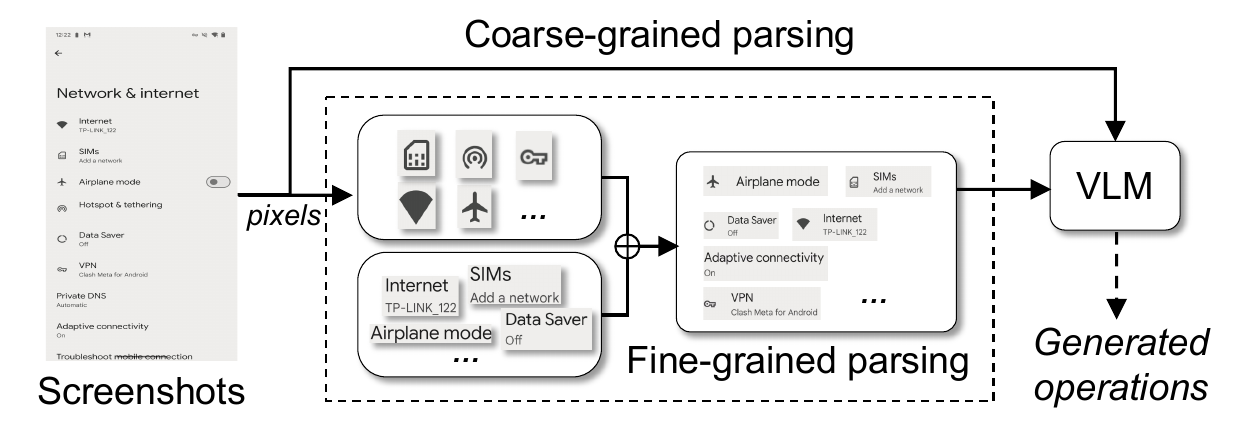}
    \label{fig:VLM-based(b)}
  }
  \vspace{-3mm}
  \caption{Comparison of LLM- and VLM-based GUI agents.} 
  \vspace{-3mm}
  \label{fig:differ}
\end{figure}

Existing GUI agents fall into two architectural paradigms based on \textit{what} they use for \textit{screen representation} (\figref{fig:differ}).
\begin{itemize}
    \item \textbf{LLM-based (UI-tree-based) agents.}
    These methods \cite{lee2024mobilegpt, wen2024autodroid, wen2024autodroidv2} rely on structured UI metadata (typically an accessibility hierarchy) as the main observation.
    They employ a GUI parsing module to map the UI tree, based on predefined rules, into a textual form (e.g., HTML-like descriptions) and use an LLM to reason about the instruction, select a target element, and generate the next action. 
    Their effectiveness depends on the availability and fidelity of UI metadata.
    \item \textbf{VLM-based (screenshot-based) agents.}
    These methods \cite{ye2025mobileagentv3, li2025mobileuse, qin2025uitar} treat a rendered screenshot as the main observation and use a vision-language model to infer the next interaction directly from pixels. 
    They avoid reliance on UI trees but require fine-grained visual grounding on small, dense mobile screens.
\end{itemize}

\begin{figure}[t]
  \centering
  \begin{minipage}{0.45\textwidth}
    \centering
    \includegraphics[width=\textwidth]{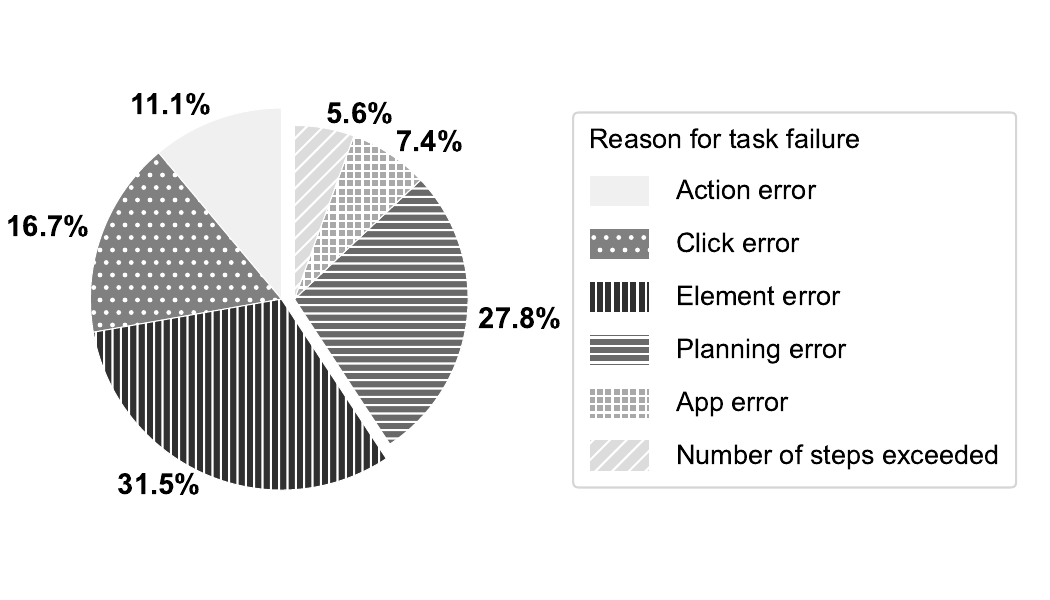}
    \vspace{-12mm}
    \caption{Distribution of failure reasons for the compact model Qwen3-VL-4B~\cite{qwen3technicalreport} and GUI-Owl-7B~\cite{ye2025mobileagentv3} on the AndroidWorld~\cite{rawles2024androidworld} and self-collected datasets.}
    \vspace{-3mm}
    \label{fig:task_failure_reason_pie}
  \end{minipage}
  \hspace{5mm}
  \begin{minipage}{0.43\textwidth}
    \centering
    \includegraphics[width=\textwidth]{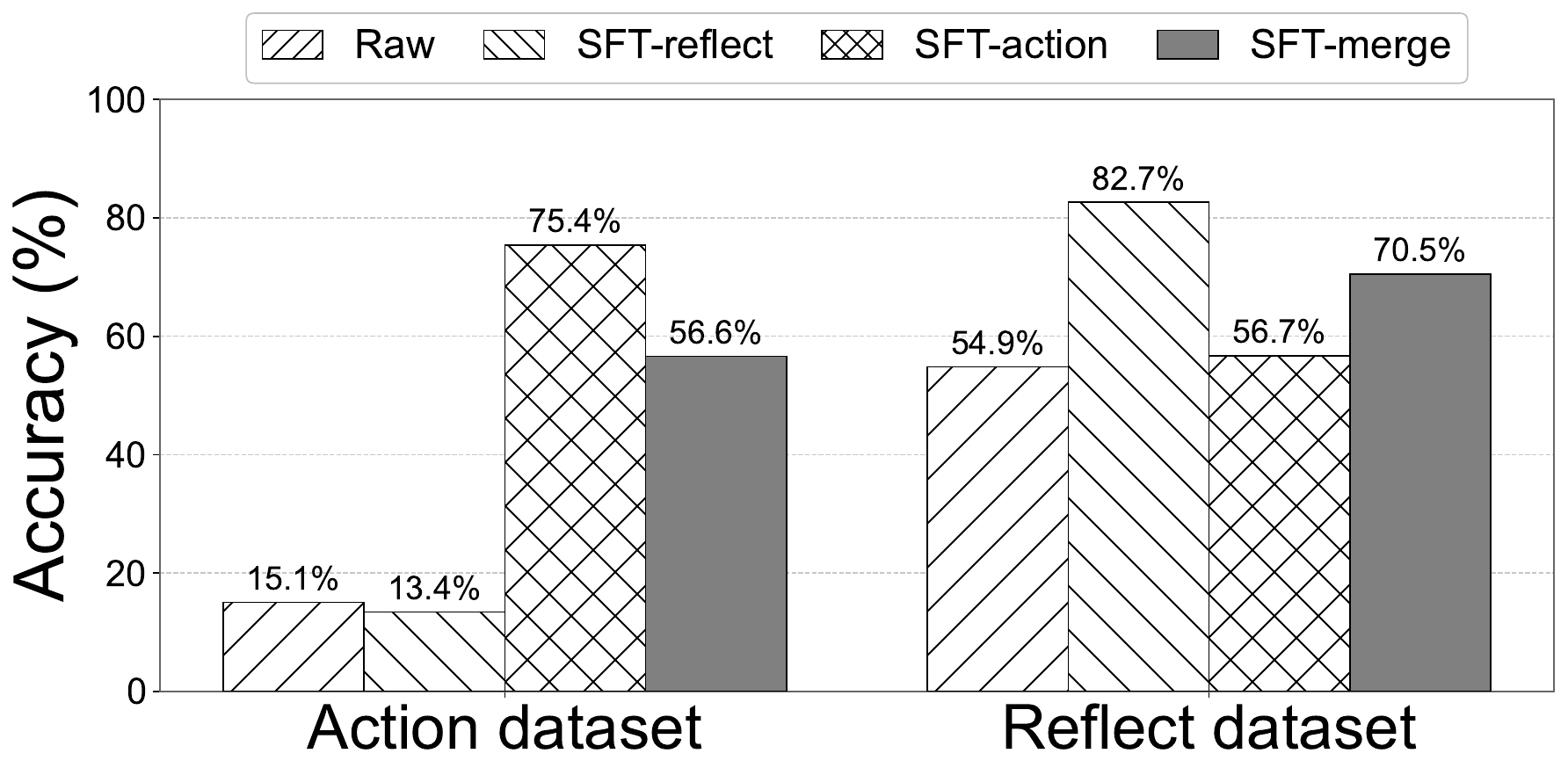}
    \caption{Accuracy of specialized vs.\ merged fine-tuning. SFT-reflect/action/merge are fine-tuned on reflection-only, action-only, and mixed data.}
    \vspace{-3mm}
    \label{fig:model_accuracy_comparison}
  \end{minipage}
\end{figure}

To understand where representative compact open-source existing GUI agents fail in real mobile interactions, we analyze error categories for the open-source model Qwen3-VL-4B~\cite{qwen3technicalreport} and GUI-Owl-7B~\cite{ye2025mobileagentv3} on AndroidWorld \cite{rawles2024androidworld} (1.1k traces, 20 apps), a dynamic benchmark for GUI agents, and our self-collected dataset (2k traces, 32 apps).
\figref{fig:task_failure_reason_pie} shows that failures fall into two patterns.
\begin{itemize}
    \item \textbf{For representative compact GUI agents, fine-grained UI perception remains the dominant bottleneck.}
    A major failure source comes from \texttt{element} (31.5\%), \texttt{click} (16.7\%), and \texttt{action} (11.1\%), which counts for 59.3\%.
    This indicates that, for representative compact open-source GUI agents, a major barrier is not high-level intent understanding, but reliably identifying \textit{what} to interact with and \textit{where} to interact on the screen. 
    While static GUI grounding benchmarks~\cite{wu2024atlas} are approaching saturation, long-horizon mobile interactions still require robust perception under changing UI states and resource constraints.
    This motivates designs that explicitly strengthen \textit{UI parsing} by improving coverage of actionable elements and disambiguating icon semantics in context. 
    \item\textbf{Cross-step coordination is fragile, leading to drift, stalls, and compounding errors.}
    A second cluster of failures reflects \textit{multi-step control issues}: \texttt{planning} (27.8\%), \texttt{App} (7.4\%) and \texttt{number of steps exceeded} (5.6\%), which counts for 40.7\%.
    This implies that even if individual actions are plausible, the agent can drift from the intended trajectory, miss transient state changes, or get stuck in loops without timely verification and recovery. 
    This motivates designs that improve \textit{cross-step coordination}, \ie deciding when to trust the current state, when to re-assess, and how to detect and correct execution anomalies before errors accumulate. 
\end{itemize}

As a separate note on \textbf{modularity}, most prior systems implement UI parsing, planning, and execution by repeatedly prompting a single LLM/VLM~\cite{ye2025mobileagentv3,li2025mobileuse}, which couples heterogeneous functions and can amplify both perception and coordination errors.
\figref{fig:model_accuracy_comparison} shows this effect under a \textit{cross-domain task} setting, where we compare a unified base model against specialized, functionally decomposed agents.
The results show that decoupling perception (action tasks) from reasoning (reflection tasks) into separate modules yields substantial accuracy gains. Specifically, SFT-action attains 75.42\% on action tasks, and SFT-reflect reaches 82.65\% on reflection tasks. In contrast, a merged model (SFT-merge) shows balanced but lower peak performance (70.52\% on reflection; 56.63\% on action), illustrating a \textit{specialization-generalization trade-off}.
It suggests that implementing the workflow as \textit{multiple agents} can improve task success. We note that this analysis is based on representative compact open-source GUI agents and is intended to motivate design choices for the compact mobile-agent regime; stronger larger-scale systems may exhibit different error distributions. More broadly, this modularity is not meant to oppose the long-term trend toward end-to-end agents. While stronger foundation agents may internalize some routine verification and correction behaviors, modularity may still remain useful for handling complex failures. In the long term, manually designed modular structures may evolve into more advanced and automated forms rather than disappear entirely.
\section{SOLUTION OVERVIEW}
\label{sec:overview}
To address the dominant grounding and cross-step coordination failures identified in \secref{sec:pre:limitation}, we propose \sysname, a \textit{multi-agent} GUI system that extends the VLM-based GUI agent paradigm by strengthening \textbf{UI parsing} and \textbf{per-action runtime coordination} within the standard \textbf{parse-plan-act} loop.

\begin{table}[]
\centering
\caption{Comparison with representative prior GUI-agent systems.}
\label{tab:prior_work_comparison}
\resizebox{0.98\linewidth}{!}{%
\begin{tabular}{|l|c|c|c|c|c|c|c|}
\hline
\textbf{Method} &
  \textbf{Architecture} &
  \textbf{Modularity} &
  \textbf{Structured parsing} &
  \textbf{Verification} &
  \textbf{Reflection/Recovery} &
  \textbf{Runtime memory} &
  \textbf{Meta-controller} \\ \hline
AutoDroid~\cite{wen2024autodroid}          & LLM-based & \cmark & \cmark(A11y tree)  & \cmark & \xmark & \cmark & \xmark \\ \hline
MobileGPT~\cite{lee2024mobilegpt}          & LLM-based & \cmark & \cmark(A11y tree)  & \xmark & \cmark & \cmark & \xmark \\ \hline
V-Droid~\cite{dai2025advancing}            & LLM-based & \cmark & \cmark(A11y tree)  & \cmark & \cmark & \cmark & \xmark \\ \hline
OmniParser~\cite{lu2024omniparser}         & VLM-based & \cmark & \cmark(Screenshot) & \xmark & \xmark & \xmark & \xmark \\ \hline
UGround~\cite{gou2024UGround}              & VLM-based & \cmark & \cmark(Screenshot) & \xmark & \xmark & \xmark & \xmark \\ \hline
MobileUse~\cite{li2025mobileuse}           & VLM-based & \cmark & \xmark             & \cmark & \cmark & \cmark & \xmark \\ \hline
Mobile-Agent-v3~\cite{ye2025mobileagentv3} & VLM-based & \cmark & \xmark             & \cmark & \cmark & \cmark & \xmark \\ \hline
\textbf{\sysname} &
  \textbf{VLM-based} &
  \textbf{\cmark} &
  \textbf{\cmark(Screenshot)} &
  \textbf{\cmark} &
  \textbf{\cmark} &
  \textbf{\cmark} &
  \textbf{\cmark} \\ \hline
\end{tabular}
}%
\end{table}

\subsection{Architecture}
\label{sec:arch}
\sysname follows the VLM-based (screenshot-based) paradigm~\cite{li2025mobileuse, ye2025mobileagentv3, qin2025uitar, cheng2024seeclick, hong2024cogagent}.
This is because it operates directly on the \textit{rendered screen}, rather than relying on an accessibility UI tree, so it can function when UI metadata is missing or unreliable (\eg dynamic layouts, or incomplete accessibility annotations).
Conceptually, \sysname follows the conventional parse-plan-act pipeline~\cite{li2025mobileuse, ye2025mobileagentv3, qin2025uitar}, but makes two functions \textit{explicit and robust}:
\textit{(i)} robust UI parsing that constructs a UI state for precise grounding, and \textit{(ii)} cross-step coordination that verifies state transitions and triggers targeted recovery around every action.

In particular, \sysname consists of three functional layers.
\begin{itemize}
    \item \textbf{Two-Stage UI Parsing Agent (\secref{sec:ui}).}
    Given the current screenshot $I_t$, the UI parsing agent produces a structured UI state $\mathcal{U}_t=\{e_k\}_{k=1}^{m}$, where each element $e_k=(b_k, t_k, \hat{y}_k, s_k)$ contains a bounding box, optional OCR text, an inferred function label, and a confidence score.
    Unlike existing approaches that directly act from pixels \cite{li2025mobileuse, ye2025mobileagentv3, cheng2024seeclick, hong2024cogagent}, \sysname separates \textit{component localization} from \textit{component-level semantic inference} stage, improving coverage of small actionable elements and reducing confusions among visually similar icons. 
    The contribution lies not in the two-stage paradigm itself (also used in parsers such as OmniParser~\cite{lu2024omniparser}), but in our parser design and its integration with anchored execution.

    \item \textbf{Planner and Executor (we follow standard components).}
    On top of $\mathcal{U}_t$, a \textit{planner} decomposes the user instruction into a sequence of sub-tasks and determines the appropriate next sub-task to address, and an \textit{executor} grounds that the current sub-task into an atomic UI action (\texttt{click}, \texttt{swipe}, \texttt{input}) on the device, based on the actionable elements identified in $\mathcal{U}_t$.
    These two modules implement the standard GUI-agent backbone.
    The focus of \sysname is to make their \textit{inputs} (UI state) and \textit{handoffs} (coordination) reliable.
    \item \textbf{Meta-Controller for Per-Action State Perception (\secref{sec:meta}).}
    To prevent drift and compounding errors in multi-step execution, \sysname introduces a \textit{meta-controller} that performs \textit{per-action state perception}.
    It maintains a contextual runtime memory (including device meta-state signals), verifies readiness and grounding confidence before executing an action, and infers outcomes and detects anomalies after execution.
\end{itemize}

As discussed in \secref{sec:pre:limitation}, these designs address two key failure modes: unreliable UI perception and unverified action execution.
\tabref{tab:prior_work_comparison} shows that, while prior systems include some components individually, \sysname uniquely integrates \textbf{structured parsing}, \textbf{verification}, \textbf{recovery}, \textbf{runtime memory}, and \textbf{meta-control} into a unified framework for \textit{service-composed} mobile workflows.

\begin{figure*}[t]
  \centering
  \includegraphics[width=0.98\textwidth]{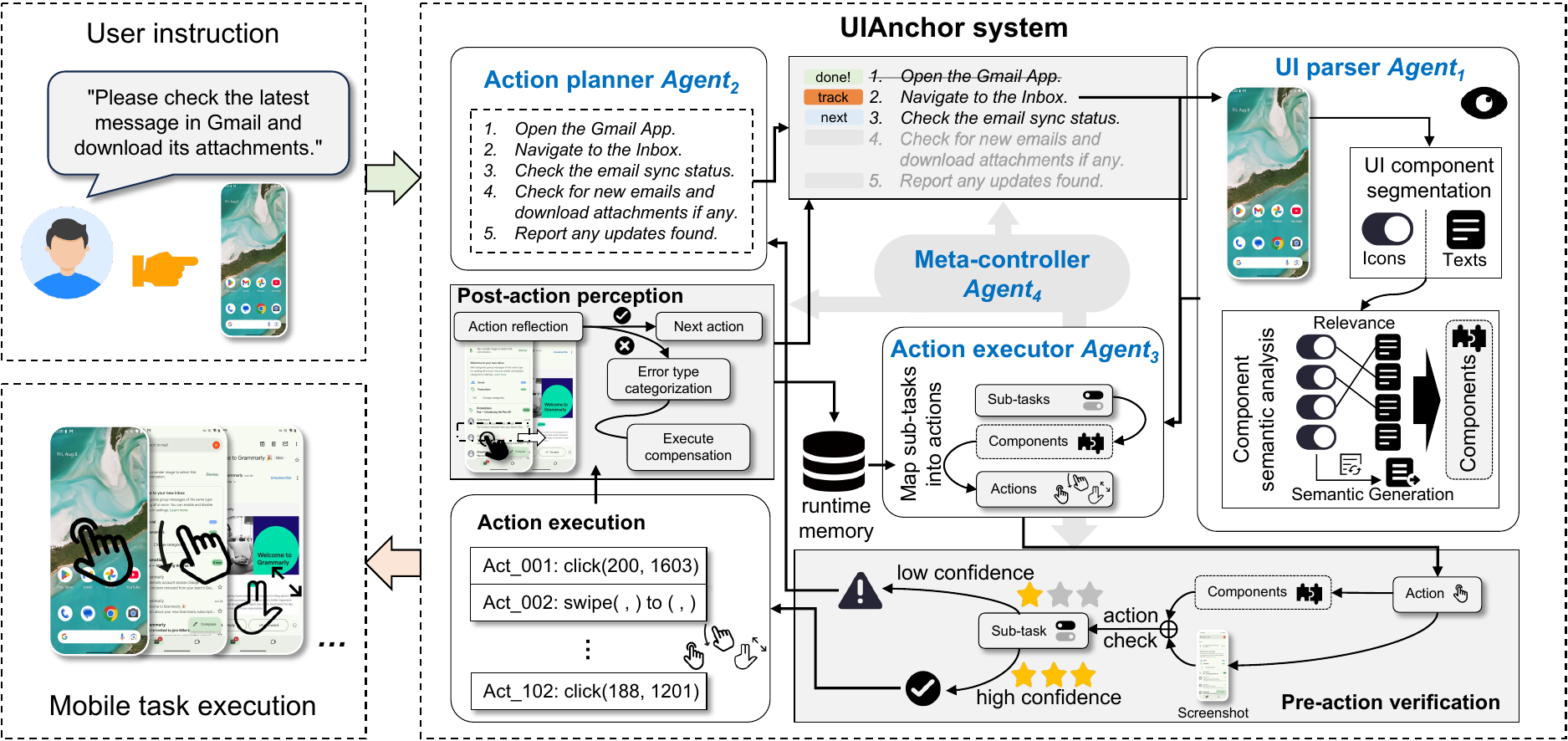}
  \caption{Overview of \sysname workflow.}
  \vspace{-3mm}
  \label{fig:overview}
\end{figure*}

\subsection{Workflow}
At each step $t$, \sysname runs the following loop (\figref{fig:overview}).
\begin{enumerate}
    \item \textbf{Decompose user instruction and plan sub-task sequence.}
    The planner decomposes the overall user instruction into a linear sequence of sub-tasks.
    \item \textbf{Determine the current sub-task.}
    The meta-controller tracks execution progress and determines which sub-task should be tackled next.
    \item \textbf{Parse the current UI on-demand.}
    The meta-controller decides whether to invoke the UI parsing agent to obtain an updated $\mathcal{U}_t$ from $I_t$ or to reuse a cached UI state when the screen is stable.
    \item \textbf{Pre-action verification and act only when confident.}
    Before executing, the meta-controller checks whether the planned behavior can be grounded to the current interface with sufficient confidence, using $\mathcal{U}_t$ and runtime context.
    If confidence is low, it intervenes (\eg re-parse, re-decide, or re-plan) instead of executing a risky action.
    \item \textbf{Execute the grounded action.}
    The executor performs the atomic UI action on-device.
    \item \textbf{Post-action perception and targeted recovery.}
    After execution, the meta-controller updates runtime memory and infers whether the intended state transition occurred.
    If anomalies are detected (no-op, unexpected pop-up, wrong navigation, stalled progress), it triggers targeted recovery.
    Only when recovery fails does it escalate to re-planning.
\end{enumerate}

We implement parsing, planning, execution, and meta-control as multiple lightweight agents rather than a single VLM-based agent.
This enables role specialization and allows the meta-controller to invoke expensive components only when needed.
We detail this multi-agent implementation in \secref{sec:implementation}.

\section{TWO-STAGE UI PARSING AGENT}
\label{sec:ui}
The UI parsing agent is the perception front-end of \sysname.
Given a raw mobile screenshot $I$, it outputs a UI state
$\mathcal{U}=\{e_k\}_{k=1}^{m}$, where each element $e_k=(b_k, t_k, \hat{y}_k, s_k)$ contains a bounding box $b_k$, optional OCR text $t_k$, an inferred function label $\hat{y}_k$ (\eg \texttt{search}, \texttt{back}, \texttt{share}), and a confidence score $s_k$.
As discussed in \secref{sec:pre:limitation}, over half of the failures arise from perception errors, \ie either \textbf{mislocalizing} the target element or \textbf{misinterpreting} an icon.
Accordingly, this agent aims to explicitly strengthen both element localization and function disambiguation.

\fakeparagraph{Challenges}
Reliable UI parsing requires \textbf{robust element localization} under heterogeneous layouts and icon styles, and \textbf{context-dependent functional inference} for visually ambiguous icons.
This challenge arises because identical icons (\eg a "heart" or "star") can represent different functions in different apps, and app-specific layouts, color schemes, and semantics vary widely across platforms.
For example, a user tapping a button expecting a “favorite” action may instead "save for later," "bookmark and share," or even "add to cart" depending on the app's context.
As a result, UI parser must go beyond visual recognition and reason about icon meaning in context.
Existing studies often satisfy only one side, leading to systematic failures.
\begin{itemize}
    \item 
    \textit{Detector-based parsers} (\eg YOLO+OCR such as OmniParser~\cite{lu2024omniparser}) rely on learned region proposals and can fail under icon-style or app-specific domain shifts.
    \item 
    \textit{Single-VLM-based parsers} (\eg CogAgent \cite{hong2024cogagent}, SeeClick \cite{cheng2024seeclick}, UI-TARS \cite{qin2025uitar}, MobileUse \cite{li2025mobileuse}) map screenshots directly to actions or coordinates.
    Their parsing is implicit, which can overlook small UI components and entangle localization with semantic reasoning in a single model prediction.
    \item 
    \textit{LLM-based parsers} (\eg AutoDroid \cite{wen2024autodroid}, MobileGPT \cite{lee2024mobilegpt}, V-Droid \cite{dai2025advancing}, AutoGLM \cite{liu2024autoglm}) rely on HTML/XML abstractions.
    They discard key visual cues (shape, style, relative layout) and thus struggle with stylized or obfuscated icons whose intent is missing or under-specified.
\end{itemize}

\fakeparagraph{Our Methods}
To address these issues, we implement UI parsing as a \textbf{two-stage} pipeline: \textit{(i)} UI component capture (\secref{sec:ui:segmentation}) aims at localization of icons and text blocks; and \textit{(ii)} component-level semantic analysis (\secref{sec:ui:context}) assigns a function label to each localized component.
This separation makes the failure points explicit, avoiding a single VLM \cite{hong2024cogagent,cheng2024seeclick,qin2025uitar} that must solve both problems implicitly.

\begin{figure}[t]
  \centering
  \includegraphics[width=0.93\textwidth]{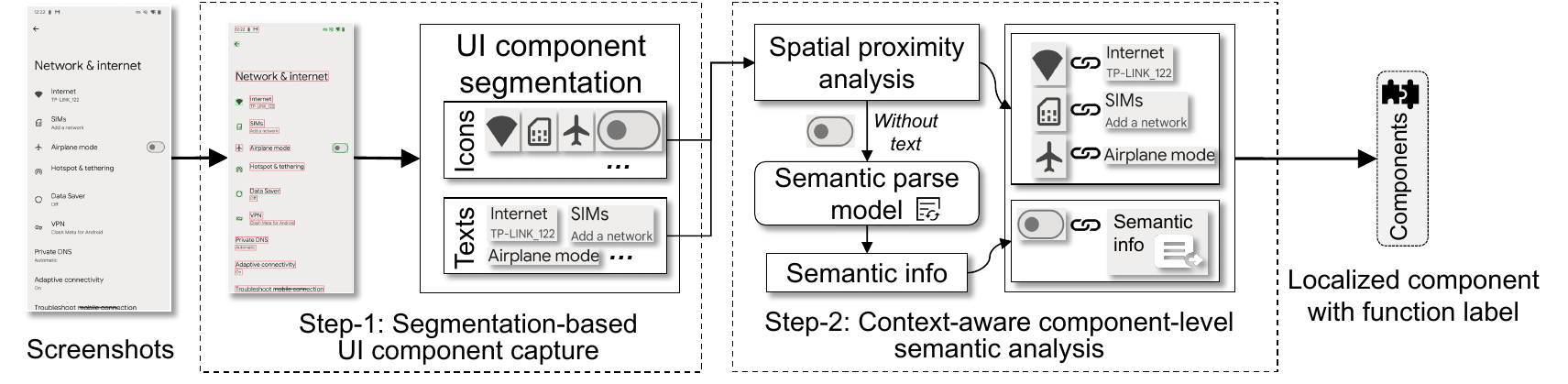}
  \caption{Illustration of the two-stage UI parser agent pipeline.}
  \label{fig:sam}
\end{figure}

\subsection{Segmentation-Based UI Component Capture}
\label{sec:ui:segmentation}
To maximize coverage across heterogeneous UIs, we adopt a \textit{segmentation-based} decomposition instead of detector-based region proposals \cite{lu2024omniparser}. Compared with detector-based region proposals\cite{lu2024omniparser}, this choice lets us start from physical UI boundaries and avoid additional icon-detector retraining.
The key observation is that many actionable UI elements exhibit clear visual boundaries (\eg icon contours, button outlines), which makes segmentation a stable primitive for zero-shot generalization when UI styles shift.

Concretely, given screenshot $I$, we apply training-free SAM2~\cite{ravi2024sam2} to generate a set of mask proposals $\{M_i\}$, which are converted into icon candidates with bounding boxes $\{b_i^{icon}\}$ after filtering tiny masks and suppressing near-duplicates (see Figure~\ref{fig:sam}).
In parallel, we run OCR to extract text spans and text boxes $\{(t_j, b_j^{txt})\}$.
The union of icon and text candidates forms a high-recall UI inventory that is passed to the semantic stage.

\begin{table}[]
\centering
\scriptsize
\caption{VLM vs Detector vs \sysname: element detection recall on screenspot-V2 datasets.}
\label{tab:Comparison-of-recall}
\begin{tabular}{|c|cc|cc|cc|c|}
\hline
\multirow{2}{*}{\textbf{Methods}} &
  \multicolumn{2}{c|}{\textbf{Mobile app}} &
  \multicolumn{2}{c|}{\textbf{Desktop}} &
  \multicolumn{2}{c|}{\textbf{Web}} &
  \multirow{2}{*}{\textbf{Average}} \\ \cline{2-7}
 &
  \multicolumn{1}{c|}{\textbf{Text}} &
  \textbf{Icon} &
  \multicolumn{1}{c|}{\textbf{Text}} &
  \textbf{Icon} &
  \multicolumn{1}{c|}{\textbf{Text}} &
  \textbf{Icon} &
   \\ \hline
\textbf{Qwen2.5VL-72B}~\cite{Qwen2.5-VL} &
  \multicolumn{1}{c|}{25.52\%} &
  9.95\% &
  \multicolumn{1}{c|}{32.47\%} &
  23.57\% &
  \multicolumn{1}{c|}{4.27\%} &
  4.43\% &
  16.50\% \\ \hline
\textbf{OmniParser-v2}~\cite{lu2024omniparser} &
  \multicolumn{1}{c|}{93.45\%} &
  85.31\% &
  \multicolumn{1}{c|}{91.75\%} &
  92.14\% &
  \multicolumn{1}{c|}{83.76\%} &
  80.79\% &
  87.90\% \\ \hline
\textbf{Ours} &
  \multicolumn{1}{c|}{\textbf{98.97\%}} &
  \textbf{91.47\%} &
  \multicolumn{1}{c|}{96.39\%} &
  85.71\% &
  \multicolumn{1}{c|}{86.32\%} &
  80.30\% &
  \textbf{90.57\%} \\ \hline
\end{tabular}
\end{table}

\tabref{tab:Comparison-of-recall} shows that our design achieves strong zero-shot generalization on ScreenSpot-V2~\cite{wu2024atlas}, a benchmark for visual element grounding in mobile app interfaces.
For example, on mobile apps, \sysname reaches 98.97\% text recall and 91.47\% icon recall without any dataset-specific tuning, outperforming the detector-based OmniParser-v2~\cite{lu2024omniparser} by 5.52\% (text) and 6.16\% (icon).
A general-purpose VLM (Qwen2.5VL-72B) attains only 25.52\% text recall and 9.95\% icon recall, indicating that model scale alone does not guarantee accurate perceptual grounding for mobile GUIs.
These results validate that our method mitigates a dominant failure mode of single-VLM GUI agents \cite{hong2024cogagent,cheng2024seeclick,qin2025uitar}, where small actionable elements are missed due to implicit parsing.

\subsection{Context-Aware Component-Level Semantic Analysis}
\label{sec:ui:context}
After reliable component localization, we resolve \textit{functional ambiguity} by combining textual cues with learned visual semantics.
We first associate each icon candidate $b_i^{icon}$ with a nearby text box using a distance-based heuristic.  The heuristic operates as follows: for each icon candidate, we identify the text block with the smallest pairwise distance, as measured by a minimum separation metric $d$, defined as the minimum Euclidean distance between the boundaries of the two bounding boxes:
\begin{equation}
d(b_i^{\text{icon}}, b_j^{\text{txt}}) = \min_{\substack{p \in \partial(b_i^{\text{icon}}) \\ q \in \partial(b_j^{\text{txt}})}} \| p - q \|_2
\end{equation}
where $\partial(\cdot)$ denotes the set of points on the boundary of a bounding box.
Let $j^*(i)=\arg\min_j d(b_i^{icon}, b_j^{txt})$ be the nearest text block.
If $d(b_i^{icon}, b_{j^*(i)}^{txt})<\tau$, where $\tau$ is a threshold set to 60 pixels based on empirical UI design conventions, we treat $t_{j^*(i)}$ as contextual metadata for the icon.
We then infer the function label via two complementary ways.
\begin{itemize}
    \item \textit{Text-guided inference.}
    When the nearby text is informative (\eg ``Favorites'', ``Share''), we directly infer the semantic meaning of the icon from its adjacent text label.
    This leverages a common UI design pattern where local text labels disambiguate visually similar icons.
    \item \textit{Visual-context inference.}
    When no reliable textual cue exists, we invoke a fine-tuned Florence-2-base model \cite{xiao2024florence} to predict actionable semantics from the icon crop together with its local layout context (\eg an expanded region around the icon).
    This yields a function label $\hat{y}_k$ with confidence $s_k$, complementing text-guided inference when UI labels are absent.
\end{itemize}
This dual strategy addresses two limitations in prior designs.
\textit{(i)} Single-VLM-based methods \cite{hong2024cogagent,cheng2024seeclick,qin2025uitar} mix localization and semantics, which can cause small elements to be ignored or misinterpreted.
\textit{(ii)} UI-tree-based methods \cite{wen2024autodroid, lee2024mobilegpt, dai2025advancing} lack sufficient context for stylized icons when metadata is missing or under-specified.

By explicitly constructing a high-recall UI inventory and then inferring functions with local context, the UI parsing agent provides the planner agent with a reliable UI state $\mathcal{U}$ for downstream decision-making.

\begin{figure}[t]
  \centering
  \includegraphics[width=0.68\textwidth]{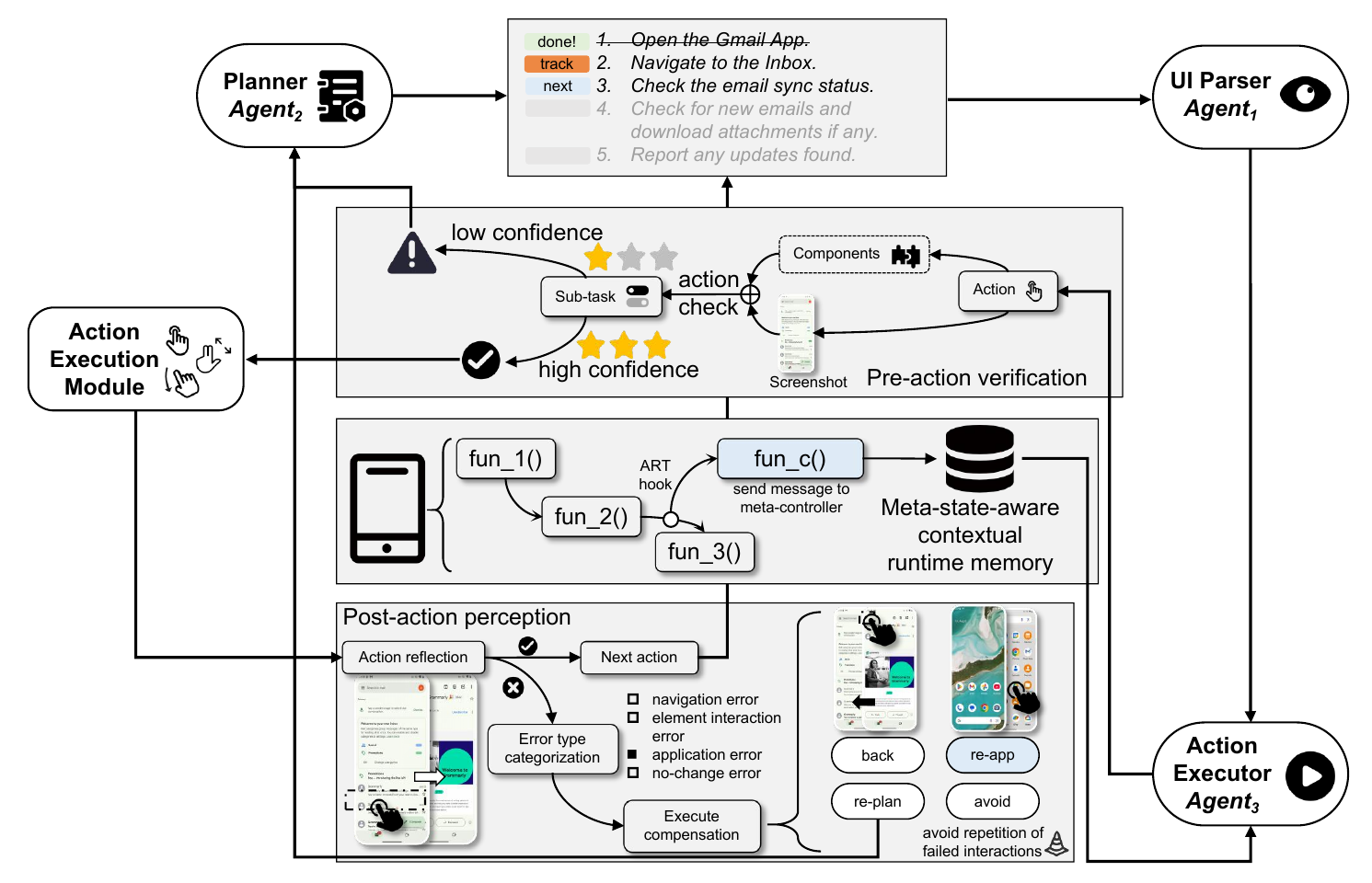}
  \caption{Illustration of the mechanism of Meta-Controller.}
  \vspace{-3mm}
  \label{fig:three-level}
\end{figure}

\section{META-CONTROLLER FOR PER-ACTION STATE PERCEPTION}
\label{sec:meta}

The meta-controller is the cross-agent coordinator of \sysname.
While agents handle \textit{what to do} (task planner), \textit{what is on screen} (UI parser), and \textit{how to do it} (action executor), the meta-controller ensures the system \textit{acts on the right state} and \textit{verifies what actually happened} at every step.
Concretely, it performs \textit{state perception} \textit{(i)} before each action to decide whether the current screen is ready, and \textit{(ii)} after each action to infer the outcome, detect anomalies, and trigger recovery when needed.

\fakeparagraph{Challenges}
Per-action state perception in mobile GUIs is difficult for two reasons.
\begin{itemize}
    \item \textbf{Action-state is partially observable from screenshots.}
    Many critical cues for deciding whether an action is feasible or successful are \textit{non-visual} (\eg foreground app, system time) or \textit{transient} (\eg toasts, short-lived pop-ups), and are easily missed by fixed-interval screen sampling.
    \item \textbf{Closed-loop control must be selective to remain efficient.}
    A naive solution is to re-plan before every action and re-parse after every action.
    However, frequent planner invocations are expensive, and over-triggering repairs can break task flow.
    Thus the controller must gate actions only when confidence is low, and  recover with targeted handlers rather than restarting the workflow.
\end{itemize}
Prior modular GUI automation frameworks, such as MobileGPT~\cite{lee2024mobilegpt}, treat error repair as a separate, user-invoked mode. This design interrupts task flow and fails to handle transient errors autonomously in a closed-loop manner. Other systems, like Mobile-Agent-v3~\cite{ye2025mobileagentv3}, attempt to mitigate the coordination problem by invoking the task planner before each action to re-determine the current sub-task. However, this design introduces significant computational overhead due to frequent planner invocations, undermining system efficiency. Moreover, existing approaches predominantly adopt an open-loop architecture~\cite{wen2024autodroid, wen2024autodroidv2, zhang2025appagent,cheng2024seeclick, liu2024autoglm}, which inherently lacks a mechanism to monitor execution validity in real time, adapt agent behavior based on runtime feedback, or autonomously trigger context-aware recovery. Our meta-controller addresses these limitations by providing a lightweight, closed-loop coordination mechanism with selective intervention.

\fakeparagraph{Our Methods}
We propose a lightweight, rule-augmented meta-controller which consists of \textit{(i)} a contextual runtime memory (\secref{sec:meta:state}) that continuously integrates \textit{device-level meta-state} and execution history, and \textit{(ii)} a per-action state perception loop (\secref{sec:meta:loop}) that performs \textit{pre-action} verification, \textit{post-action} outcome inference, and targeted \textit{error handling}.

\begin{figure}[t]
  \centering
  \includegraphics[width=0.98\textwidth]{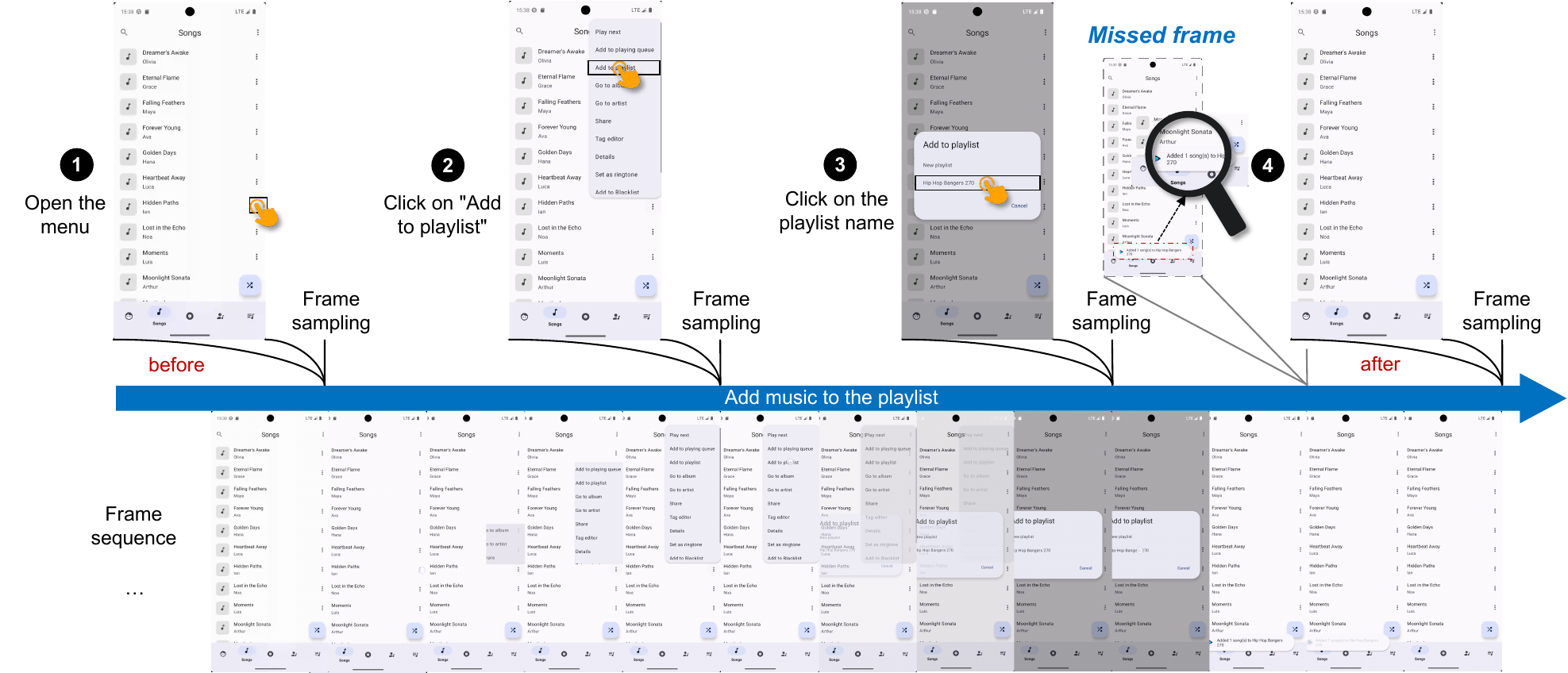}
  \caption{Illustration of missed critical frames due to fixed-interval sampling in VLM-based GUI agents.}
  \label{fig:missed-frame}
\end{figure}

\subsection{Meta-State-Aware Contextual Runtime Memory}
\label{sec:meta:state}
We define the \textbf{meta-state} $M_t$ as a real-time stream of contextual information beyond pixels, including persistent device state (\eg current foreground application, system time) and ephemeral interface signals (\eg toast notifications, transient pop-ups).

Access to $M_t$ is crucial for reliable action-state perception because visual frames captured only before/after an operation can be identical even when the action succeeded (\eg a success toast appears briefly and disappears), and time-dependent tasks (\eg ``next Thursday'') require accurate device time.
As illustrated in~\figref{fig:missed-frame}, VLM-based GUI agents typically capture the device state by sampling the frames at fixed intervals after actions. These intervals cannot be arbitrarily short as the UI often requires time to complete rendering and settle into a stable visual state. Such sampling often misses critical frames containing transient notifications (e.g., success toasts), which are vital for accurately assessing task progress. For example, sampled frames taken before (Frame-1) and after (Frame-4) adding a song to a playlist may appear identical, depriving the model of reliable cues to determine whether the addition succeeded.

\subsubsection{Meta-State Monitor Implementation}
To capture such ephemeral signals, we implement a \textbf{meta-state monitor} by leveraging the LSPosed framework~\cite{LSPosed_Framework} to hook key Android runtime (ART) methods in real-time. 
This allows us to intercept system-level events without modifying the target applications. 
Specifically, we hook:
\begin{itemize}
    \item \texttt{Toast.makeText()} to capture transient notification text and duration.
    \item \texttt{View.performClick()} to detect UI click events and extract view metadata.
    \item Additional hooks for system intents, activity lifecycle events, and permission dialogs.
\end{itemize}
These hooks are registered during the application loading phase (\texttt{XC\_LoadPackage}), ensuring comprehensive coverage of both system and application-level interactions. 
The intercepted meta-information (including event type, content, timestamp, and source application) is immediately serialized and streamed to the meta-controller agent via a lightweight IPC channel.

\subsubsection{Contextual Runtime Memory}
The collected meta-state feed is continuously integrated into a structured \textbf{contextual runtime memory} $H_c$, which maintains the controller’s execution context. 
As detailed in \tabref{tab:contextual_memory_structure}, $H_c$ is organized as a temporal sequence of memory entries, each containing:
\begin{itemize}
    \item \textbf{Timestamp} and \textbf{Action Type/Content}: When and what action was executed.
    \item \textbf{Screenshot Paths}: Visual records before and after execution.
    \item \textbf{Meta-state}: Runtime device and interface context.
    \item \textbf{Execution Context}: Task progress and action history.
\end{itemize}
By unifying the real-time meta-state with execution history, $H_c$ provides a coherent view of the evolving environment and supports consistent decision-making between agents. 
This design enables the system to reason about events that are visually imperceptible but semantically critical for task progress assessment.

\begin{table}[t]
\centering
\scriptsize
\caption{Structure of the Contextual Runtime Memory ($H_c$).}
\label{tab:contextual_memory_structure}
\begin{tabular}{|p{15mm}| p{18mm} |p{75mm}|}
\hline
\textbf{Field} & \textbf{Data Type} & \textbf{Description} \\
\hline
Timestamp & \texttt{DateTime} & The precise system time when the action was executed. Provides temporal ordering for memory entries. \\ \hline
Action Type & \texttt{String (Enum)} & Type of the executed action. \eg \texttt{CLICK} (click at coordinates), \texttt{TYPE} (type text into input), \texttt{SWIPE} (swipe between coordinates). \\ \hline
Action Content & \texttt{Serializable Object} & Specific parameters of the executed action. \eg \texttt{CLICK}: \{"coordinate": [x, y]\}, \texttt{TYPE}: \{"text": "example", "clear\_text": true\}. \\ \hline
Screenshot Paths & \texttt{List<String>} & File paths of screenshots capturing UI states: pre-action (state before execution) and post-action (state after execution). \\ \hline
Meta-state & \texttt{Context Object} & Real-time contextual information, including persistent device state (e.g., current foreground application, system time) and ephemeral interface signals (e.g., toast notifications, pop-up dialogs). \\ \hline
Execution Context & \texttt{Context Object} & Structured snapshot containing: current sub-task goal, recent action history with outcomes, encountered error states and recovery attempts. \\
\hline
\end{tabular}
\end{table}

\subsection{Per-Action State Perception and Closed-Loop Control}
\label{sec:meta:loop}

We use \textbf{per-action state perception} to decide \textit{when to act} and \textit{how to react}.
For each planned step, the controller runs a pre-action verification stage (\secref{sec:meta:loop:pre}) and a post-action perception stage (\secref{sec:meta:loop:post}), and invokes a targeted handler (\secref{sec:meta:loop:error}) when anomalies are detected.

\subsubsection{Pre-Action Verification via Grounding Confidence}
\label{sec:meta:loop:pre}
Before executing any planned action, the controller estimates whether the executor can ground the prescribed sub-task to the current interface with high confidence.

We compute a grounding confidence score $C_{\text{ground}}\in[0,1]$ as a weighted combination of semantic alignment and contextual feasibility:
\begin{equation}
C_{\text{ground}}=\alpha\cdot S_{\text{sem}}(d_{\text{plan}},d_{\text{UI}})+(1-\alpha)\cdot F_{\text{ctx}}(M_t),
\end{equation}
where $S_{\text{sem}}$ measures the semantic similarity between the planner's target description $d_{\text{plan}}$ and the textual attributes $d_{\text{UI}}$ of candidate UI components extracted by the UI parser (see \secref{sec:ui}). Specifically, we compute the cosine similarity between their corresponding embeddings generated by a lightweight pre-trained sentence transformer (\eg all-MiniLM-L6-v2~\cite{wang2020minilm}). 
$F_{\text{ctx}}(M_t)\in[0,1]$ encodes context feasibility derived from the real-time meta-state $M_t$.
It is a heuristic function that evaluates factors such as the visibility of target UI elements (via accessibility tree), the absence of blocking overlays (\eg permission dialogs detected via the meta-state monitor), and navigational readiness (\eg whether the correct application screen is foreground). 
The weighting factor $\alpha$ is empirically set to 0.6 to balance the contributions of semantic and contextual signals.

We empirically set thresholds $\tau_{\text{low}}=0.3$ and $\tau_{\text{high}}=0.7$ based on an ablation study detailed in \secref{sec:evaluation}.
If $C_{\text{ground}}\ge\tau_{\text{high}}$, execution proceeds with high certainty.
If $C_{\text{ground}}\le\tau_{\text{low}}$, the controller triggers on-demand re-planning by supplying the current $M_t$ and relevant excerpts from $H_c$ to the planner. 
When the score falls within the intermediate range $(\tau_{\text{low}}, \tau_{\text{high}})$, the executor proceeds but logs the action for potential future verification. 
This two-threshold mechanism prevents the execution of low-certainty actions, which often propagate errors, while avoiding the computational overhead of re-planning at every step.

\subsubsection{Post-Action Outcome Perception via State Differencing}
\label{sec:meta:loop:post}
After executing an action, the controller performs a differential analysis between the \textit{pre-action} state $S_{t-1}$ and the \textit{post-action} state $S_t$, supplemented by meta-state transition $\Delta M=M_t-M_{t-1}$.
It classifies the action outcome into one of four categories $O\in\{O_1,O_2,O_3,O_4\}$:
\[
O=
\begin{cases}
O_1, & \text{action successful and sub-task completed},\\
O_2, & \text{action successful but sub-task incomplete},\\
O_3, & \text{erroneous action leading to an error page/application},\\
O_4, & \text{no detectable interface change}.
\end{cases}
\]
This outcome classification is produced by the \textit{action reflector}, a dedicated module that leverages a multi-modal model. The reflector implements the decision function $f$ by constructing a comparative prompt that encapsulates the observed changes and contextual information, and then calls a VLM to obtain the outcome classification:
\begin{equation}
O=f\Big(\phi_{\text{visual}}(S_{t-1},S_t),\ \phi_{\text{textual}}(S_{t-1},S_t),\ \Delta M,\ H_c\Big),
\end{equation}
where $\phi_{\text{visual}}$ captures pixel/structure changes, $\phi_{\text{textual}}$ compares semantic content (OCR text and UI labels), $\Delta M$ captures meta-state transitions, and $H_c$ provides task-progress context.
The controller updates $H_c$ with the pre/post observations and the inferred $O$.
If $O\in\{O_1,O_2\}$, execution continues (advance the sub-task for $O_1$, or continue within the sub-task for $O_2$).
If $O\in\{O_3,O_4\}$, the controller invokes targeted recovery.

\subsubsection{Targeted Error Handling}
\label{sec:meta:loop:error}
When $O\in\{O_3,O_4\}$, the controller refines diagnosis by inferring an error type:
\begin{equation}
E_{\text{type}}=\arg\max_{e\in E_{\text{set}}}P\big(e\mid O,\phi_{\text{visual}},\phi_{\text{textual}},\Delta M,H_c\big),
\end{equation}
where $E_{\text{set}}$ includes error types such as \textit{navigation error}, \textit{element interaction error}, \textit{application error}, and \textit{no-change error}.
Based on $E_{\text{type}}$ and context in $H_c$, the controller selects a recovery policy $\pi_{\text{rec}}$ from a small set $\Pi_{\text{rec}}$:
\[
\pi_{\text{rec}}=g(E_{\text{type}},H_c)\in\Pi_{\text{rec}}=\{\pi_{\text{back}},\pi_{\text{replan}},\pi_{\text{reapp}},\pi_{\text{avoid}}\},
\]
where:
$\pi_{\text{back}}$ performs system back navigation;
$\pi_{\text{redecide}}$ re-selects the current action from candidates, marking the previous as erroneous;
$\pi_{\text{reapp}}$ re-launches or switches to the target application;
$\pi_{\text{avoid}}$ re-decides while explicitly avoiding repetition of failed interactions.
In particular, for $\pi_{\text{redecide}}$ and $\pi_{\text{avoid}}$, the executor re-evaluates the candidate set $A_{\text{candidate}}$ under updated constraints:
\[
A_{\text{valid}}=\{a\in A_{\text{candidate}}\mid a\neq a_{\text{failed}},\ \text{feasibility}(a,S_t,H_c)>\theta\},
\]
and selects the highest-confidence action from $A_{\text{valid}}$.
All recovery steps are guided by $H_c$ to maintain task continuity without restarting the workflow.

\section{MULTI-AGENT IMPLEMENTATION}
\label{sec:implementation}
\begin{figure}[t]
  \centering
  \includegraphics[width=0.75\textwidth]{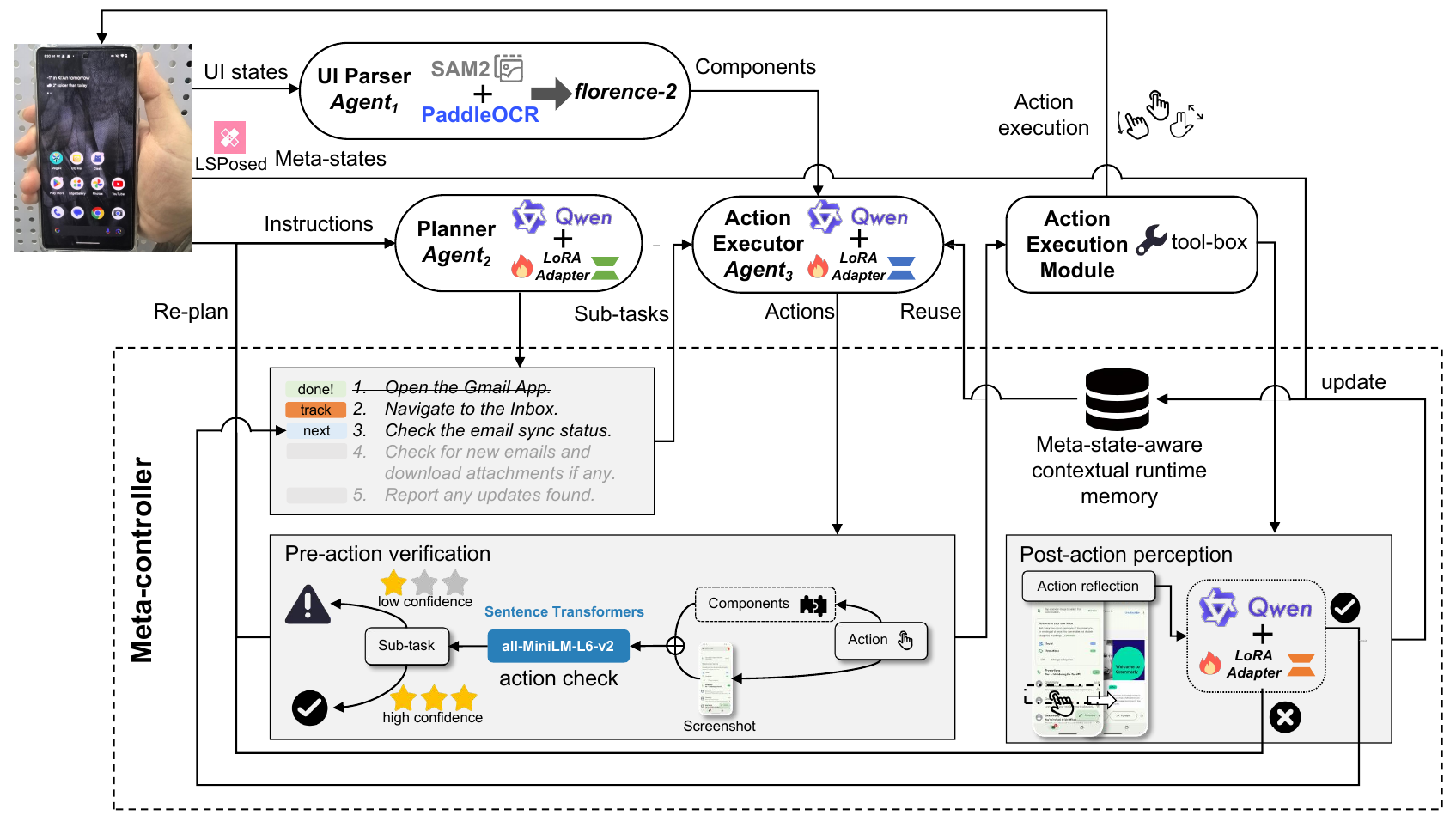}
  \caption{Implementation of \sysname: system integration, data flow, and model deployment in the multi-agent framework.}
  \label{fig:implement}
\end{figure}

This section describes the implementation of \sysname (see \figref{fig:implement}), including our Multi-LoRA agent instantiation (\secref{sec:implementation:multi}), the training pipeline for role-specific adapters (\secref{sec:implementation:training}), and key deployment details for execution and evaluation (\secref{sec:implementation:deployment}).

\subsection{Multi-LoRA Specialized Agent Implementation}
\label{sec:implementation:multi}
\begin{figure}[t]
  \centering
  \includegraphics[width=0.5\textwidth]{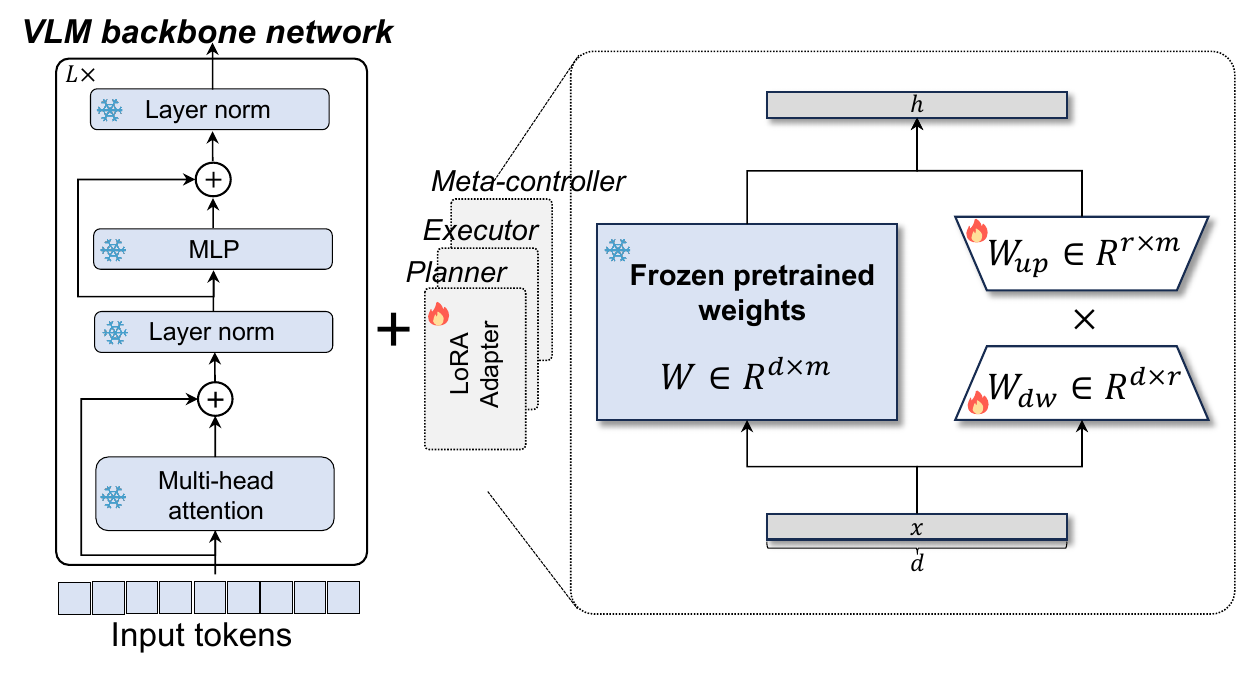}
  \caption{Illustration of the specialized multi-agent (\ie Planner, Executor, Meta-controller), which is implemented by fine-tuning lightweight LoRA adapters on a shared VLM backbone.}
  \vspace{-3mm}
  \label{fig:lora}
\end{figure}

We implement \sysname as a modular multi-agent system with four components: a \textbf{UI Parser}, a \textbf{Planner}, an \textbf{Executor}, and a \textbf{Meta-Controller}.
These agents follow two implementation styles that correspond to their roles.
The \textbf{UI Parser} is implemented as a lightweight perception pipeline to maximize recall and structured UI extraction.
The remaining three agents (\textbf{Planner/Executor/Meta-Controller}) require flexible reasoning over screenshots and UI states, and we implement them using a shared Large Vision-Language Model (LVLM) backbone with \textbf{Multi-LoRA} specialization for efficient deployment.
Specifically, as shown in \figref{fig:lora}, the Planner, Executor, and Meta-Controller share a single frozen LVLM backbone (\texttt{Qwen3-VL-4B-Instruct} \cite{qwen3technicalreport}), while each role is instantiated by activating a small, task-specific LoRA adapter~\cite{hu2022lora}.
At runtime, we switch roles by loading only the corresponding adapter, avoiding multiple full model instances.
We inject low-rank updates into the transformer linear layers while keeping the visual encoder frozen, ensuring consistent visual grounding across agents.
Unless otherwise specified, we use rank $r{=}16$ and scaling $\alpha{=}32$.

\begin{itemize}
    \item \textbf{UI Parser (perception pipeline).}
    Given a screenshot, the UI Parser segments UI elements using SAM2~\cite{ravi2024sam2}, extracts on-screen text via PaddleOCR~\cite{cui2025paddleocr30technicalreport}, and associates nearby icon-text pairs by spatial proximity.
    It then uses a fine-tuned Florence-2 model~\cite{xiao2024florence} to produce concise semantic labels (\eg ``send button''), yielding a structured UI component list (bounding boxes, OCR text, types, and labels) for downstream grounding.
    \item \textbf{Planner (LoRA adapter on \texttt{Qwen3-VL-4B-Instruct}~\cite{qwen3technicalreport}.}
    The Planner converts a user instruction into an ordered sub-task sequence.
    \item \textbf{Executor (LoRA adapter on \texttt{Qwen3-VL-4B-Instruct}~\cite{qwen3technicalreport}).}
    The Executor grounds each sub-task into a concrete UI action (\eg \texttt{click}, \texttt{input\_text}, \texttt{scroll}) and its target component, conditioned on the parsed UI state and runtime memory.
    \item
    \textbf{Meta-Controller (hybrid verifier).}
    The Meta-Controller improves reliability through pre-action verification and post-action outcome perception.
    It uses a lightweight verifier (\texttt{all-MiniLM-L6-v2}~\cite{wang2020minilm}) for pre-action confidence estimation, and a LoRA-specialized \texttt{Qwen3-VL-4B-Instruct}~\cite{qwen3technicalreport} verifier for post-action perception.
\end{itemize}

\subsection{Training Pipeline for Specialized Agents}
\label{sec:implementation:training}
We train each LoRA-based agent (Planner/Executor/Meta-Controller) using a three-stage pipeline inspired by prior mobile GUI agent training recipes (\eg MobileRL~\cite{xu2025mobilerl}, Mobile-Agent-v3~\cite{ye2025mobileagentv3}, Mobile-R1~\cite{gu2025mobile_R1}).
The UI parser, based on pre-trained models, remains fixed and is not fine-tuned throughout the training process.
\begin{itemize}
    \item \textit{Stage 1: Action-Only SFT.}
    We first fine-tune each agent’s LoRA adapter on expert action trajectories (20k steps across 32 apps, see Section~\ref{sec:implementation:datasets} for dataset details) without reasoning traces.
    This stage teaches basic mobile GUI interaction patterns and substantially reduces exploration cost for subsequent reinforcement learning.
    \item \textit{Stage 2: Reasoning-Augmented SFT.}
    We then enrich the training corpus with intermediate reasoning steps generated by a strong instruction-following model through iterative refinement  (see Section~\ref{sec:implementation:datasets} for the construction process).
    Each agent is further fine-tuned on the enriched data with role emphasis: the Planner focuses on subgoal decomposition and structured plans, while the Executor focuses on action grounding and state transitions.
    \item \textit{Stage 3: GRPO with Failure-Aware Curriculum.}
    Building on the SFT-initialized adapters, we perform online multi-turn reinforcement learning using Group Relative Policy Optimization (GRPO)~\cite{shao2024deepseekmath} with curriculum learning.
    We implement the curriculum via \textit{Failure Curriculum Filtering (FCF)} ~\cite{bengio2009curriculum,xu2025mobilerl,ye2025mobileagentv3}, which dynamically re-weights task sampling based on per-task failure statistics, \ie tasks that consistently yield zero reward are gradually down-weighted to avoid wasteful exploration and stabilize training.
\end{itemize}

\subsection{Training and Evaluation Datasets}
\label{sec:implementation:datasets}

We split the dataset into disjoint training and evaluation sets, with no overlap in applications or instantiated task descriptions. The training set covers 32 applications, including the 20 AndroidWorld apps and additional real-world mobile apps.
Following AndroidWorld, we instantiate task descriptions by randomizing entities, input contents, and field values from shared templates.

\textbf{Training Set Construction:} We collect over 20k action trajectories from 200 tasks across 32 applications using GPT-4o and other state-of-the-art models.
Human annotators review the collected samples to remove failed, ambiguous, or inconsistent trajectories and correct obvious annotation errors.
The training data includes three components:
\begin{itemize}
\item \textit{Action Trajectories:} The foundational data comprises raw, multi-step action sequences (\eg tap, type, swipe) paired with corresponding screen states and task instructions.
\item \textit{Plan Data:} To train the Planner agent, we employ GPT-4o to analyze the temporal and logical relevance within action sequences, summarizing them into structured, high-level subgoal plans. These generated plans undergo rigorous manual review to ensure correctness and coherence.
\item \textit{Reflect \& Action Data:} To train the Executor and Meta-Controller, we first manually annotate the Reflect data based on the generated plans and current execution state. The action-level training data is then constructed by determining the appropriate atomic action for the current sub-task, guided by the outcome of the reflection process.
\end{itemize}

\textbf{Testing Set:} The testing set is specifically designed to assess the agent's capability in meta-status-aware scenarios. 
It comprises 51 tasks spanning 9 applications across 5 categories: \textit{Writing, Science Education, Food, Office,} and \textit{Health}. These 9 applications are held out from training and do not overlap with the 32 training applications. The task types are diverse and challenging, including: handling transient system \textit{toasts}, performing multi-field \textit{data entry}, making dynamic \textit{state judgments}, completing \textit{compositional tasks}, and operating within \textit{complex UI layouts} that require advanced visual understanding. Although the evaluation tasks are designed to be similar in task type to the training set (e.g., \textit{data entry}, \textit{edit}, and \textit{delete}), their concrete task descriptions are independently instantiated with different entity names and field values, and do not overlap with those used in training. Each evaluation task is manually reviewed and validated to ensure that the goal condition is executable and unambiguous on the target app version.

\subsection{Prototype Deployment}
\label{sec:implementation:deployment}
We implement \sysname as a hybrid on-device and edge/cloud-assisted system. 
On-device inference runs via \texttt{llama.cpp}~\cite{llama_cpp_2024} with 8-bit quantization to meet latency and memory constraints, while mobile UI interactions are executed through a Python-based automation library that exposes action APIs and UI querying utilities.
\begin{itemize}
    \item \textbf{Training and Serving.}
    We train all LoRA adapters using the \texttt{Swift} framework~\cite{zhao2024swiftascalablelightweightinfrastructure} and serve models with \texttt{vLLM}~\cite{kwon2023efficient_vllm} for high-throughput inference with dynamic batching.
    \item \textbf{Devices and Infrastructure.}
    We test \sysname on physical devices (Google Pixel~7, Xiaomi~17~Ultra, Huawei Mate~70~Pro) and Android Virtual Device (AVD) emulators to cover diverse hardware/software configurations.
    We use an NVIDIA Jetson AGX Orin for edge assistance and an NVIDIA RTX~3090 server (with 2$\times$3090 GPUs) for cloud assistance.
    Training is conducted on a server with 8$\times$NVIDIA RTX~3090 GPUs and 4$\times$NVIDIA L20 GPUs.
\end{itemize}
\section{EVALUATION}
\label{sec:evaluation}

\subsection{Experiment Setup}
\label{sec:experiment:setup}
 
\textbf{Tasks and Datasets.} We evaluate \sysname on four complementary benchmarks that cover dynamic interaction, long-horizon workflows, cross-platform grounding, and real-world mobile complexity:

\begin{itemize}
    \item \textbf{AndroidWorld}~\cite{rawles2024androidworld} evaluates end-to-end interactive execution under dynamic UI states, with 116 tasks across 20 apps and millions of randomized task variants.

    \item \textbf{MobileWorld} \cite{kong2025mobileworld} is a challenging mobile-use benchmark with 201 cross-app tasks across 20 applications, featuring agent-user interaction and MCP-augmented hybrid tool-use scenarios.

    \item \textbf{ScreenSpot-v2}~\cite{wu2024atlas} focuses on cross-platform GUI grounding, containing 1.2k re-annotated tasks spanning mobile, desktop, and web interfaces.

    \item \textbf{MMBench-GUI}~\cite{wang2025mmbenchgui} is a multimodal benchmark for evaluating AI agents' understanding and reasoning on graphical user interfaces (GUIs). It assesses both static screen comprehension and dynamic task execution across mobile, desktop, and web platforms.

    \item \textbf{Self-collected MobileTask Dataset} targets practical mobile edge cases, including complex multi-layer interfaces, modal dialogs, toast pop-ups, and dynamic UI changes, with 200 tasks across 30+ apps.
\end{itemize}

\textbf{Baselines.}
To ensure a comprehensive and fair comparison, we select representative baselines that cover the mainstream approaches to screen understanding, action generation, and execution control.

\begin{itemize}
    \item \textbf{Holistic VLM-driven GUI Agents} directly leverage VLM to interpret screenshots and generate UI actions in an end-to-end manner. They emphasize strong visual grounding and generalization.
    \begin{itemize}
        \item \textbf{UI-TARS}~\cite{qin2025uitar}: a fine-tuned VLM designed for end-to-end GUI interaction.
        \item \textbf{GUI-Owl}~\cite{ye2025mobileagentv3}: a foundational GUI agent model (7B/32B) built upon Qwen2.5-VL.
        \item \textbf{MAI-UI}~\cite{zhou2025maiuitechnicalreportrealworld}: a family of foundation GUI agents (2B/8B/32B) built upon Qwen3-VL, with native support for agent-user interaction and MCP-style tool calls.
    \end{itemize}

    \item \textbf{Step-wise GUI Agents} introduce structured intermediate representations or multi-step decision policies to improve execution stability and long-horizon reasoning.
    They explicitly model state transitions or verification but typically rely on predefined structures or UI representations.
    \begin{itemize}
        \item \textbf{AutoDroid}~\cite{wen2024autodroid}: a memory-augmented agent with step-wise policies and explicit state tracking.
        \item \textbf{MobileUse}~\cite{li2025mobileuse}: employs hierarchical reflection for autonomous mobile operation.
        \item \textbf{V-Droid}~\cite{dai2025advancing}: a verifier-driven approach that parses XML-based UI representations.
        \item \textbf{Mobile-Agent-v3}~\cite{ye2025mobileagentv3}: a multi-agent framework endowed with capabilities for knowledge evolution, task planning, sub-task execution, and reflective reasoning.
    \end{itemize}
\end{itemize}

\textbf{Metrics.}
We evaluate the GUI agent performance along complementary dimensions of effectiveness, efficiency, and on-device cost:

\begin{itemize}
    \item \textbf{Success Rate (SR)} measures end-to-end task completion and serves as the primary effectiveness metric.
    Task completion follows the criteria of DroidTask and LlamaTouch, requiring correct action subsequences or state transitions.

    \item \textbf{Action Type Accuracy} evaluates the precision of high-level action category prediction by comparing the predicted action type (\eg `click`, `swipe`, `input\_text`) against the ground-truth type for each step. 

    \item \textbf{Action Accuracy} measures the exact match of both the action type and its associated parameters (e.g., coordinates, text content, direction) with the ground truth.

    \item \textbf{LLM Inference Latency} captures end-to-end response time, reflecting real-time usability in on-device settings.

    \item \textbf{Token Consumption} measures computational cost and scalability by counting input and output tokens.

\end{itemize}

\begin{table}[]
\centering
\scriptsize
\caption{Performance comparison of \sysname and baselines across task complexity levels (L1-L5). The “Meta-state” denotes contextual signals including foreground app, system time, and LSPosed-hooked toast notifications and transient pop-ups. “-” denotes undisclosed (typically very large) model sizes.}
\label{tab:performance_comparison_L1_L5}
\begin{tabular}{|l|l|c|c|c|c|c|c|c|c|c|}
\hline
\textbf{} &
  \textbf{Methods} &
  \multicolumn{1}{l|}{\textbf{Size}} &
  \textbf{Screenshot} &
  \textbf{Ally tree} &
  \textbf{Meta-state} &
  \textbf{L1} &
  \textbf{L2} &
  \textbf{L3} &
  \textbf{L4} &
  \textbf{L5} \\ \hline
\multirow{4}{*}{Models} &
  Claude-Sonnet-4-20250514-thinking~\cite{anthropic2025claude4} &
  - &
  \cmark &
  \cmark &
  \xmark &
  45.8\% &
  44.9\% &
  13.0\% &
  20.4\% &
  0.0\% \\ \cline{2-11} 
 &
  GPT-4o-2024-11-20~\cite{hurst2024gpt} &
  - &
  \cmark &
  \cmark &
  \xmark &
  37.5\% &
  46.9\% &
  16.7\% &
  20.4\% &
  7.7\% \\ \cline{2-11} 
 &
  Qwen3-VL-4B-Instruct~\cite{qwen3technicalreport} &
  4B &
  \cmark &
  \xmark &
  \xmark &
  66.7\% &
  59.2\% &
  31.5\% &
  5.6\% &
  0.0\% \\ \cline{2-11} 
 &
  Qwen3-VL-2B-Instruct~\cite{qwen3technicalreport} &
  2B &
  \cmark &
  \xmark &
  \xmark &
  62.5\% &
  50.0\% &
  12.3\% &
  3.7\% &
  0.0\% \\ \hline
\multirow{2}{*}{Agents} &
  Mobile-Agent-v3 (GUI-Owl-7B)~\cite{ye2025mobileagentv3} &
  7B &
  \cmark &
  \xmark &
  \xmark &
  75.0\% &
  87.8\% &
  49.6\% &
  35.2\% &
  0.0\% \\ \cline{2-11} 
 &
  UIAnchor (Qwen3-VL-4B-Instruct-grpo) &
  \textbf{4B} &
  \cmark &
  \xmark &
  \cmark &
  \textbf{87.5\%} &
  \textbf{87.8\%} &
  \textbf{52.8\%} &
  \textbf{51.9\%} &
  3.8\% \\ \hline
\end{tabular}
\end{table}

\subsection{Main Result}
\label{Success_Rate}
\subsubsection{Performance Across Complexity Levels (L1-L5)}
\label{sec_task_level}
We evaluate UIAnchor and baselines under five progressive task-complexity levels (L1--L5), defined by three factors:
\textbf{task length} (average execution steps), \textbf{apps} involved, and \textbf{UI perception granularity} (pixel size of the critical UI component at the failure step for unsuccessful tasks).
\noindent\textbf{Level definitions.}
\begin{itemize}
    \item \textbf{L1}: $<10$ steps, single app, large critical components ($>100\times100$ pixel), clear semantics, no distractors.
    \item \textbf{L2}: $<10$ steps, single app, large components ($>100\times100$ pixel), with distractors.
    \item \textbf{L3}: $10-20$ steps, single app, small components ($<100\times100$ pixel), with distractors.
    \item \textbf{L4}: $20-30$ steps, multi-app, small components ($<100\times100$ pixel), with distractors.
    \item \textbf{L5}: $\geq30$ steps, multi-app, small components ($<100\times100$ pixel), with distractors.
\end{itemize}
On a typical smartphone screen, 100 pixel corresponds to roughly \textbf{mm-level} in physical size. Both AndroidWorld (L1--L5: 20, 42, 31, 14, 9; total 116) and our MobileTask testing dataset (L1--L5: 4, 7, 23, 13, 4; total 51) are partitioned accordingly, and aggregated success rates are reported in \tabref{tab:performance_comparison_L1_L5}.
\textit{First}, \sysname is robust across practical levels.
    \sysname achieves the best results in four of five levels: 87.5\% (L1), 87.8\% (L2, tied with Mobile-Agent-v3), 52.8\% (L3), and 51.9\% (L4).
    The gains at L3/L4 highlight the effect of our \textbf{UI parser} for small/occluded targets and the \textbf{meta-controller} for long-horizon planning and cross-app transitions.
\textit{Second}, standalone lightweight VLMs fail to scale with horizon and app transitions.
    Qwen3-VL-4B-Instruct remains competitive on L1-L3, but drops sharply on L4 (5.6\%) and L5 (0.0\%),
    indicating that fine-grained UI grounding and long-horizon state management require dedicated agent modules beyond a single VLM.
\textit{Third}, extreme long-horizon tasks remain hard for current LLMs.
    On L5 ($\geq$30 steps), UIAnchor reaches 3.8\% and even GPT-4o only 7.7\%, while other lightweight VLMs drop to 0\%.
    Despite this, UIAnchor outperforms GPT-4o on L1/L3/L4 and surpasses Mobile-Agent-v3 (GUI-Owl-7B) by +16.7\% on L4.
\textbf{Summary}. L5 remains a hard bottleneck for today’s V/LLM modes and agents.
On L4, the hardest attainable tier aligned with users' service-composed usage trends, \sysname improves success by \textbf{31.5\%} over GPT-4o and \textbf{16.6\%} over Mobile-Agent-v3, showing that a meta-controller plus a fine-grained UI parser scales better than brute-force scaling.

\begin{figure}[t]
  \centering
  \subfloat[AndroidWorld testing
    \label{fig:bound_plot}]{
    \includegraphics[width=0.35\textwidth]{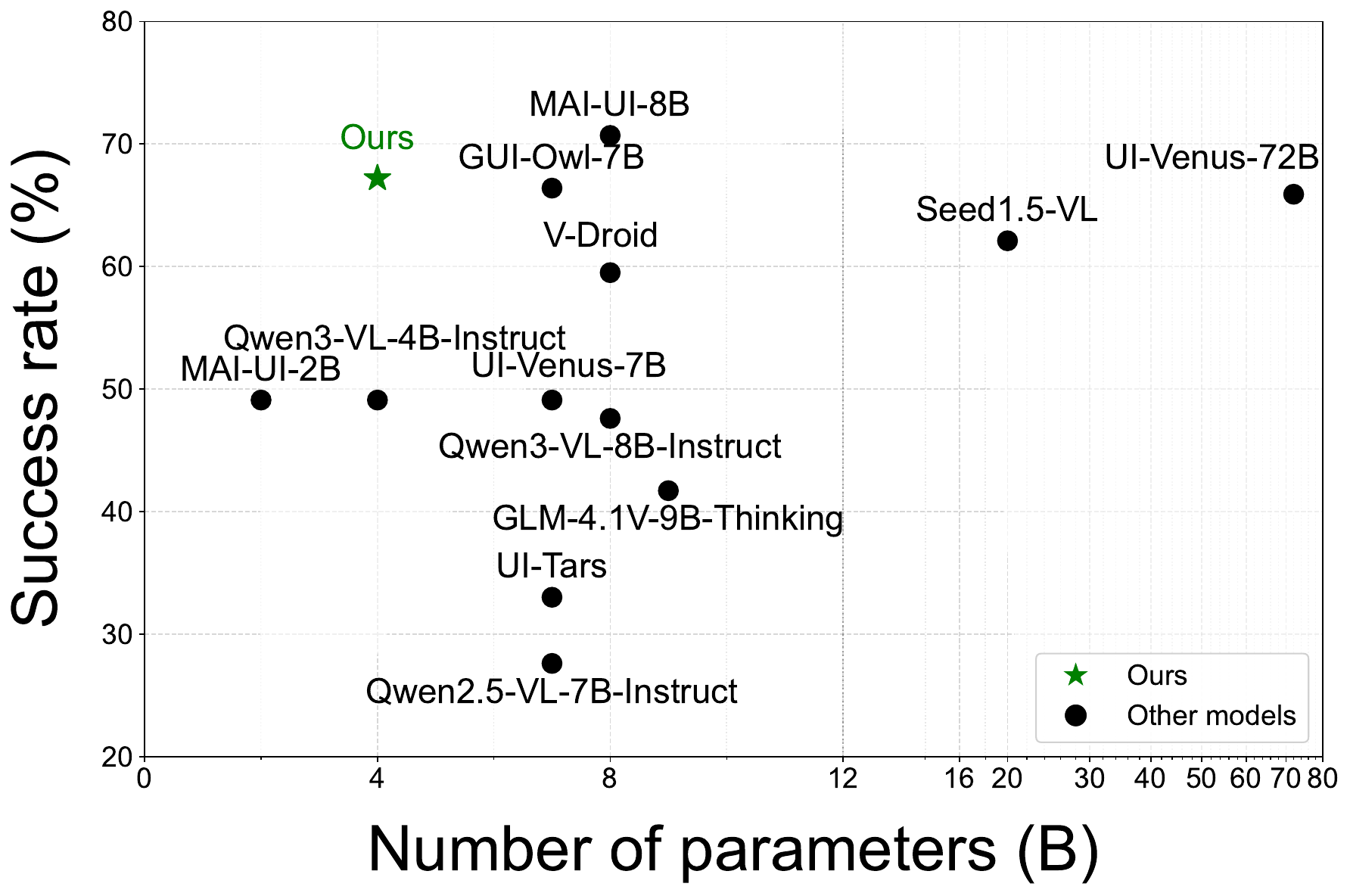}
  }
  \hspace{4mm}
  \subfloat[Self-collected MobileTask testing
    \label{fig:self_bound_plot}]{
    \includegraphics[width=0.35\textwidth]{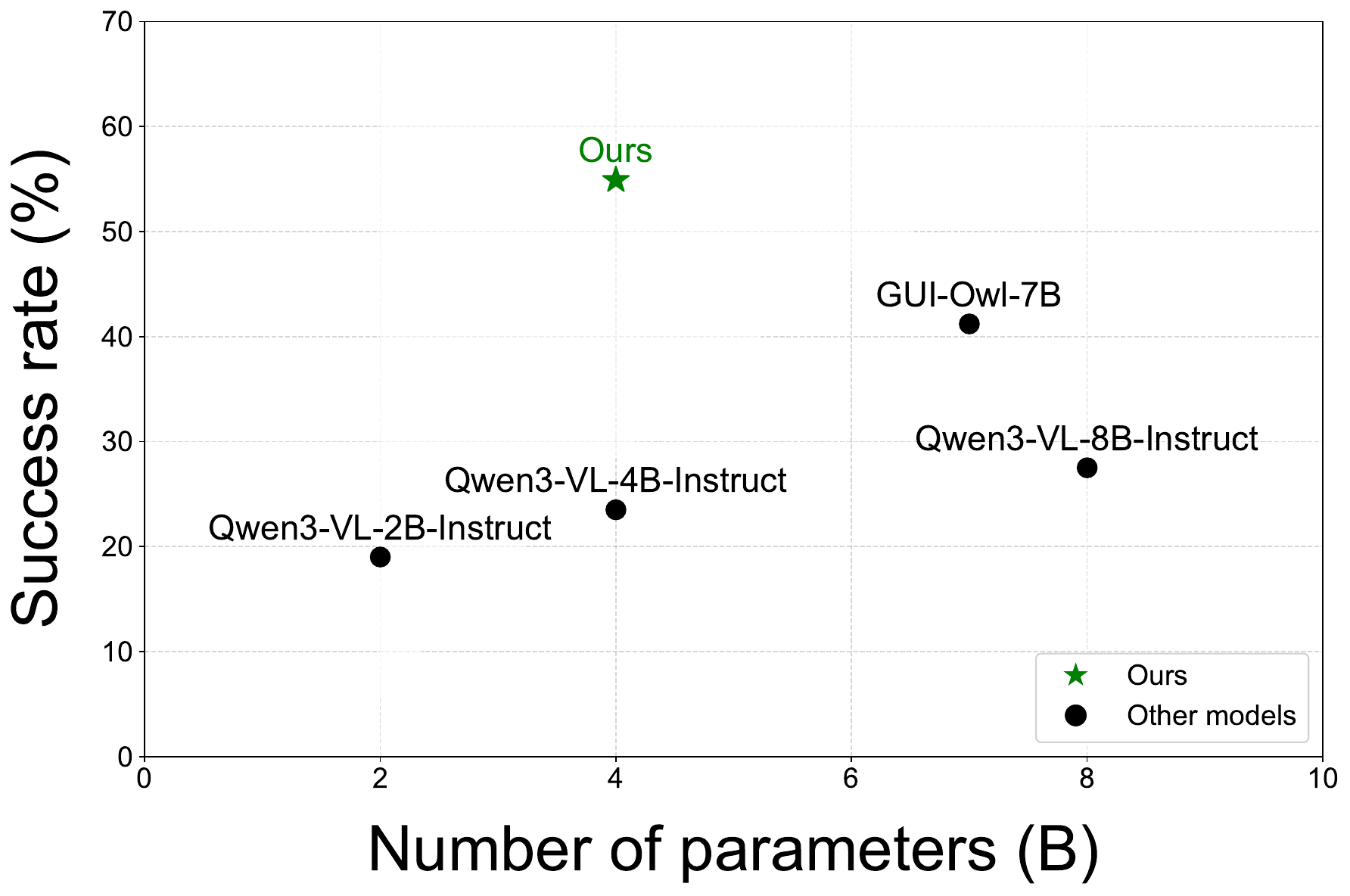}
  }
  \caption{Comparison of \sysname and baselines in terms of success rate-parameter size trade-off. }
  \label{fig:bound_plot_all}
\end{figure}

\begin{table}[]
\centering
\caption{Success rates (\%) on AndroidWorld (mobile GUI tasks). The “Meta-state” denotes contextual signals including foreground app, system time, and LSPosed-hooked toast notifications and transient pop-ups. “-” denotes undisclosed (typically very large) model sizes. Notably, AndroidWorld contains many challenging tasks (\eg multi-step data entry, pop-ups/detours), where even careful human execution achieves only $\sim$80\% success under the benchmark protocol. AndroidWorld spans \textbf{L1-L5} complexity levels.}
\label{tab:SR-models}
\resizebox{0.98\linewidth}{!}{%
\begin{tabular}{|l|l|c|c|c|c|c|c|}
\hline
 &
  \textbf{Methods} &
  \textbf{Size} &
  \textbf{Screenshot} &
  \textbf{Ally tree} &
  \multicolumn{1}{l|}{\textbf{Meta-state}} &
  \textbf{Mobile-deployable} &
  \textbf{Success rates (\%)} \\ \hline
                         & Human                                                         & -   & \cmark & \xmark & \cmark & -      & 80.0          \\ \hline
\multirow{15}{*}{Models} & GPT-4o-2024-11-20~\cite{hurst2024gpt}                         & -   & \cmark & \xmark & \xmark & \xmark & 34.5          \\ \cline{2-8} 
                         & Claude-Sonnet-4-20250514-thinking~\cite{anthropic2025claude4} & -   & \cmark & \xmark & \xmark & \xmark & 41.0          \\ \cline{2-8} 
                         & UI-Tars-1.5~\cite{qin2025uitar}                               & -   & \cmark & \xmark & \xmark & \xmark & 64.2          \\ \cline{2-8} 
                         & UI-Venus-72B~\cite{gu2025uivenus}                             & 72B & \cmark & \xmark & \xmark & \xmark & 65.9          \\ \cline{2-8} 
                         & Seed1.5-VL~\cite{guo2025seed1}                                                    & 20B & \cmark & \cmark & \xmark & \xmark & 62.1          \\ \cline{2-8} 
                         & GLM-4.1V-9B-Thinking~\cite{hong2025glm}                       & 9B  & \cmark & \cmark & \xmark & \xmark & 41.7          \\ \cline{2-8} 
                         & Qwen3-VL-8B-Instruct~\cite{qwen3technicalreport}              & 8B  & \cmark & \xmark & \xmark & \xmark & 47.6          \\ \cline{2-8} 
                         & MAI-UI~\cite{zhou2025maiuitechnicalreportrealworld}                                                        & 8B  & \cmark & \xmark & \xmark & \xmark & \textbf{70.7} \\ \cline{2-8} 
                         & GUI-Owl-7B~\cite{ye2025mobileagentv3}                         & 7B  & \cmark & \xmark & \xmark & \cmark & 66.4          \\ \cline{2-8} 
                         & UI-Venus-7B~\cite{gu2025uivenus}                              & 7B  & \cmark & \xmark & \xmark & \cmark & 49.1          \\ \cline{2-8} 
                         & UI-Tars-7B~\cite{qin2025uitar}                                & 7B  & \cmark & \xmark & \xmark & \cmark & 33.0          \\ \cline{2-8} 
                         & Qwen2.5-VL-7B-Instruct~\cite{Qwen2.5-VL}                      & 7B  & \cmark & \xmark & \xmark & \cmark & 27.6          \\ \cline{2-8} 
                         & Qwen3-VL-4B-Instruct~\cite{qwen3technicalreport}              & 4B  & \cmark & \xmark & \xmark & \cmark & 45.3          \\ \cline{2-8} 
                         & MAI-UI~\cite{zhou2025maiuitechnicalreportrealworld}                                                        & 2B  & \cmark & \xmark & \xmark & \cmark & 49.1          \\ \cline{2-8} 
                         & Qwen3-VL-2B-Instruct~\cite{qwen3technicalreport}              & 2B  & \cmark & \xmark & \xmark & \cmark & 36.4          \\ \hline
\multirow{9}{*}{Agents}  & Agent S2 (Agent S2)~\cite{Agent-S2}                           & -   & \cmark & \xmark & \xmark & \xmark & 54.3          \\ \cline{2-8} 
                         & GUI-Explorer (GPT-4o)~\cite{xie2025gui_explorer}              & -   & \cmark & \cmark & \xmark & \xmark & 47.4          \\ \cline{2-8} 
                         & AndroidGen (GPT-4o)~\cite{lai2025androidgen}                  & -   & \xmark & \cmark & \xmark & \xmark & 46.8          \\ \cline{2-8} 
                         & UGround (GPT-4o)~\cite{gou2024UGround}                        & -   & \cmark & \xmark & \xmark & \xmark & 32.8          \\ \cline{2-8} 
                         & AndroidWorld (GPT-4o Turbo)~\cite{rawles2024androidworld}     & -   & \xmark & \cmark & \xmark & \xmark & 30.6          \\ \cline{2-8} 
                         & AutoDroid (GPT-4o)~\cite{wen2024autodroid}                    & -   & \xmark & \cmark & \xmark & \xmark & 15.7          \\ \cline{2-8} 
                         & MobileUse (Qwen2.5-VL-72B)~\cite{li2025mobileuse}             & 72B & \cmark & \xmark & \xmark & \xmark & 62.9          \\ \cline{2-8} 
                         & V-Droid (V-Droid)~\cite{dai2025advancing}                     & 8B  & \xmark & \cmark & \xmark & \cmark & 59.5          \\ \cline{2-8} 
 &
  \textbf{\sysname (Qwen3-VL-4B-Instruct-grpo)} &
  \textbf{4B} &
  \textbf{\cmark} &
  \textbf{\xmark} &
  \textbf{\cmark} &
  \textbf{\cmark} &
  \textbf{67.2} \\ \hline
\end{tabular}
}%
\end{table}

\begin{table}[h]
\centering
\scriptsize
\caption{Success rates (\%) on MobileWorld GUI-only. The “Meta-state” denotes contextual signals including foreground app, system time, and LSPosed-hooked toast notifications and transient pop-ups.}
\label{tab:SR-models-mobileworld}
\begin{tabular}{|l|c|c|c|c|c|c|}
\hline
\textbf{Methods} &
  \textbf{Parameter size} &
  \textbf{Screenshot} &
  \textbf{Ally tree} &
  \multicolumn{1}{l|}{\textbf{Meta-state}} &
  \multicolumn{1}{l|}{\textbf{Mobile-deployable}} &
  \textbf{GUI-only (\%)} \\ \hline
GUI-Owl-7B~\cite{ye2025mobileagentv3}                   & 7B  & \cmark & \xmark & \xmark & \cmark & 7.7           \\ \hline
GUI-Owl-32B~\cite{ye2025mobileagentv3}                  & 32B & \cmark & \xmark & \xmark & \xmark & 8.5           \\ \hline
UI-Venus-7B~\cite{gu2025uivenus}                        & 7B  & \cmark & \xmark & \xmark & \cmark & 8.5           \\ \hline
UI-Venus-72B~\cite{gu2025uivenus}                       & 72B & \cmark & \xmark & \xmark & \xmark & 16.4          \\ \hline
Qwen3-VL-4B-Instruct~\cite{qwen3technicalreport}        & 4B  & \cmark & \xmark & \xmark & \cmark & 7.7           \\ \hline
Qwen3-VL-8B-Instruct~\cite{qwen3technicalreport}        & 8B  & \cmark & \xmark & \xmark & \xmark & 9.4           \\ \hline
Qwen3-VL-32B-Instruct~\cite{qwen3technicalreport}       & 32B & \cmark & \xmark & \xmark & \xmark & 11.9          \\ \hline
MAI-UI-2B~\cite{zhou2025maiuitechnicalreportrealworld}  & 2B  & \cmark & \xmark & \xmark & \cmark & 10.3          \\ \hline
MAI-UI-8B~\cite{zhou2025maiuitechnicalreportrealworld}  & 8B  & \cmark & \xmark & \xmark & \xmark & 27.5          \\ \hline
MAI-UI-32B~\cite{zhou2025maiuitechnicalreportrealworld} & 32B & \cmark & \xmark & \xmark & \xmark & \textbf{36.2} \\ \hline
GELab-Zero-4B~\cite{yan2025step}                        & 4B  & \cmark & \xmark & \xmark & \cmark & 16.1          \\ \hline
Qwen3-VL-4B-Instruct-grpo                               & 4B  & \cmark & \xmark & \xmark & \cmark & 12.0          \\ \hline
\textbf{\sysname (Qwen3-VL-4B-Instruct-grpo)} &
  \textbf{4B} &
  \textbf{\cmark} &
  \textbf{\xmark} &
  \textbf{\cmark} &
  \textbf{\cmark} &
  \textbf{23.9} \\ \hline
\end{tabular}
\end{table}

\subsubsection{Performance Comparison} 
We evaluate \sysname, built upon the \textit{Qwen3-VL-4B-Instruct} backbone, on two benchmarks: the standard \textit{AndroidWorld}, the challenging \textit{MobileWorld}~\cite{kong2025mobileworld}, and our \textit{self-collected MobileTask testing dataset}. Since \sysname focuses on GUI-based automation and does not natively support MCP tool invocation, we evaluate on its GUI-only subset (117 tasks) for fair comparison.
MobileTask targets \textit{meta-state-aware} interactions, including transient toast handling, multi-step information input, and error recovery. 
We compare \sysname against a broad set of baselines, including proprietary models (GPT-4o~\cite{hurst2024gpt}, Claude-Sonnet-4~\cite{anthropic2025claude4}), open-source VLMs (V-Droid~\cite{dai2025advancing}, UI-TARS-7B~\cite{qin2025uitar}, UI-Venus-7B/72B~\cite{gu2025uivenus}, Seed1.5-VL~\cite{guo2025seed1}, GLM-4.1V-9B-Thinking~\cite{hong2025glm}, MAI-UI~\cite{zhou2025maiuitechnicalreportrealworld}, GUI-Owl-7B~\cite{ye2025mobileagentv3}, and the Qwen3-VL/Qwen2.5-VL series~\cite{qwen3technicalreport,Qwen2.5-VL}), as well as recent agent frameworks (AndroidWorld~\cite{rawles2024androidworld}, AutoDroid~\cite{wen2024autodroid}, UGround~\cite{gou2024UGround}, AndroidGen~\cite{lai2025androidgen}, GUI-Explorer~\cite{xie2025gui_explorer}, MobileUse~\cite{li2025mobileuse}, Agent S2~\cite{Agent-S2}).
\tabref{tab:SR-models} and \figref{fig:bound_plot_all} show the results.
\textit{First}, on AndroidWorld, \sysname achieves the best success-size tradeoff (\figref{fig:bound_plot}) with 67.2\% success, surpassing larger models such as UI-Venus-72B (65.9\%) and MobileUse (62.9\%), and improving over the base Qwen3-VL-4B-Instruct (45.3\%). 
\textit{Second}, on MobileTask, \sysname reaches 54.9\% accuracy, significantly outperforming Qwen3-VL-4B-Instruct (23.5\%) and other comparable or larger models (\eg Qwen3-VL-8B-Instruct at 27.5\%, GUI-Owl-7B at 41.2\%), highlighting strong robustness in state-aware scenarios. 
\textit{Third}, on MobileWorld GUI-only subset, \sysname achieves 23.9\% success rate (Table~\ref{tab:SR-models-mobileworld}), substantially outperforming the base Qwen3-VL-4B-Instruct (7.7\%) and surpassing larger models including GUI-Owl-32B (8.5\%) and UI-Venus-72B (16.4\%). \sysname remains competitive with MAI-UI-8B (27.5\%) despite using only a 4B backbone. 
\textit{Fourth}, \sysname is highly parameter-efficient: it contains 4.083B parameters in total, adding only 83M LoRA (2.1\% overhead) while delivering a >2.3$\times$ improvement over the base model. 
\textbf{Summary}. \sysnameposs \textbf{meta-controller} coordination beats brute-force scaling, enabling reliable mobile GUI automation with a 4B backbone.
It relies on screenshots (no A11y tree), remaining robust when accessibility trees are unavailable.

\begin{table}[]
\centering
\scriptsize
\caption{Success rate (\%) on MMBench-GUI-L1 test tasks across diverse platforms. "-" denotes undisclosed (typically very large) model sizes. MMBench-GUI-L1 contains only single-step tasks, thus covering only \textbf{L1-L2} in our taxonomy.}
\label{tab:mmbench_L1}
\begin{tabular}{|lcccccccc|}
\hline
\multicolumn{1}{|l|}{\textbf{Model}} &
  \multicolumn{1}{c|}{\textbf{Parameter size}} &
  \multicolumn{1}{c|}{\textbf{Windows}} &
  \multicolumn{1}{c|}{\textbf{MacOS}} &
  \multicolumn{1}{c|}{\textbf{Linux}} &
  \multicolumn{1}{c|}{\textbf{iOS}} &
  \multicolumn{1}{c|}{\textbf{Android}} &
  \multicolumn{1}{c|}{\textbf{Web}} &
  \textbf{Overall} \\ \hline
\multicolumn{9}{|c|}{\textit{Task difficulty (easy)}} \\ \hline
\multicolumn{1}{|l|}{GPT-4o~\cite{hurst2024gpt}} &
  \multicolumn{1}{c|}{-} &
  \multicolumn{1}{c|}{62.47} &
  \multicolumn{1}{c|}{67.89} &
  \multicolumn{1}{c|}{62.38} &
  \multicolumn{1}{c|}{58.52} &
  \multicolumn{1}{c|}{56.41} &
  \multicolumn{1}{c|}{58.51} &
  60.16 \\ \hline
\multicolumn{1}{|l|}{Claude-3.5-Sonnet~\cite{anthropic2024claude35sonnet}} &
  \multicolumn{1}{c|}{-} &
  \multicolumn{1}{c|}{41.34} &
  \multicolumn{1}{c|}{50.04} &
  \multicolumn{1}{c|}{41.61} &
  \multicolumn{1}{c|}{42.03} &
  \multicolumn{1}{c|}{38.96} &
  \multicolumn{1}{c|}{41.79} &
  41.54 \\ \hline
\multicolumn{1}{|l|}{Claude-3.7-Sonnet~\cite{anthropic2026claude37sonnet}} &
  \multicolumn{1}{c|}{-} &
  \multicolumn{1}{c|}{34.66} &
  \multicolumn{1}{c|}{49.05} &
  \multicolumn{1}{c|}{39.37} &
  \multicolumn{1}{c|}{42.76} &
  \multicolumn{1}{c|}{37.45} &
  \multicolumn{1}{c|}{40.80} &
  39.08 \\ \hline
\multicolumn{1}{|l|}{Qwen-Max-VL~\cite{Qwen-VL}} &
  \multicolumn{1}{c|}{-} &
  \multicolumn{1}{c|}{69.05} &
  \multicolumn{1}{c|}{72.51} &
  \multicolumn{1}{c|}{69.91} &
  \multicolumn{1}{c|}{70.82} &
  \multicolumn{1}{c|}{63.09} &
  \multicolumn{1}{c|}{69.46} &
  68.15 \\ \hline
\multicolumn{1}{|l|}{Qwen2.5-VL-72B~\cite{Qwen2.5-VL}} &
  \multicolumn{1}{c|}{72B} &
  \multicolumn{1}{c|}{65.86} &
  \multicolumn{1}{c|}{75.23} &
  \multicolumn{1}{c|}{73.02} &
  \multicolumn{1}{c|}{67.24} &
  \multicolumn{1}{c|}{58.09} &
  \multicolumn{1}{c|}{72.08} &
  66.98 \\ \hline
\multicolumn{1}{|l|}{UI-TARS-72B-DPO~\cite{qin2025uitar}} &
  \multicolumn{1}{c|}{72B} &
  \multicolumn{1}{c|}{41.59} &
  \multicolumn{1}{c|}{28.52} &
  \multicolumn{1}{c|}{35.16} &
  \multicolumn{1}{c|}{31.08} &
  \multicolumn{1}{c|}{52.25} &
  \multicolumn{1}{c|}{35.33} &
  40.18 \\ \hline
\multicolumn{1}{|l|}{InternVL3-72B~\cite{zhu2025internvl3}} &
  \multicolumn{1}{c|}{72B} &
  \multicolumn{1}{c|}{74.67} &
  \multicolumn{1}{c|}{78.72} &
  \multicolumn{1}{c|}{79.16} &
  \multicolumn{1}{c|}{83.57} &
  \multicolumn{1}{c|}{80.10} &
  \multicolumn{1}{c|}{81.18} &
  79.15 \\ \hline
\multicolumn{1}{|l|}{GUI-Owl-7B~\cite{ye2025mobileagentv3}} &
  \multicolumn{1}{c|}{7B} &
  \multicolumn{1}{c|}{82.96} &
  \multicolumn{1}{c|}{84.52} &
  \multicolumn{1}{c|}{85.57} &
  \multicolumn{1}{c|}{82.61} &
  \multicolumn{1}{c|}{83.28} &
  \multicolumn{1}{c|}{88.13} &
  84.50 \\ \hline
\multicolumn{1}{|l|}{Qwen3-VL-2B~\cite{qwen3technicalreport}} &
  \multicolumn{1}{c|}{2B} &
  \multicolumn{1}{c|}{69.26} &
  \multicolumn{1}{c|}{62.94} &
  \multicolumn{1}{c|}{55.49} &
  \multicolumn{1}{c|}{61.71} &
  \multicolumn{1}{c|}{64.54} &
  \multicolumn{1}{c|}{72.14} &
  65.15 \\ \hline
\multicolumn{1}{|l|}{Qwen3-VL-4B~\cite{qwen3technicalreport}} &
  \multicolumn{1}{c|}{4B} &
  \multicolumn{1}{c|}{88.91} &
  \multicolumn{1}{c|}{86.82} &
  \multicolumn{1}{c|}{86.56} &
  \multicolumn{1}{c|}{84.40} &
  \multicolumn{1}{c|}{84.17} &
  \multicolumn{1}{c|}{84.89} &
  85.98 \\ \hline
\multicolumn{1}{|l|}{Ours (Qwen3-VL-2B-Instruct-grpo)} &
  \multicolumn{1}{c|}{2B} &
  \multicolumn{1}{c|}{83.04} &
  \multicolumn{1}{c|}{77.10} &
  \multicolumn{1}{c|}{78.24} &
  \multicolumn{1}{c|}{76.70} &
  \multicolumn{1}{c|}{81.60} &
  \multicolumn{1}{c|}{86.08} &
  81.41 \\ \hline
\multicolumn{1}{|l|}{\textbf{Ours (Qwen3-VL-4B-Instruct-grpo})} &
  \multicolumn{1}{c|}{4B} &
  \multicolumn{1}{c|}{92.40} &
  \multicolumn{1}{c|}{91.74} &
  \multicolumn{1}{c|}{92.24} &
  \multicolumn{1}{c|}{88.82} &
  \multicolumn{1}{c|}{88.57} &
  \multicolumn{1}{c|}{90.91} &
  \textbf{90.72} \\ \hline
\multicolumn{9}{|c|}{\textit{Task difficulty (medium)}} \\ \hline
\multicolumn{1}{|l|}{GPT-4o~\cite{hurst2024gpt}} &
  \multicolumn{1}{c|}{-} &
  \multicolumn{1}{c|}{56.33} &
  \multicolumn{1}{c|}{63.13} &
  \multicolumn{1}{c|}{59.70} &
  \multicolumn{1}{c|}{54.06} &
  \multicolumn{1}{c|}{57.69} &
  \multicolumn{1}{c|}{54.98} &
  57.24 \\ \hline
\multicolumn{1}{|l|}{Claude-3.5-Sonnet~\cite{anthropic2024claude35sonnet}} &
  \multicolumn{1}{c|}{-} &
  \multicolumn{1}{c|}{39.28} &
  \multicolumn{1}{c|}{47.63} &
  \multicolumn{1}{c|}{45.97} &
  \multicolumn{1}{c|}{44.57} &
  \multicolumn{1}{c|}{42.03} &
  \multicolumn{1}{c|}{34.33} &
  41.26 \\ \hline
\multicolumn{1}{|l|}{Claude-3.7-Sonnet~\cite{anthropic2026claude37sonnet}} &
  \multicolumn{1}{c|}{-} &
  \multicolumn{1}{c|}{39.34} &
  \multicolumn{1}{c|}{39.23} &
  \multicolumn{1}{c|}{42.28} &
  \multicolumn{1}{c|}{39.45} &
  \multicolumn{1}{c|}{36.05} &
  \multicolumn{1}{c|}{36.17} &
  38.39 \\ \hline
\multicolumn{1}{|l|}{Qwen-Max-VL~\cite{Qwen-VL}} &
  \multicolumn{1}{c|}{-} &
  \multicolumn{1}{c|}{63.40} &
  \multicolumn{1}{c|}{73.85} &
  \multicolumn{1}{c|}{66.90} &
  \multicolumn{1}{c|}{68.02} &
  \multicolumn{1}{c|}{63.66} &
  \multicolumn{1}{c|}{64.59} &
  65.44 \\ \hline
\multicolumn{1}{|l|}{Qwen2.5-VL-72B~\cite{Qwen2.5-VL}} &
  \multicolumn{1}{c|}{72B} &
  \multicolumn{1}{c|}{66.29} &
  \multicolumn{1}{c|}{72.73} &
  \multicolumn{1}{c|}{72.63} &
  \multicolumn{1}{c|}{59.27} &
  \multicolumn{1}{c|}{66.24} &
  \multicolumn{1}{c|}{68.24} &
  67.45 \\ \hline
\multicolumn{1}{|l|}{UI-TARS-72B-DPO~\cite{qin2025uitar}} &
  \multicolumn{1}{c|}{72B} &
  \multicolumn{1}{c|}{38.83} &
  \multicolumn{1}{c|}{41.60} &
  \multicolumn{1}{c|}{37.14} &
  \multicolumn{1}{c|}{41.72} &
  \multicolumn{1}{c|}{54.74} &
  \multicolumn{1}{c|}{31.55} &
  41.77 \\ \hline
\multicolumn{1}{|l|}{InternVL3-72B~\cite{zhu2025internvl3}} &
  \multicolumn{1}{c|}{72B} &
  \multicolumn{1}{c|}{71.46} &
  \multicolumn{1}{c|}{78.58} &
  \multicolumn{1}{c|}{79.88} &
  \multicolumn{1}{c|}{78.43} &
  \multicolumn{1}{c|}{81.36} &
  \multicolumn{1}{c|}{78.67} &
  77.89 \\ \hline
\multicolumn{1}{|l|}{GUI-Owl-7B~\cite{ye2025mobileagentv3}} &
  \multicolumn{1}{c|}{7B} &
  \multicolumn{1}{c|}{88.89} &
  \multicolumn{1}{c|}{88.10} &
  \multicolumn{1}{c|}{91.24} &
  \multicolumn{1}{c|}{84.35} &
  \multicolumn{1}{c|}{85.25} &
  \multicolumn{1}{c|}{83.56} &
  86.86 \\ \hline
\multicolumn{1}{|l|}{Qwen3-VL-2B~\cite{qwen3technicalreport}} &
  \multicolumn{1}{c|}{2B} &
  \multicolumn{1}{c|}{81.01} &
  \multicolumn{1}{c|}{66.45} &
  \multicolumn{1}{c|}{67.88} &
  \multicolumn{1}{c|}{52.25} &
  \multicolumn{1}{c|}{70.26} &
  \multicolumn{1}{c|}{65.12} &
  69.35 \\ \hline
\multicolumn{1}{|l|}{Qwen3-VL-4B~\cite{qwen3technicalreport}} &
  \multicolumn{1}{c|}{4B} &
  \multicolumn{1}{c|}{93.32} &
  \multicolumn{1}{c|}{89.30} &
  \multicolumn{1}{c|}{90.14} &
  \multicolumn{1}{c|}{77.40} &
  \multicolumn{1}{c|}{84.97} &
  \multicolumn{1}{c|}{82.58} &
  86.84 \\ \hline
\multicolumn{1}{|l|}{Ours (Qwen3-VL-2B-Instruct-grpo)} &
  \multicolumn{1}{c|}{2B} &
  \multicolumn{1}{c|}{90.22} &
  \multicolumn{1}{c|}{77.18} &
  \multicolumn{1}{c|}{85.57} &
  \multicolumn{1}{c|}{67.84} &
  \multicolumn{1}{c|}{84.00} &
  \multicolumn{1}{c|}{76.19} &
  82.18 \\ \hline
\multicolumn{1}{|l|}{\textbf{Ours (Qwen3-VL-4B-Instruct-grpo)}} &
  \multicolumn{1}{c|}{4B} &
  \multicolumn{1}{c|}{93.32} &
  \multicolumn{1}{c|}{84.53} &
  \multicolumn{1}{c|}{95.40} &
  \multicolumn{1}{c|}{85.23} &
  \multicolumn{1}{c|}{90.97} &
  \multicolumn{1}{c|}{83.01} &
  \textbf{89.75} \\ \hline
\multicolumn{9}{|c|}{\textit{Task difficulty (hard)}} \\ \hline
\multicolumn{1}{|l|}{GPT-4o~\cite{hurst2024gpt}} &
  \multicolumn{1}{c|}{-} &
  \multicolumn{1}{c|}{60.69} &
  \multicolumn{1}{c|}{60.38} &
  \multicolumn{1}{c|}{52.42} &
  \multicolumn{1}{c|}{45.27} &
  \multicolumn{1}{c|}{50.93} &
  \multicolumn{1}{c|}{50.83} &
  53.49 \\ \hline
\multicolumn{1}{|l|}{Claude-3.5-Sonnet~\cite{anthropic2024claude35sonnet}} &
  \multicolumn{1}{c|}{-} &
  \multicolumn{1}{c|}{37.40} &
  \multicolumn{1}{c|}{42.70} &
  \multicolumn{1}{c|}{34.07} &
  \multicolumn{1}{c|}{40.86} &
  \multicolumn{1}{c|}{36.96} &
  \multicolumn{1}{c|}{38.11} &
  37.55 \\ \hline
\multicolumn{1}{|l|}{Claude-3.7-Sonnet~\cite{anthropic2026claude37sonnet}} &
  \multicolumn{1}{c|}{-} &
  \multicolumn{1}{c|}{32.99} &
  \multicolumn{1}{c|}{34.48} &
  \multicolumn{1}{c|}{31.97} &
  \multicolumn{1}{c|}{39.20} &
  \multicolumn{1}{c|}{36.99} &
  \multicolumn{1}{c|}{38.92} &
  35.65 \\ \hline
\multicolumn{1}{|l|}{Qwen-Max-VL~\cite{Qwen-VL}} &
  \multicolumn{1}{c|}{-} &
  \multicolumn{1}{c|}{66.64} &
  \multicolumn{1}{c|}{67.59} &
  \multicolumn{1}{c|}{65.80} &
  \multicolumn{1}{c|}{60.23} &
  \multicolumn{1}{c|}{58.78} &
  \multicolumn{1}{c|}{65.34} &
  63.69 \\ \hline
\multicolumn{1}{|l|}{Qwen2.5-VL-72B~\cite{Qwen2.5-VL}} &
  \multicolumn{1}{c|}{72B} &
  \multicolumn{1}{c|}{70.68} &
  \multicolumn{1}{c|}{68.91} &
  \multicolumn{1}{c|}{70.98} &
  \multicolumn{1}{c|}{57.59} &
  \multicolumn{1}{c|}{53.94} &
  \multicolumn{1}{c|}{68.10} &
  64.56 \\ \hline
\multicolumn{1}{|l|}{UI-TARS-72B-DPO~\cite{qin2025uitar}} &
  \multicolumn{1}{c|}{72B} &
  \multicolumn{1}{c|}{31.48} &
  \multicolumn{1}{c|}{35.87} &
  \multicolumn{1}{c|}{24.19} &
  \multicolumn{1}{c|}{36.33} &
  \multicolumn{1}{c|}{58.13} &
  \multicolumn{1}{c|}{19.94} &
  35.78 \\ \hline
\multicolumn{1}{|l|}{InternVL3-72B~\cite{zhu2025internvl3}} &
  \multicolumn{1}{c|}{72B} &
  \multicolumn{1}{c|}{75.08} &
  \multicolumn{1}{c|}{77.44} &
  \multicolumn{1}{c|}{76.19} &
  \multicolumn{1}{c|}{70.37} &
  \multicolumn{1}{c|}{75.73} &
  \multicolumn{1}{c|}{78.11} &
  75.70 \\ \hline
\multicolumn{1}{|l|}{GUI-Owl-7B~\cite{ye2025mobileagentv3}} &
  \multicolumn{1}{c|}{7B} &
  \multicolumn{1}{c|}{87.78} &
  \multicolumn{1}{c|}{96.43} &
  \multicolumn{1}{c|}{94.33} &
  \multicolumn{1}{c|}{87.83} &
  \multicolumn{1}{c|}{88.85} &
  \multicolumn{1}{c|}{94.06} &
  90.90 \\ \hline
\multicolumn{1}{|l|}{Qwen3-VL-2B~\cite{qwen3technicalreport}} &
  \multicolumn{1}{c|}{2B} &
  \multicolumn{1}{c|}{78.49} &
  \multicolumn{1}{c|}{81.88} &
  \multicolumn{1}{c|}{91.30} &
  \multicolumn{1}{c|}{73.81} &
  \multicolumn{1}{c|}{82.30} &
  \multicolumn{1}{c|}{83.36} &
  82.25 \\ \hline
\multicolumn{1}{|l|}{Qwen3-VL-4B~\cite{qwen3technicalreport}} &
  \multicolumn{1}{c|}{4B} &
  \multicolumn{1}{c|}{91.91} &
  \multicolumn{1}{c|}{93.02} &
  \multicolumn{1}{c|}{84.41} &
  \multicolumn{1}{c|}{93.84} &
  \multicolumn{1}{c|}{92.48} &
  \multicolumn{1}{c|}{89.54} &
  91.29 \\ \hline
\multicolumn{1}{|l|}{Ours (Qwen3-VL-2B-Instruct-grpo)} &
  \multicolumn{1}{c|}{2B} &
  \multicolumn{1}{c|}{88.60} &
  \multicolumn{1}{c|}{87.05} &
  \multicolumn{1}{c|}{94.41} &
  \multicolumn{1}{c|}{81.03} &
  \multicolumn{1}{c|}{88.59} &
  \multicolumn{1}{c|}{90.34} &
  89.02 \\ \hline
\multicolumn{1}{|l|}{\textbf{Ours (Qwen3-VL-4B-Instruct-grpo)}} &
  \multicolumn{1}{c|}{4B} &
  \multicolumn{1}{c|}{94.79} &
  \multicolumn{1}{c|}{95.35} &
  \multicolumn{1}{c|}{94.89} &
  \multicolumn{1}{c|}{87.91} &
  \multicolumn{1}{c|}{92.85} &
  \multicolumn{1}{c|}{92.68} &
  \textbf{93.29} \\ \hline
\end{tabular}
\end{table}

\subsubsection{Cross-Platform Performance}
We evaluate single-step GUI understanding on \textit{MMBench-GUI-L1}, a QA benchmark spanning six platforms (\ie Windows, MacOS, Linux, iOS, Android, Web). 
Importantly, \sysname is trained \emph{only} on our \textit{MobileTask Train Dataset} (trajectory data; \secref{sec:implementation:datasets}), and the MMBench test set is strictly held out (no overlap with training data).
As shown in \tabref{tab:mmbench_L1}, \sysname achieves state-of-the-art accuracy across Easy/Medium/Hard: 90.72\%/89.75\%/93.29\%, outperforming all baselines including InternVL3-72B and GUI-Owl-7B. 
Gains are consistent across platforms, with strong results on Windows and Linux. 
Moreover, our 2B variant remains competitive (81.41/82.18/89.02), surpassing multiple larger models (\eg Qwen2.5-VL-72B and both Claude variants).
\textbf{Summary}. Trajectory-only training yields transferable GUI understanding, enabling strong cross-platform generalization without benchmark-specific supervision.

\begin{figure}[t]
  \centering
  \subfloat[Qwen3-VL
    \label{fig:ui_parse_SR}]{
    \includegraphics[width=0.3\textwidth]{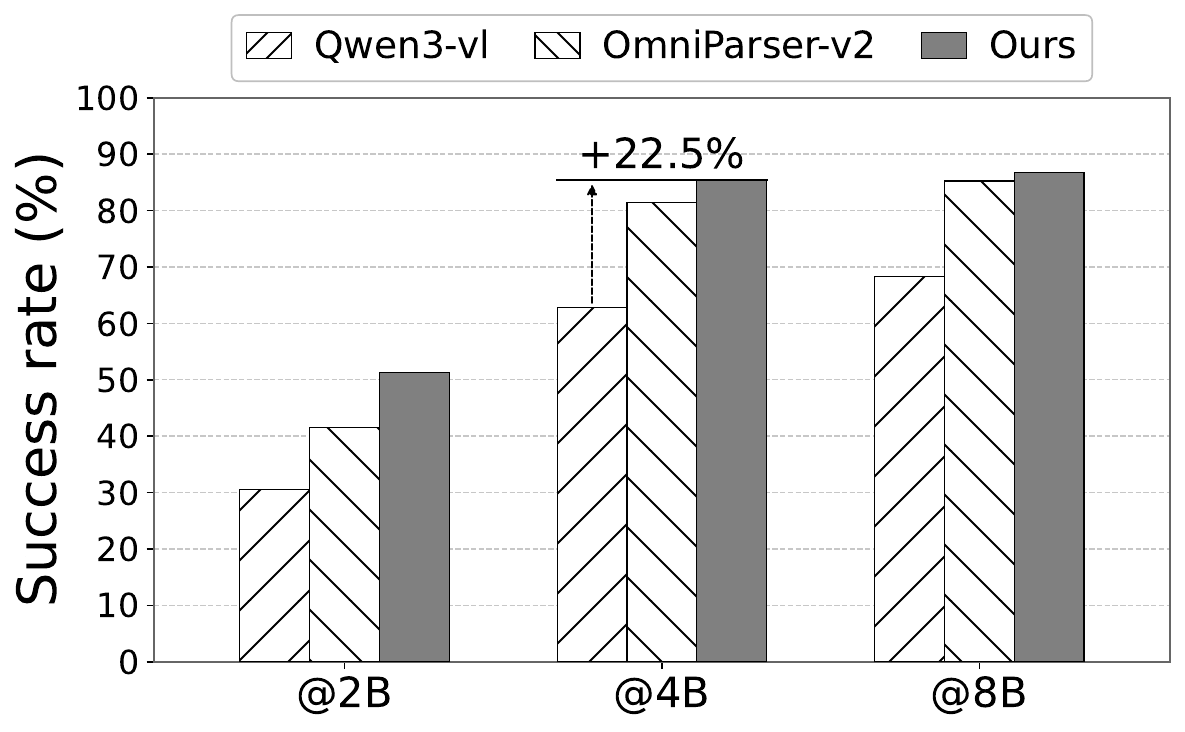}
  }
  \subfloat[Qwen2.5-VL
    \label{fig:ui_parse_SR_qwen2_5}]{
    \includegraphics[width=0.3\textwidth]{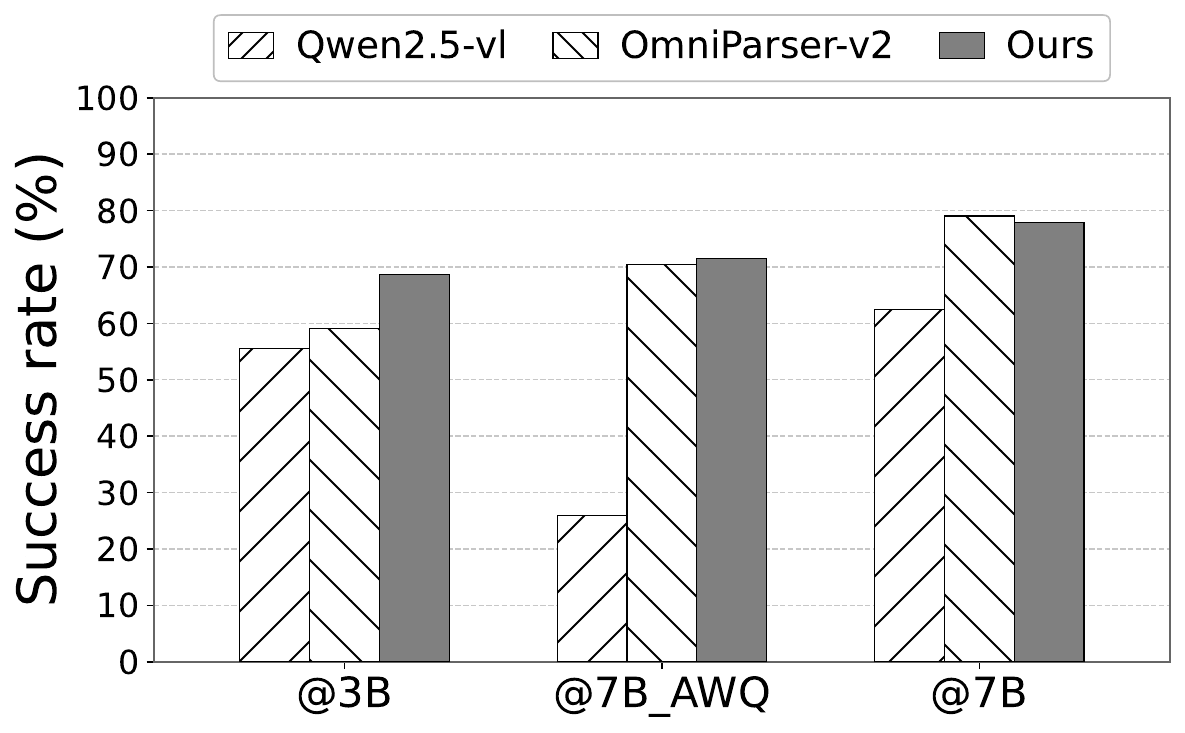}
  }
  \subfloat[MiniCPM\_V\_4\_5
    \label{fig:ui_parse_SR_minicpm_4_5}]{
    \includegraphics[width=0.3\textwidth]{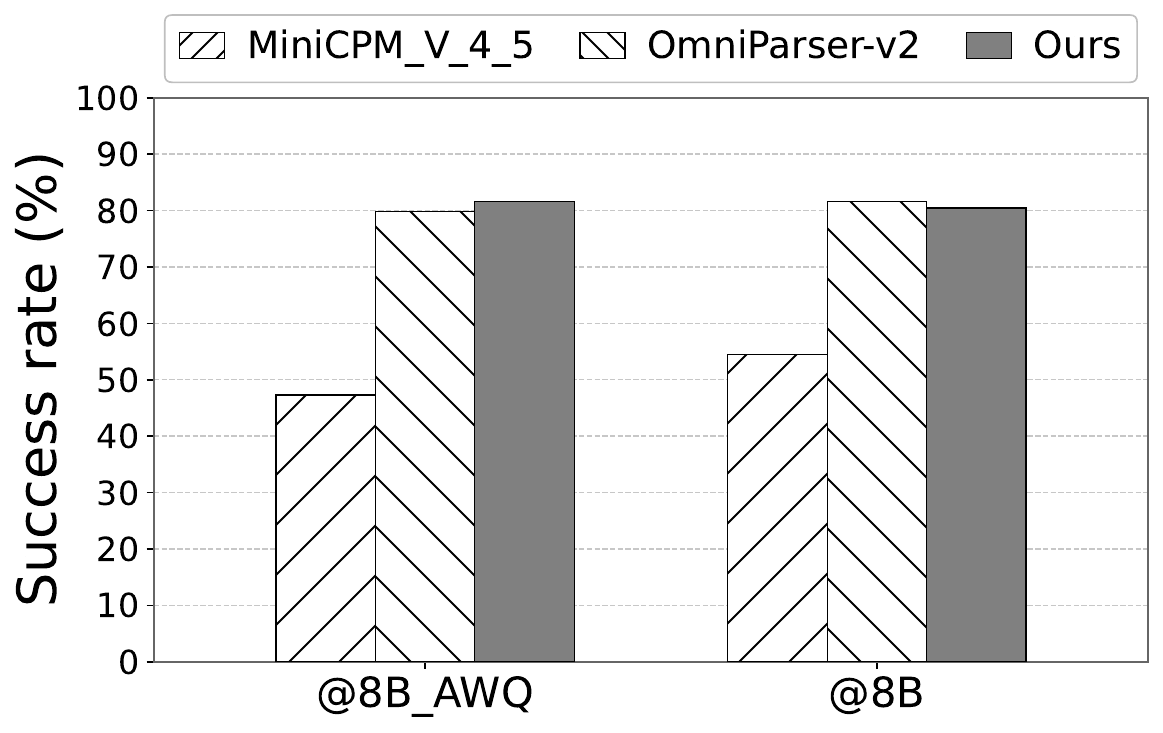}
  }
  \caption{Success rate of Qwen3-VL/Qwen2.5-VL/MiniCPM\_V\_4\_5 with supplemental UI element context on screenSpotV2-mobile
(Ablation of the baseline model versus its performance when augmented with pre-processed UI element information from OmniParser-v2 and our method, across 2B/3B/4B/7B/8B scales). }
  \label{fig:ui_parse_result}
\end{figure}

\subsection{Performance of \sysnameposs UI Parser}
We test \sysname’s UI parser by augmenting \textit{Qwen3-VL} models (2B/4B/8B) with structured UI elements and testing on \textit{ScreenSpotV2-Mobile}~\cite{wu2024atlas} \emph{without} any fine-tuning on the benchmark data. 
We compare the vanilla VLM~\cite{qwen3technicalreport}, OmniParser-v2~\cite{lu2024omniparser}, and \sysname’s parser (\figref{fig:ui_parse_result}).
\textit{First}, \sysname consistently improves success rate across all scales, with the largest gain on 4B: +22.53\%(62.87\%$\rightarrow$85.40\%). 
\textit{Second}, the augmented 4B model nearly matches the 8B baseline (85.40\% \textit{vs.} 86.8\%; 1.4\% gap), indicating structured UI context can offset reduced capacity. 
\textit{Third}, \sysname outperforms OmniParser-v2 at every scale (up to +9.78\% on 2B). 
\textit{Finally}, our segmentation-based parser achieves 91.47\% icon recall (\tabref{tab:Comparison-of-recall}), +6.16\% over YOLO-based methods, consistent with the accuracy gains.
\textbf{Summary}. \sysnameposs high-recall structured UI parsing substantially boosts visual grounding, reducing dependence on model scale for mobile GUI interaction.

\begin{figure}[t]
  \centering
  \begin{minipage}{0.60\textwidth}
    \centering
    \includegraphics[width=\textwidth]{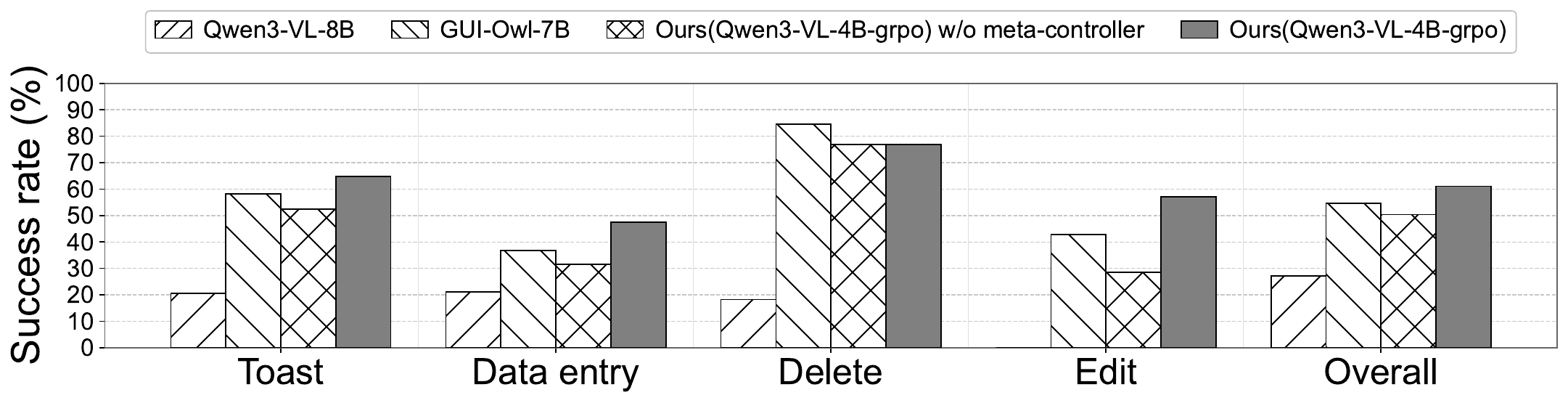}
    \caption{Success rate (\%) across different models and task categories on unseen applications.}
    \label{fig:unseen_app_performance}
  \end{minipage}
  \hfill
  \begin{minipage}{0.36\textwidth}
    \centering
    \includegraphics[width=\textwidth]{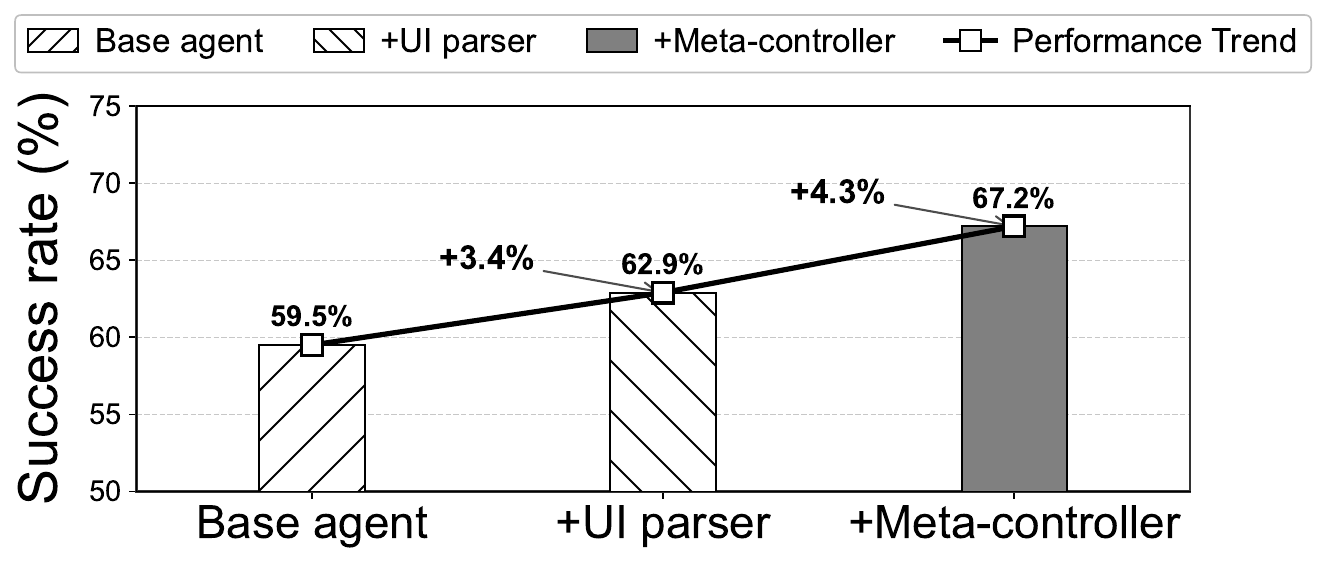}
    \caption{Success rate (\%) ablation study of UI parser and meta-controller.}
    \label{fig:agent_success_rate_ablation}
  \end{minipage}
\end{figure}

\subsection{Performance on Unseen Apps/Traces}
\label{subsubsec:app-generalization}
We test \sysname on \textit{37} held-out tasks across \textit{9} unseen applications. 
None of these tasks appear in training.  \figref{fig:unseen_app_performance} shows the resullts.
\textit{First}, \sysname with the meta-controller achieves the best overall success of 61.1\%, outperforming Qwen3-VL-8B (27.1\%) and GUI-Owl-7B (54.7\%), demonstrating strong generalization with a compact 4B backbone. 
\textit{Second}, removing the meta-controller drops success to 50.41\% (-10.69\% points), with the largest degradations on state-sensitive categories: \textit{toast handling} (64.7\%$\rightarrow$52.35\%), \textit{data entry} (47.4\%$\rightarrow$31.58\%), and \textit{edit operations} (57.1\%$\rightarrow$28.6\%). 
\textit{Third}, even without the meta-controller, \sysname remains competitive and often exceeds Qwen3-VL-8B, indicating the benefit of specialized multi-agents, while meta-controller is key to consistently surpassing all baselines on unseen tasks.
\textbf{Summary}. \sysname generalizes effectively to unseen apps, and the meta-controller is essential for robust state tracking and recovery in complex interactions.

\subsection{Ablation Study and Micro-Benchmarks}
\subsubsection{Impact of UI Parser}
We quantify the benefit of structured UI parsing by augmenting a base agent (Planner+Executor) with element-level screen representations. 
On AndroidWorld, this raises success from 59.5\% to 62.9\% (+3.4\%). 
The gain indicates that converting raw screenshots into structured UI elements reduces mobile UI ambiguity and improves grounding, especially for \textit{localization-critical} actions (\eg tapping specific buttons and filling designated fields). 
Overall, accurate parsing strengthens downstream planning and execution by providing a more reliable UI state.

\begin{figure}[t]
  \centering
  \captionsetup[subfloat]{labelformat=empty}
  \subfloat[
    \label{fig:action_type_accuracy_comparison}]{
    \includegraphics[width=0.3\textwidth]{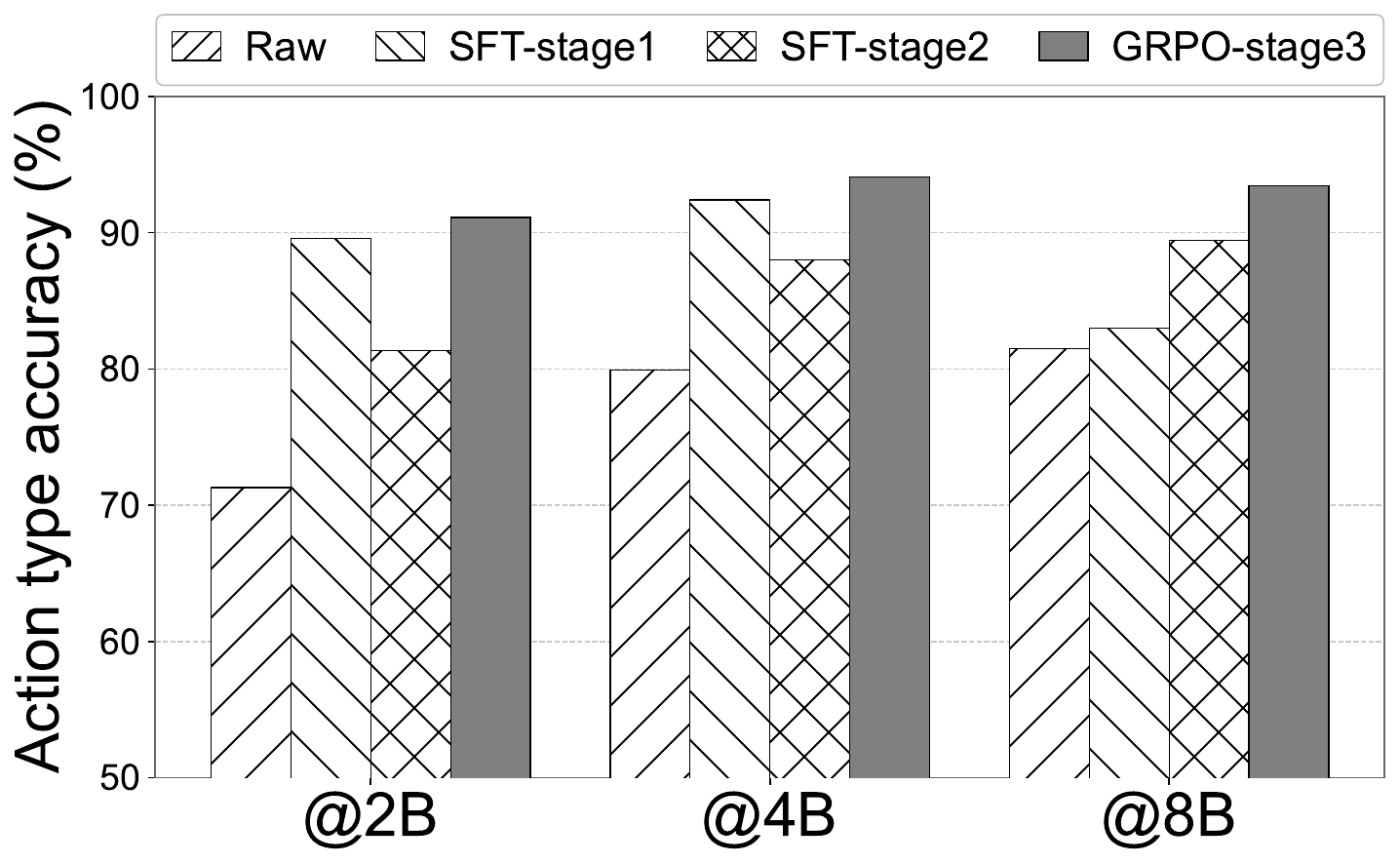}
  } \hfill
  \subfloat[
    \label{fig:action_accuracy_comparison}]{
    \includegraphics[width=0.3\textwidth]{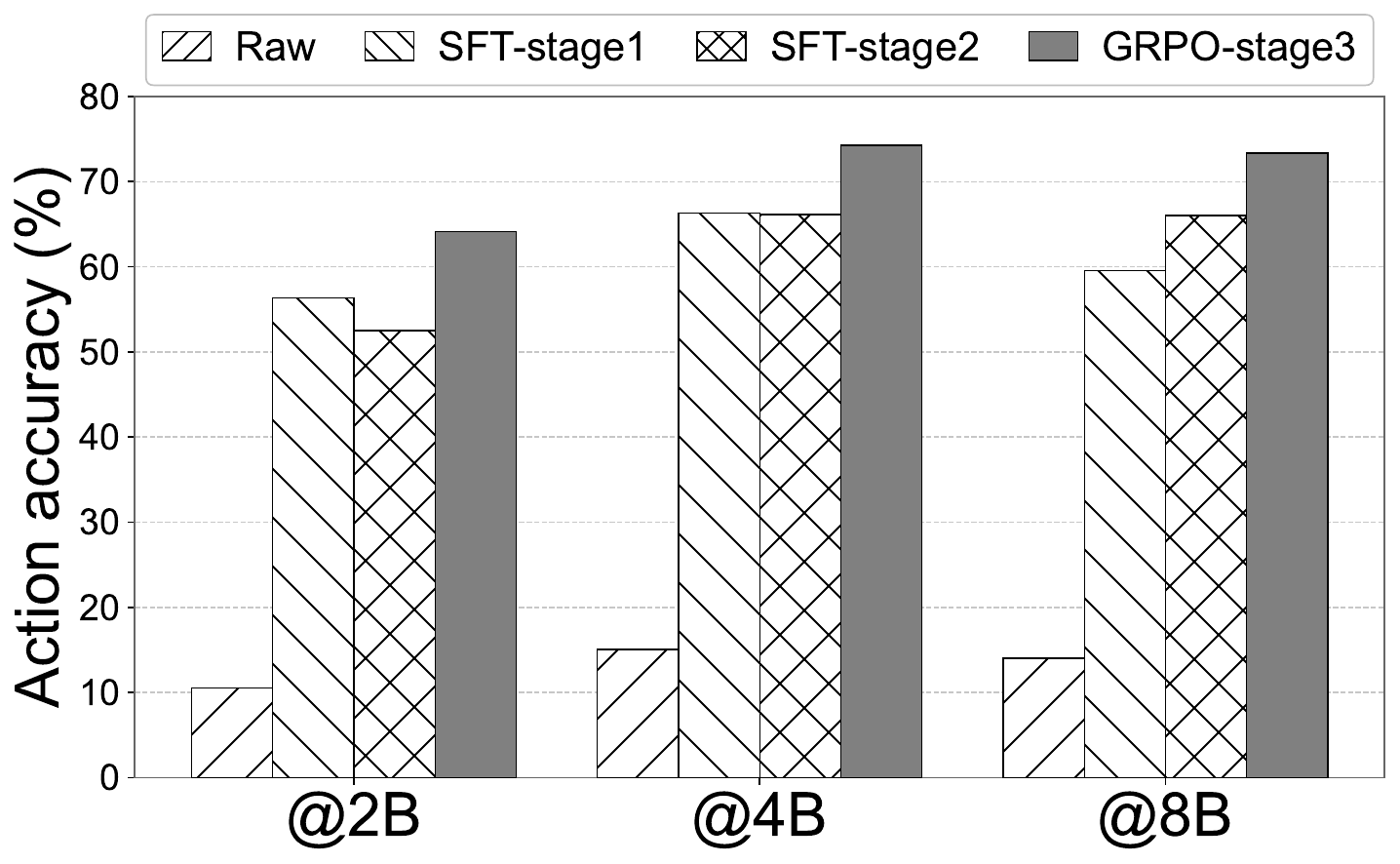}
  } \hfill
  \subfloat[
    \label{fig:success_rate_comparison}]{
    \includegraphics[width=0.3\textwidth]{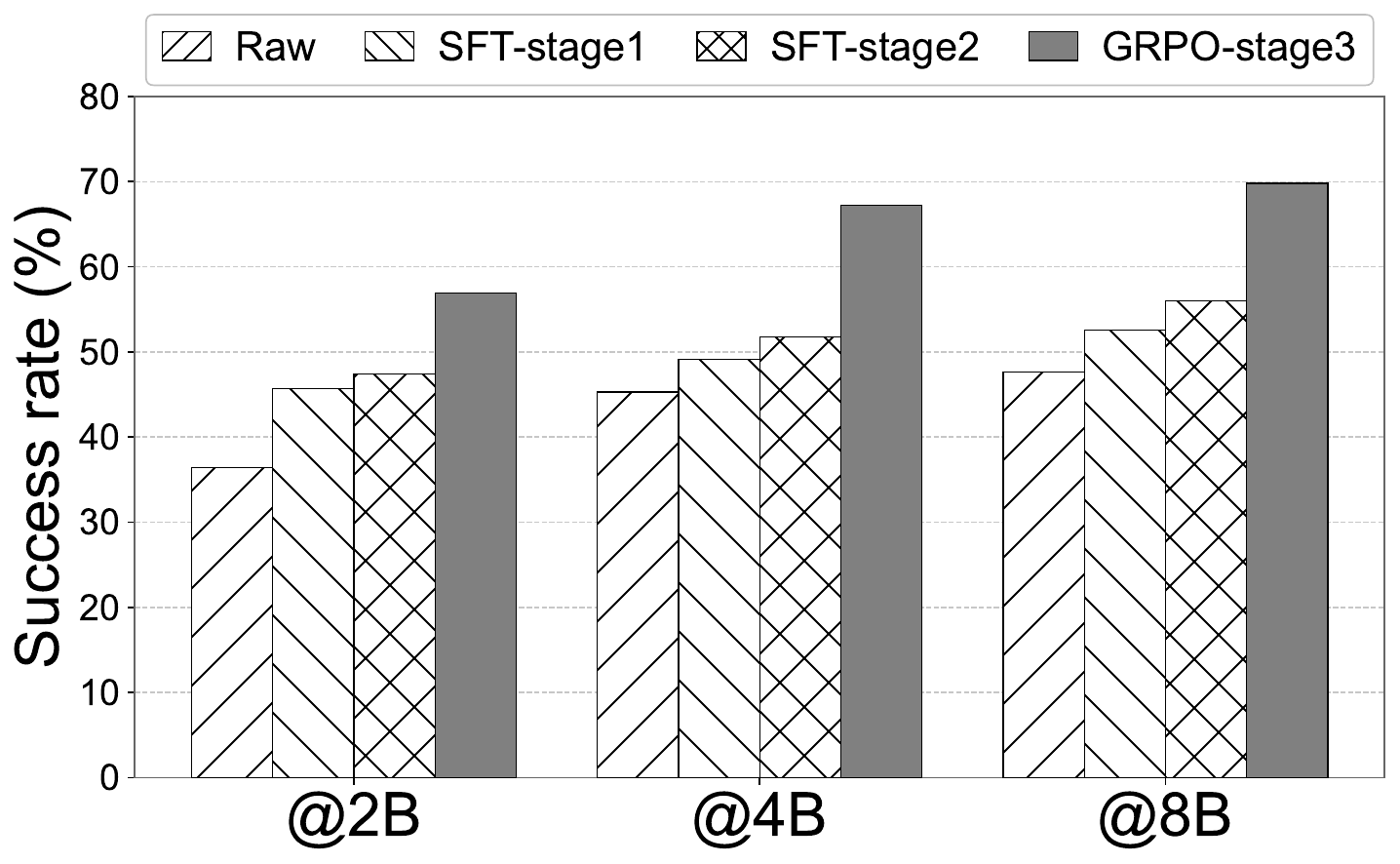}
  } \hfill
  \vspace{-3mm}
  \caption{Improvements of task success rates by incrementally applying SFT and GRPO to the base models. }
  \vspace{-2mm}
  \label{fig:sft-grpo-ablation}
\end{figure}

\begin{figure}[t]
  \centering
  \begin{minipage}{0.40\textwidth}
      \vspace{3mm}
      \includegraphics[height=0.50\textwidth]{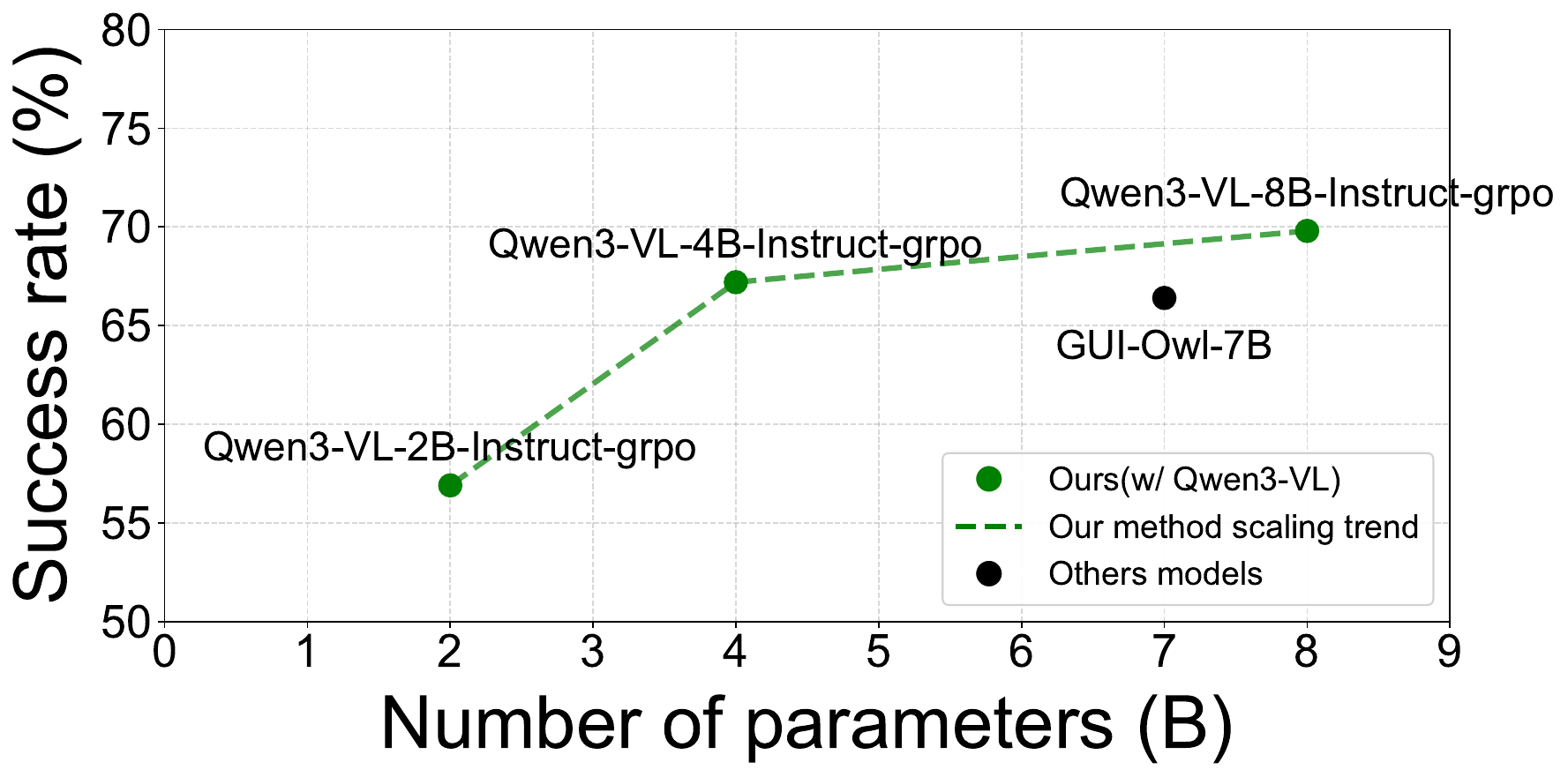}
      \caption{Performance-parameter trade-off in vision-language models.}
      \vspace{-3mm}
      \label{fig:bound_plot_explore}
    
  \end{minipage}
  \hfill
  \begin{minipage}{0.58\textwidth}
    \centering
    \subfloat[
      \label{fig:latency-on-server}]{
      \includegraphics[width=0.49\textwidth]{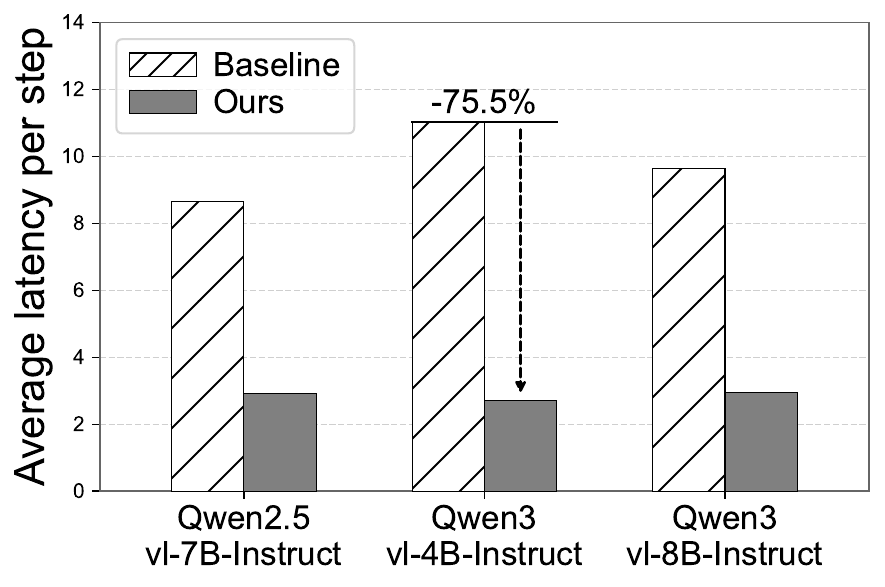}
    }
    \subfloat[
      \label{fig:avg_prompt_tokens}]{
      \includegraphics[width=0.49\textwidth]{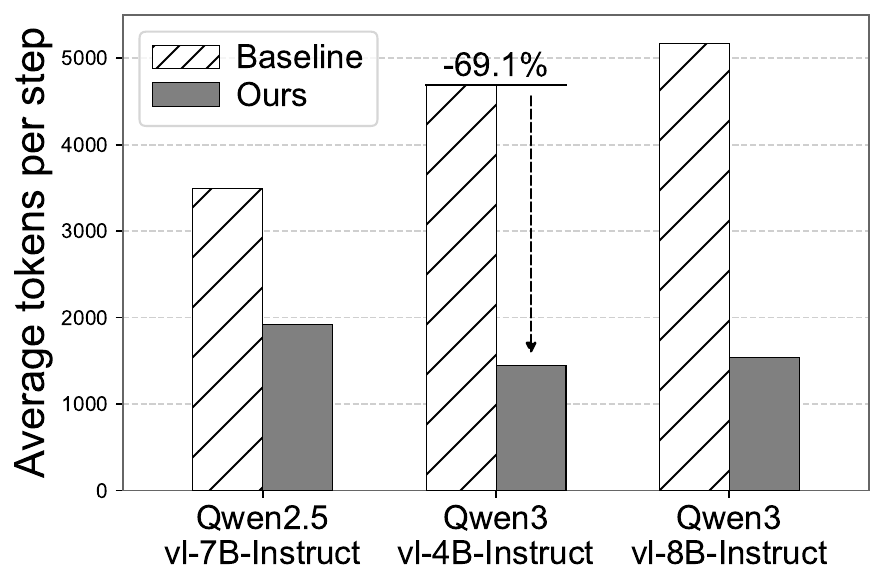}
    }
    \vspace{-3mm}
    \caption{System overhead comparison: \sysname~vs. M3A implementation in AndroidWorld (evaluated on 2×3090 GPUs).}
    \vspace{-3mm}
    \label{fig:system-overhead}
  \end{minipage}
  \hfill
\end{figure}

\subsubsection{Impact of Meta-Controller}
We then add the meta-controller on top of the parser-augmented agent, improving success from 62.9\% to 67.2\% (+4.3\%), the largest single-module gain. 
The meta-controller enables closed-loop coordination via pre-action feasibility checks and post-action outcome perception (including transient signals such as toast messages), and supports targeted recovery without restarting workflows. 
Together, these mechanisms mitigate compounding errors and are central to \sysnameposs reliability on multi-step GUI tasks.

\subsubsection{Benefits of Multi-Stage Training}
We evaluate our three-stage training pipeline (SFT-Stage1, SFT-Stage2, GRPO-Stage3) on Qwen3-VL models at three scales (2B/4B/8B). 
All variants use the same pipeline (UI parser + meta-controller) and are evaluated on AndroidWorld. 
Across scales, performance improves monotonically with each stage; we highlight the 4B model as it best balances capability and efficiency.
\figref{fig:sft-grpo-ablation} shows the results.
\textit{First}, \textbf{Action-only SFT (Stage 1)} grounds the base 4B model in mobile GUI semantics, improving success rate from 45.3\% to 49.1\%. 
Although Stage 1 achieves strong action-level metrics (92.41\% action-type accuracy; 66.28\% action accuracy), task success remains limited, suggesting that accurate low-level actions alone do not ensure end-to-end completion. 
\textit{Second}, \textbf{Reasoning-augmented SFT (Stage 2)} further increases success to 51.8\%, indicating that supervising intermediate reasoning better supports multi-step task execution, even with slightly lower action-level metrics (88.03\% type; 66.15\% action). 
\textit{Third}, \textbf{GRPO with a failure-aware curriculum (Stage 3)} yields the largest gain, boosting success to 67.2\% (+15.4\% over SFT-variant), demonstrating the value of preference-based optimization for aligning behavior with task completion.
\textbf{Summary}. The staged pipeline delivers compounding benefits. SFT provides essential grounding, while GRPO drives substantial end-to-end improvements. Notably, the final 4B model reaches 67.2\% success, surpassing several larger baselines on AndroidWorld.

\subsubsection{Trade-Offs between Model Size and Task Performance}
We study the model size-performance trade-off in the edge-deployable regime (\(\leq\)8B) to identify (i) the lower bound of usable capacity and (ii) the best performance achievable under a fixed parameter budget.
\textit{First}, varying the backbone size shows a clear capacity threshold: the 4B \sysname reaches 67.2\% success with only 2.1\% LoRA overhead, while shrinking to 2B causes an 15\% absolute drop, indicating that \textit{4B} is a practical inflection point for supporting the semantic understanding and multi-step reasoning required by GUI tasks. 
\textit{Second}, under the \(\leq\)8B budget, \sysname is highly parameter-efficient: it exceeds GUI-Owl-7B (66.4\%) by 0.8 points with \textit{41.6\%} fewer parameters, and outperforms V-Droid (59.5\%) by 7.7 points with roughly half the parameter budget. 
\textit{Third}, gains are driven more by specialization than scaling: the additional \textit{83M} LoRA parameters (2.1\% of total) enable task-specific planning/grounding/reflection behaviors that deliver disproportionate improvements.
\textbf{Summary}. \sysname suggests \textit{4B} as a competitive lower bound for edge GUI automation, and shows that architectural specialization can improve performance-per-parameter more effectively than increasing model size within edge constraints.

\begin{figure}[t]
  \centering
    \scriptsize
    \subfloat[
      \label{fig:prefilling_speed_comparison}]{
      \includegraphics[width=0.3\textwidth]{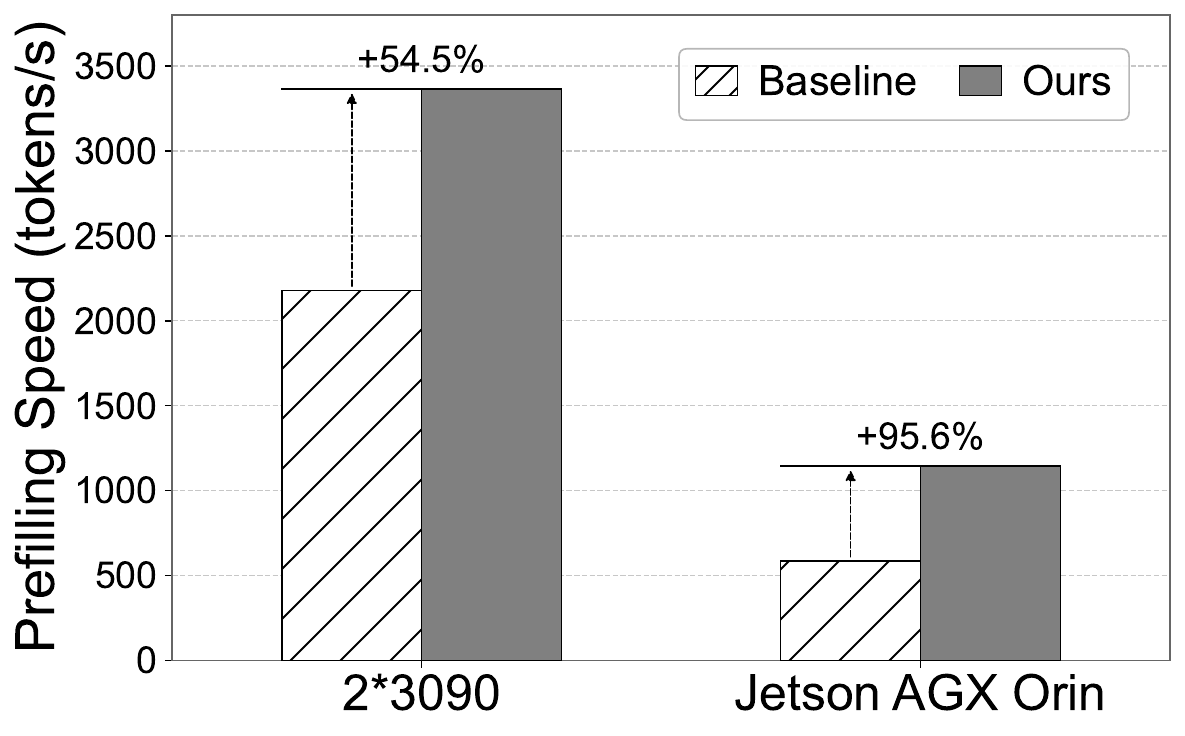}
    }
    \hspace{4mm}
    \subfloat[
      \label{fig:decoding_speed_comparison}]{
      \includegraphics[width=0.3\textwidth]{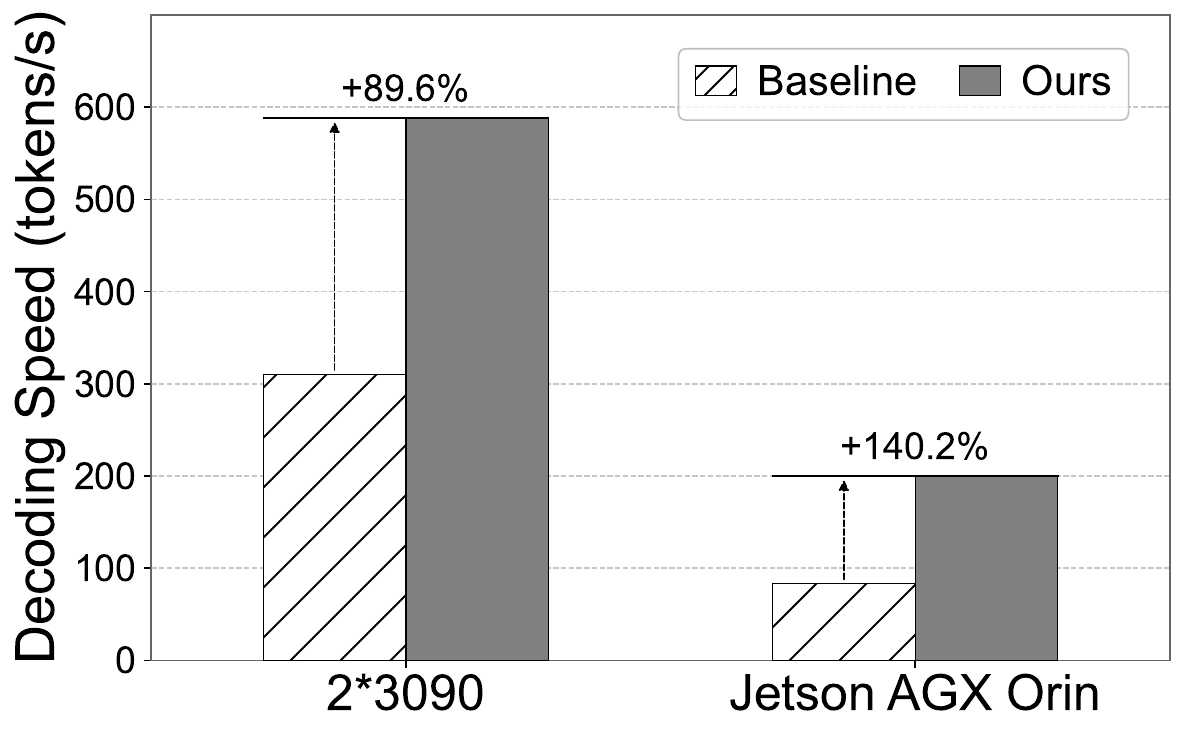}
    }
    \hspace{4mm}
    \subfloat[
      \label{fig:latency_comparison}]{
      \includegraphics[width=0.3\textwidth]{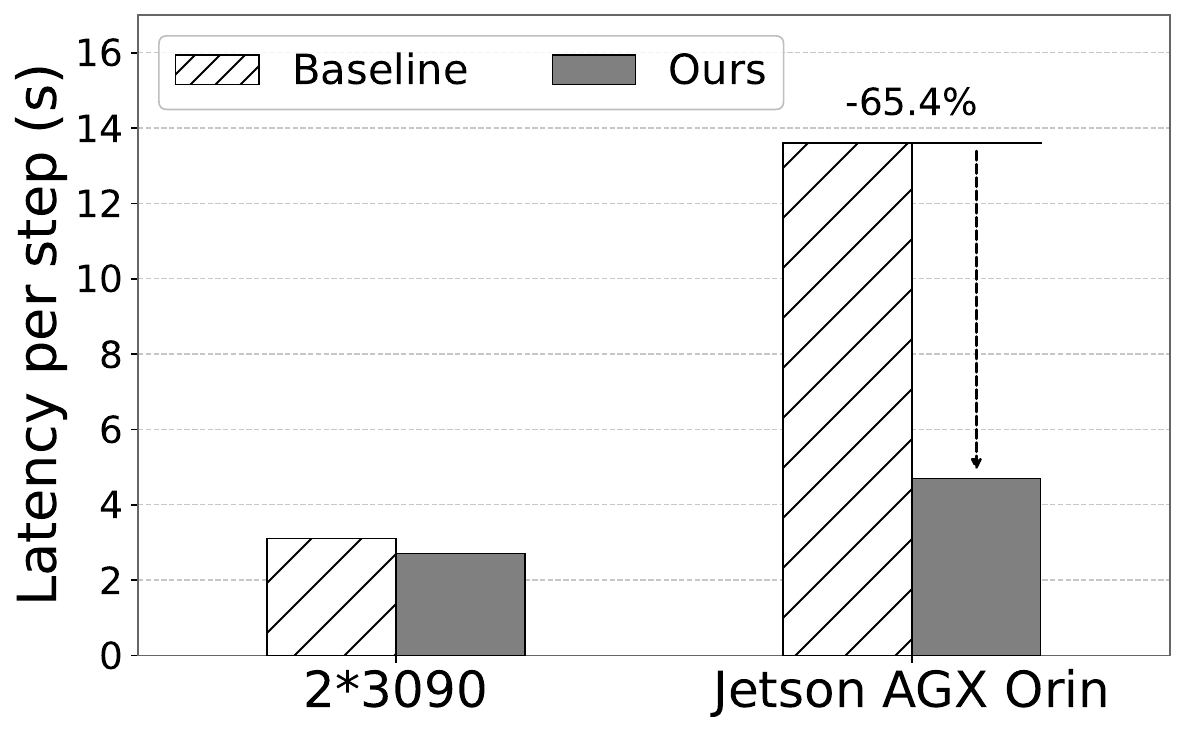}
    }
    \vspace{-3mm}
    \caption{System performance comparison on cloud and edge platforms.}
    \vspace{-2mm}
    \label{fig:cloud_edge_comparison}

\end{figure}

\subsection{System Overhead}

\subsubsection{Latency}
\label{subsubsec:latency-analysis}
We evaluate \sysname against the M3A implementation under both \textit{cloud} and \textit{edge} deployments.
On cloud GPUs, we compare three backbones: Qwen3-VL-4B-Instruct (ours), Qwen2.5-VL-7B-Instruct (a common base for many GUI agents such as GUI-Owl-7B), and Qwen3-VL-8B-Instruct (\figref{fig:system-overhead}).
Under the \sysname framework, we further compare Qwen3-VL-4B-Instruct with the Qwen2.5-VL-7B-Instruct baseline across edge and cloud settings (\figref{fig:cloud_edge_comparison}).
The edge server is an NVIDIA Jetson AGX Orin; the cloud server uses 2$\times$RTX~3090.

\textit{First}, as shown in \figref{fig:latency_comparison}, \sysname achieves 4.7\,s/step under weak edge assistance, reducing latency by 65.4\% compared to the Qwen2.5-VL-7B-Instruct.
With cloud assistance, \sysname further reduces per-step latency to 2.7\,s. The two-stage UI Parser completes within 100 ms per step on average, which is negligible compared to the VLM inference time.
Second, to attribute latency gains to model-side efficiency, we benchmark prefill and decoding using vLLM under identical runtime settings~\cite{kwon2023efficient_vllm}.
On the cloud server (2$\times$RTX~3090, \texttt{--gpu-memory-utilization 0.85}), \sysname reaches 3365.5\,tokens/s prefill (+54.5\% over 2178.8) and 588.6\,tokens/s decoding (+89.7\% over 310.4) (\figref{fig:prefilling_speed_comparison}, \figref{fig:decoding_speed_comparison}).
On the edge server (Jetson AGX Orin, \texttt{--gpu-memory-utilization 0.7}), \sysname achieves 1145.5\,tokens/s prefill (+95.6\% over 585.5) and 200.3\,tokens/s decoding (+140.2\% over 83.4).
These show that our 4B backbone materially accelerates both prompt processing and token generation, with the largest gains appearing on VRAM/bandwidth-constrained hardware.
\textit{Third}, as shown in \figref{fig:avg_prompt_tokens}, removing A11y-tree inputs reduces per-step tokens from 4.6k to 1.4k (-69\%),
directly shrinking decoding workload and translating into lower step latency.
Together with the throughput gains above, this explains \sysname's consistent end-to-end speedup.
Fourth, on high-end cloud GPUs, end-to-end serving can become memory/overhead-bound, so 4B and 7B/8B models may exhibit similar wall-clock speed.
In contrast, on resource-limited servers (\eg Jetson-class devices or GPUs with limited VRAM/bandwidth), 7B/8B models hit memory constraints earlier (tighter batch/context budgets and heavier quantization/offload), making parameter count a first-order bottleneck.
Accordingly, our 4B design cuts end-to-end latency by 65\% on such deployments, while enabling cheaper tiers and higher concurrency without hardware upgrades.
Fifth, \sysname targets hands-busy and safety-/distraction-sensitive workflows, where verified reliability matters more than sub-second reactions.
At 2.7\,s/step, \sysname is human-comparable: humans spend $\sim$20\,s of \emph{active} interaction to finish an 8-12 step task, whereas \sysname reduces user effort to one spoken instruction and bears the remaining latency itself.

\textbf{Summary}. Across edge and cloud, \sysname achieves human-like speed (2.7\,s/step) while substantially reducing latency versus the baseline.
The gains come from (i) a compact 4B backbone that improves prefill/decoding throughput, especially on constrained devices, and (ii) token-efficient screenshot-only UI representations (-69\% tokens without A11y trees), alongside a self-verifying meta-control loop that prioritizes reliable execution.

\begin{figure}[t]
  \centering
  \captionsetup[subfloat]{labelformat=empty}
  \subfloat[
    \label{fig:prefilling_device}]{
    \includegraphics[width=0.24\textwidth]{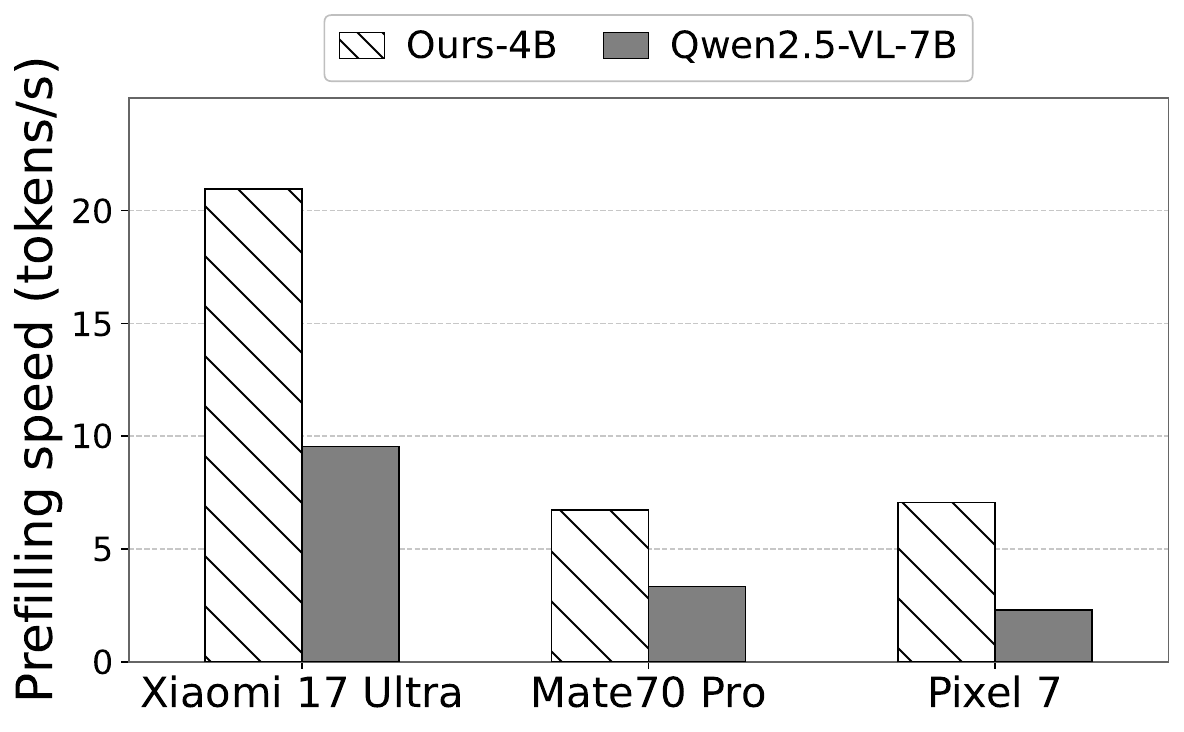}
  }
  \subfloat[
    \label{fig:decoding_device}]{
    \includegraphics[width=0.24\textwidth]{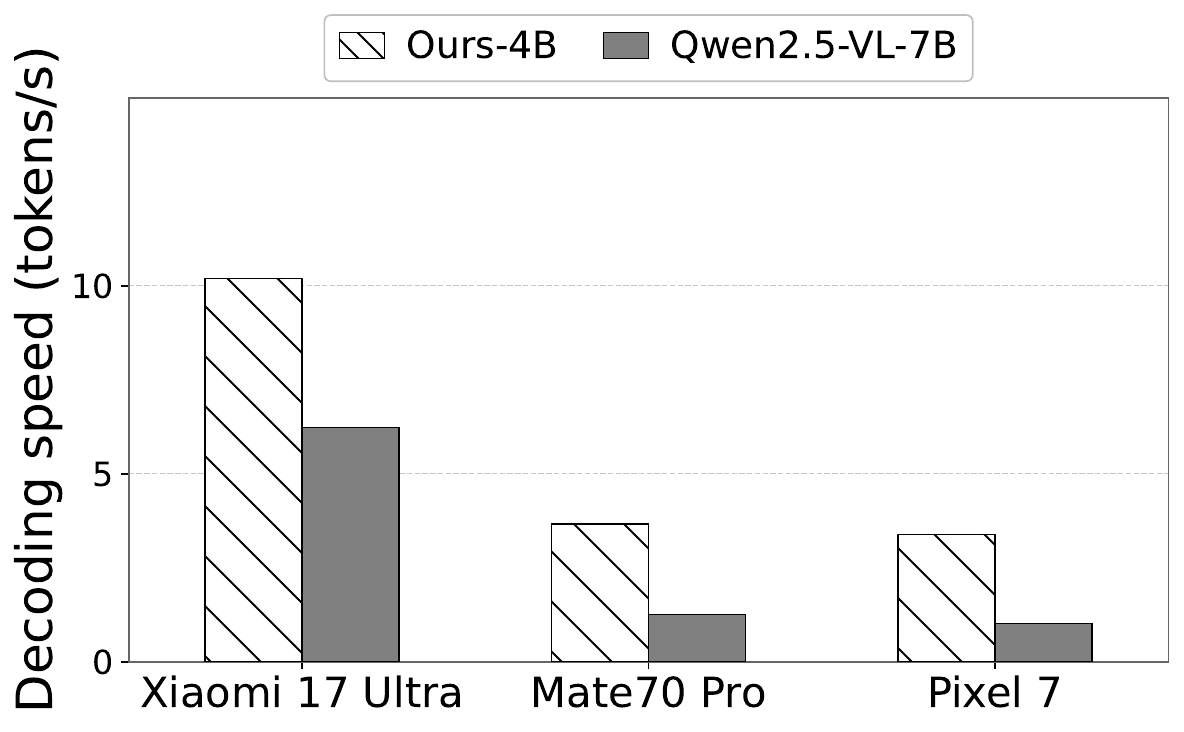}
  }
  \subfloat[
    \label{fig:latency_decice}]{
    \includegraphics[width=0.24\textwidth]{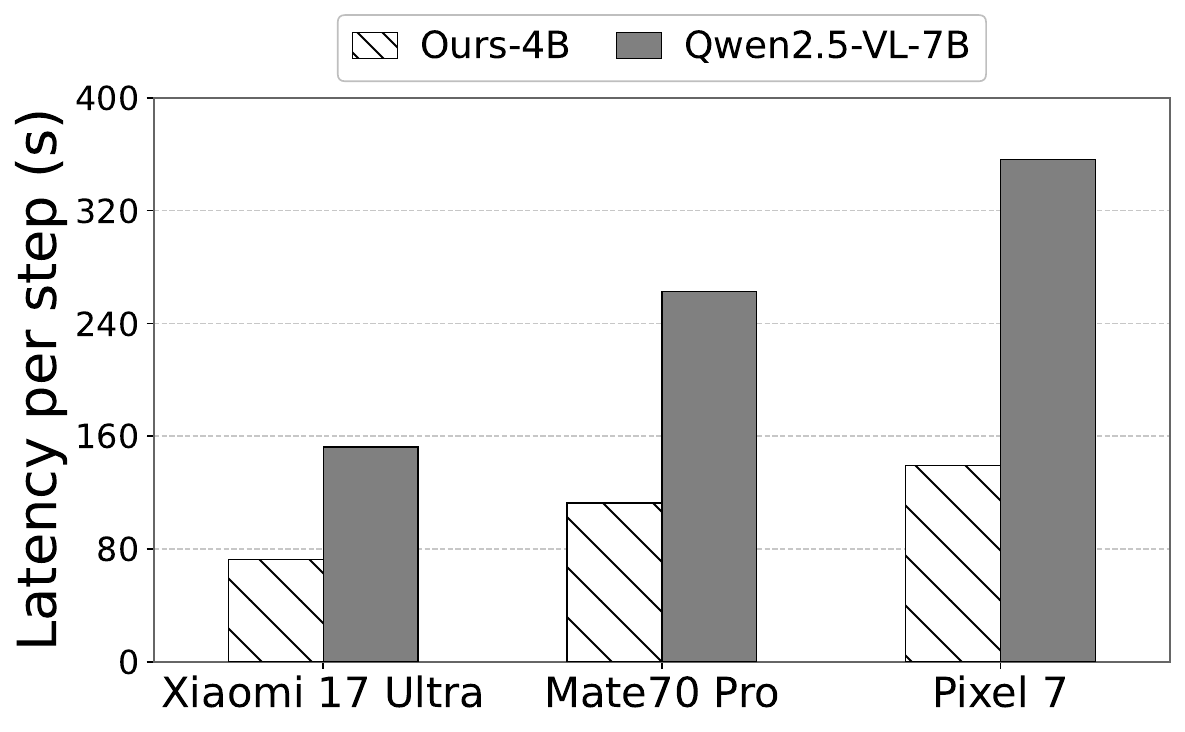}
  }
  \subfloat[
    \label{fig:energy}]{
    \includegraphics[width=0.24\textwidth]{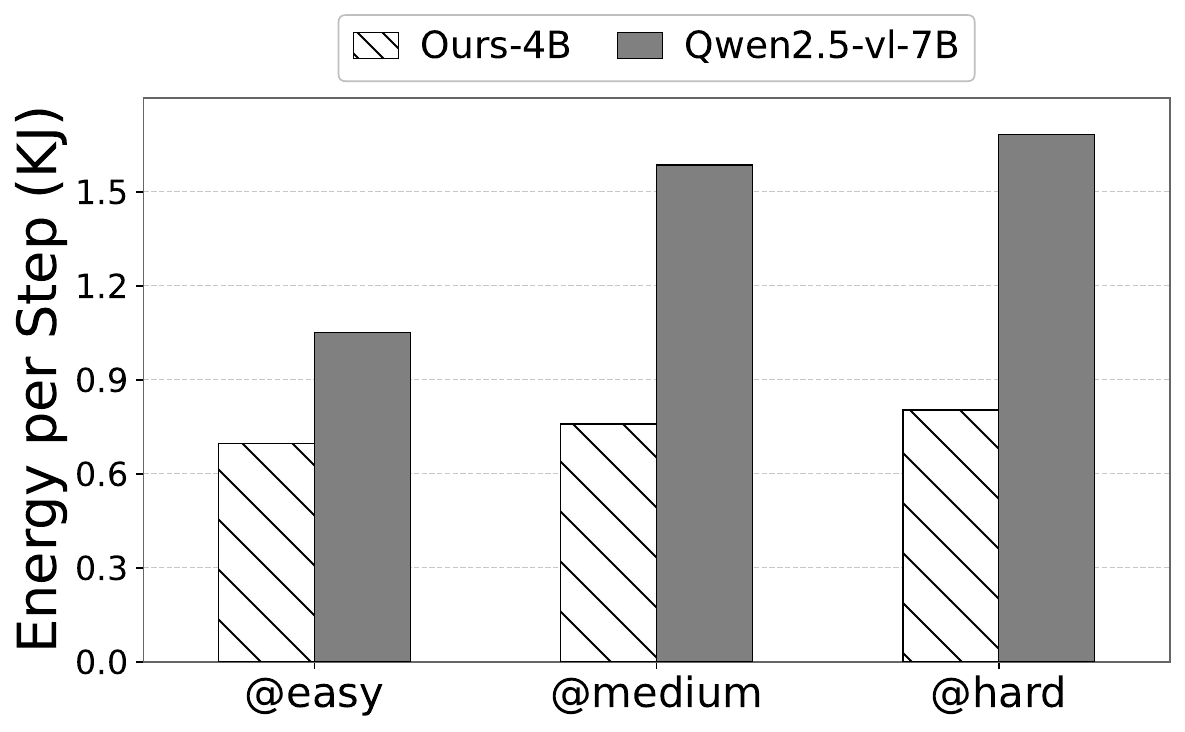}
  }
  \vspace{-2mm}
  \caption{On-device efficiency analysis of \sysname compared with baselines across diverse mobile platforms (\ie Xiaomi 17 Ultra, Huawei Mate70 Pro, Google Pixel 7).}
  \vspace{-4mm}
  \label{fig:system-overhead-device}
\end{figure}

\subsubsection{Reasoning Throughput}
\label{subsubsec:reasoning-efficiency}
We evaluate reasoning efficiency by comparing token throughput and per-step latency of \sysname (Qwen3-VL-4B-Instruct) against prevalent vision-based agents built on Qwen2.5-VL-7B-Instruct (\eg GUI-Owl-7B and UI-Tars-7B) on identical multi-step tasks. 
We use three smartphones: Xiaomi 17 Ultra, Huawei Mate70 Pro, and Google Pixel 7.
\textit{First}, \sysname achieves higher prefilling throughput (20.96/6.73/7.06 token/s) than the 7B baseline (9.53/3.35/2.29 tokens/s), improving by 119.7\% on average. 
\textit{Second}, the advantage persists in decoding: \sysname reaches 10.19/3.66/3.39 tokens/s versus 6.22/1.25/1.02 tokens/s, a 161.5\% average increase. 
\textit{Third}, reduced token count further lowers end-to-end latency (\eg 72.39s vs. 152.26s on Xiaomi 17 Ultra), though real-time interaction remains future work.
\textbf{Summary}. Higher throughput and fewer input tokens jointly reduce per-step cost and overall task time.

\subsubsection{Energy Efficiency}
\label{subsubsec:energy-cost}
We measure energy cost on AndroidWorld (Easy/Medium/Hard) using a controlled setup on Google Pixel 7: WiFi-connected, no charging interference, real-time current/voltage logging, and energy computed by integrating power over execution time. We compare against Qwen2.5-VL-7B-Instruct based agents (GUI-Owl-7B and UI-Tars-7B).
\textit{First}, \sysname consumes less energy across all difficulties: Easy 0.70 \textit{vs.} 1.05 kJ (-33.3\%). 
\textit{Second}, energy savings increase with complexity: Medium 0.76 \textit{vs.} 1.59 kJ (-52.2\%), Hard 0.80 vs. 1.68 kJ (-52.4\%). 
\textit{Third}, reductions stem from the compact 4B backbone, token-efficient UI parsing, and targeted execution that avoids redundant computation.
\textbf{Summary}. \sysname reduces energy by 33$\sim$52\%, enabling more practical sustained on-device automation.

\begin{figure}[t]
  \centering
  \subfloat[Automation preference]{
    \centering
    \includegraphics[width=0.22\textwidth]{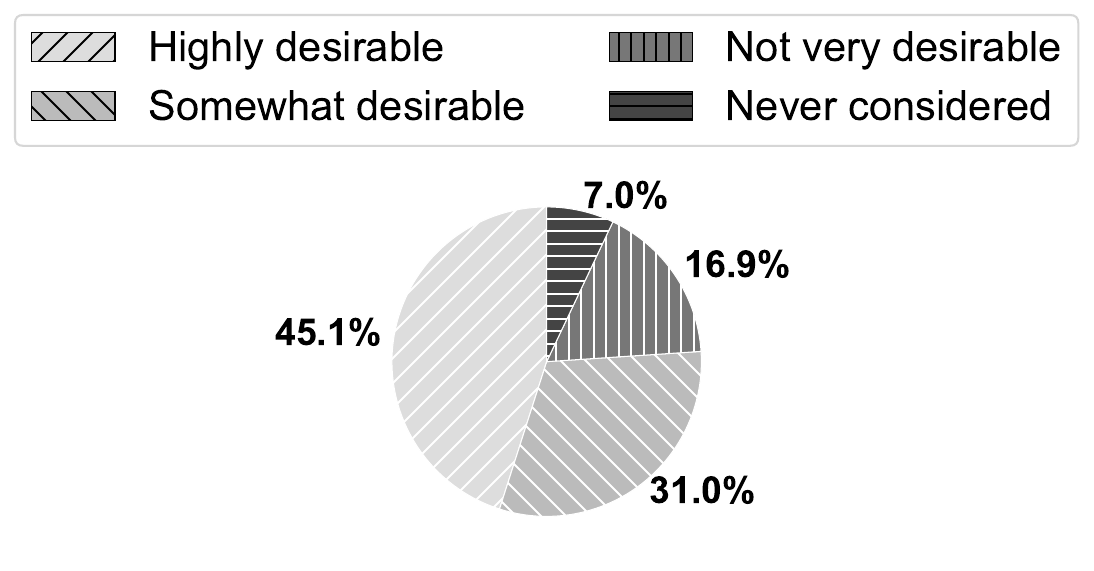}
    \label{fig:GS_Q1_Desirability_of_task_automation}
  }\hfill
  \subfloat[Tool limitations]{
    \includegraphics[width=0.22\textwidth]{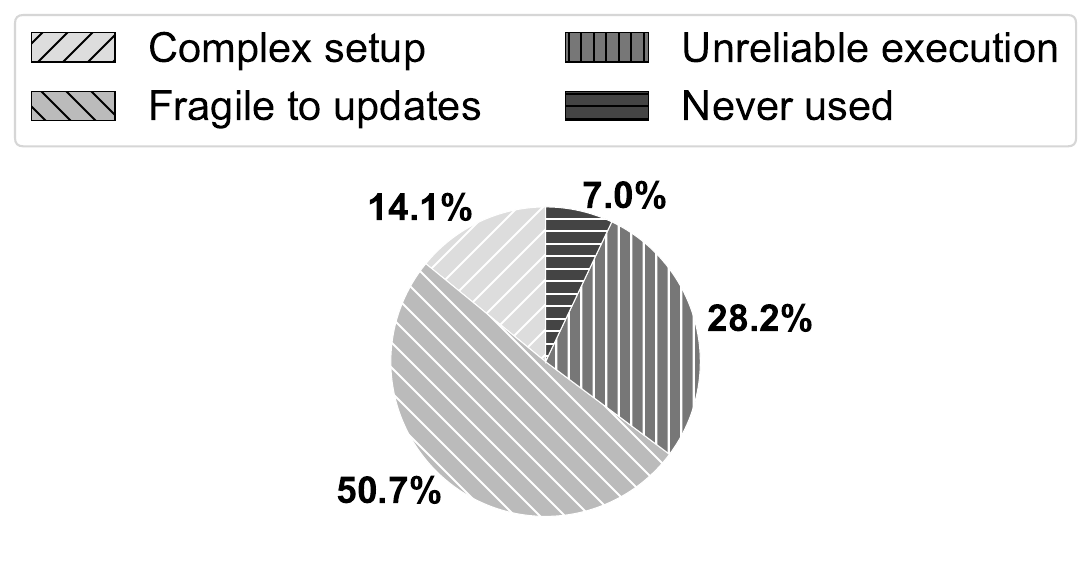}
    \label{fig:GS_Q2_Limitations_of_existing_tools}
  }\hfill
  \subfloat[Feedback burden]{
    \includegraphics[width=0.20\textwidth]{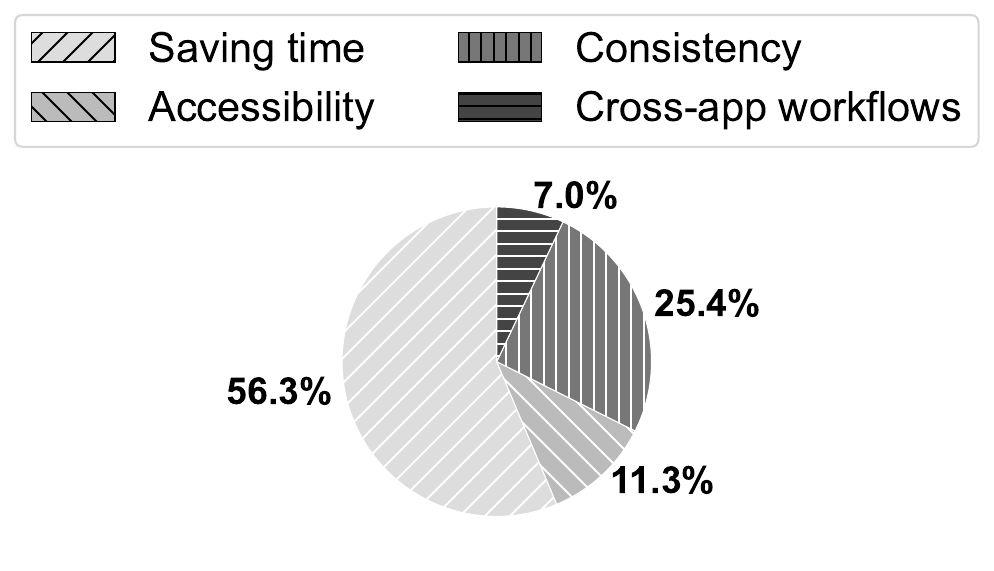}
    \label{fig:GS_Q3_Most_significant_benefit}
  }\hfill
  \subfloat[Primary benefit]{
    \includegraphics[width=0.22\textwidth]{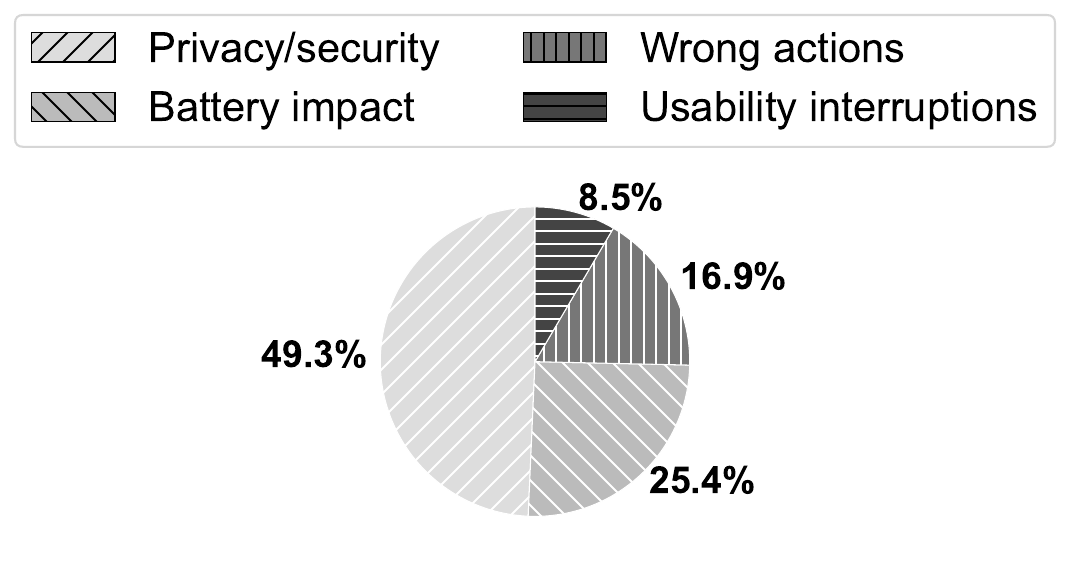}
    \label{fig:GS_Q4_Primary_concern}
  }
  \vspace{-2mm}
  \caption{Survey results highlighting user demands for reliable mobile GUI automation.} 
  \vspace{-3mm}
  \label{fig:general_survery}
\end{figure}

\begin{figure}[t]
  \centering
  \subfloat[Navigation alignment]{
    \centering
    \includegraphics[width=0.22\textwidth]{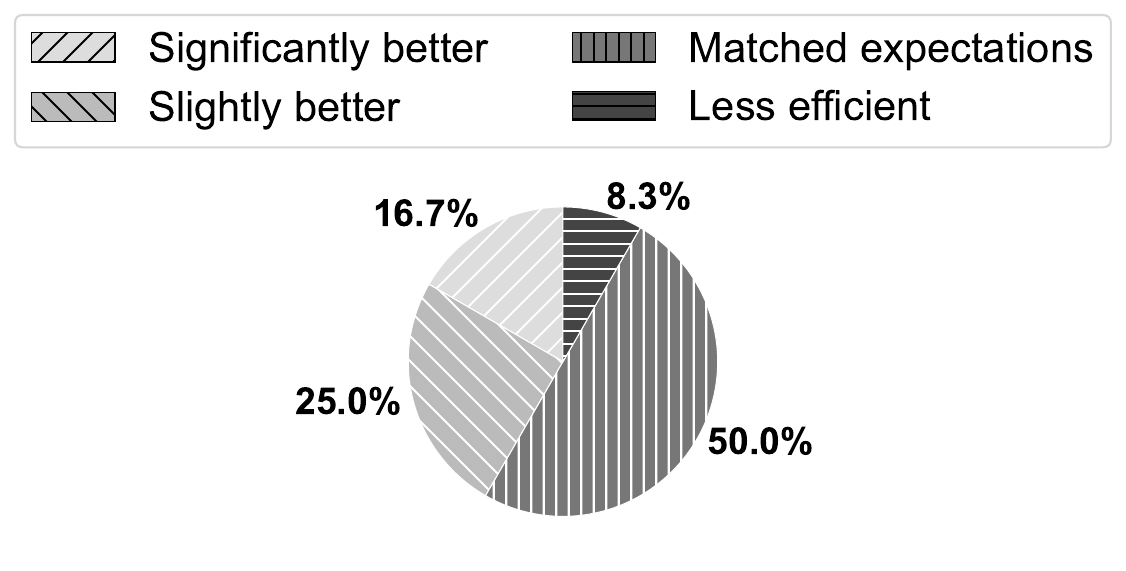}
    \label{fig:PD_Q1_Navigation_vs_expectations}
  }\hfill
  \subfloat[Reliability perception]{
    \includegraphics[width=0.21\textwidth]{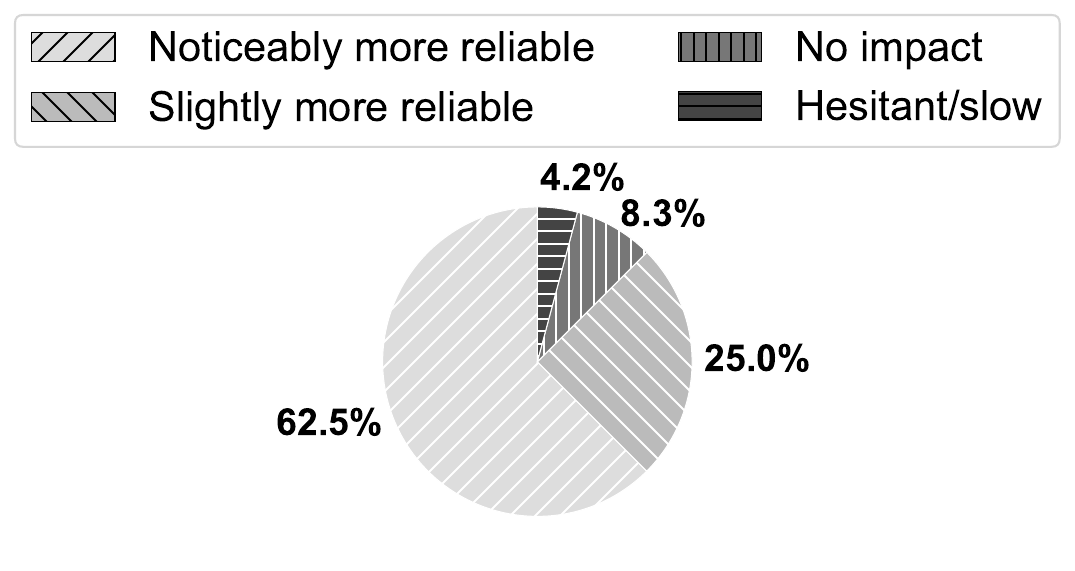}
    \label{fig:PD_Q2_Perception_of_reliability}
  }\hfill
  \subfloat[Robustness to pop-ups]{
    \includegraphics[width=0.22\textwidth]{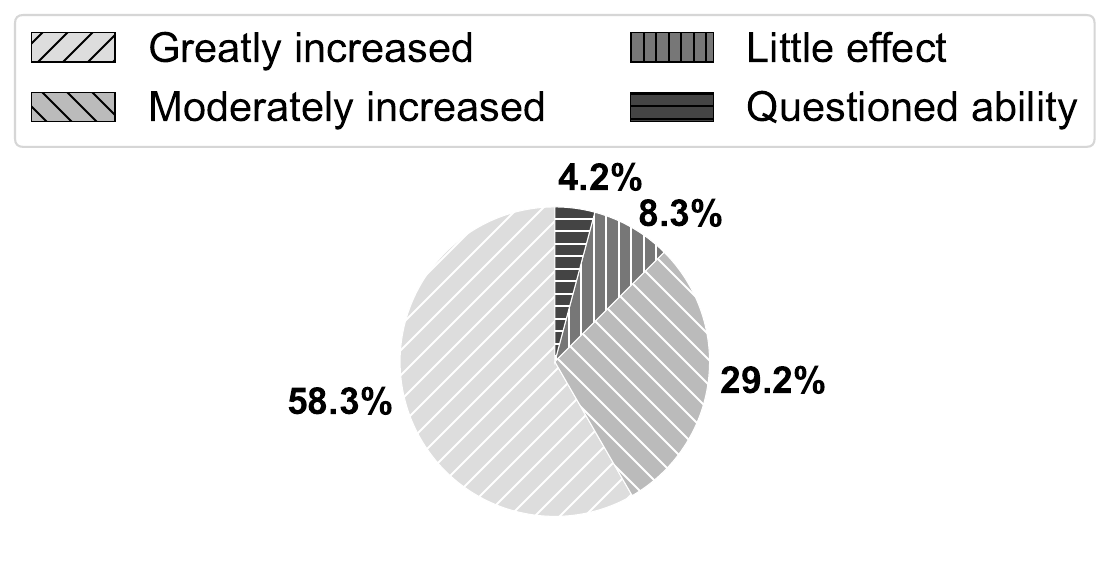}
    \label{fig:PD_Q3_Adapting_to_pop}
  }\hfill
  \subfloat[Need for self-recovery]{
    \includegraphics[width=0.22\textwidth]{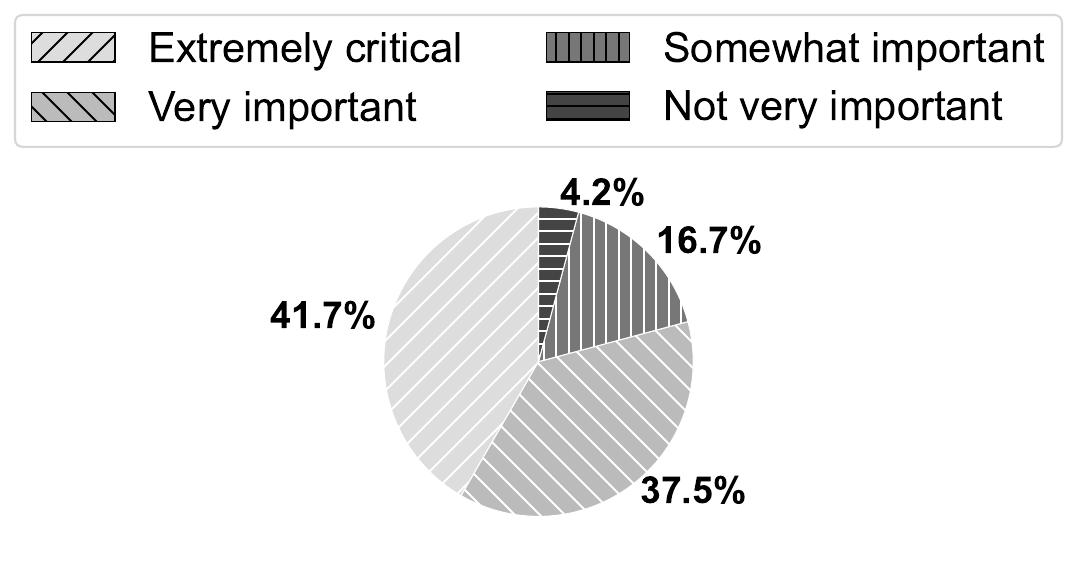}
    \label{fig:PD_Q4_Importance_of_self-recovery}
  }
  \vspace{-3mm}
  \caption{Post-demo user evaluation of \sysname.}
  \vspace{-3mm}
  \label{fig:demo_survery}
\end{figure}

\subsection{User Study}
We conduct a two-phase user study to assess the perceived utility and reliability of \sysname for mobile task automation.
We recruited 71 participants via online and in-person channels to ensure diverse technical backgrounds and usage patterns.
Participants included 48 males and 23 females, aged 19--38 (\(\mu\)=26.7, \(\sigma\)=4.8), all being regular smartphone users.
The study consists of: (i) an online \textit{general survey} completed by all 71 participants, and (ii) an in-lab \textit{post-demonstration evaluation} with 24 participants after observing live \sysname demos.
Both phases used single-choice questions (one option per question).

\subsubsection{User Demand Study}
Results (\figref{fig:general_survery}) indicate strong demand for GUI automation but clear dissatisfaction with current tools.
\SI{45.1}\% report high desirability for task automation (\figref{fig:GS_Q1_Desirability_of_task_automation}), while \SI{50.7}\% identify fragility to app/UI updates as the primary limitation (\figref{fig:GS_Q2_Limitations_of_existing_tools}).
Participants primarily value efficiency gains: \SI{56.3}\% select time saving and reduced cognitive load as the most significant benefit (\figref{fig:GS_Q3_Most_significant_benefit}).
Meanwhile, privacy/security is the top adoption concern (\SI{49.3}\%, \figref{fig:GS_Q4_Primary_concern}), highlighting the importance of on-device execution.

\subsubsection{Post-Demo User Evaluation}
After observing \sysname (\figref{fig:PD_Q1_Navigation_vs_expectations}-\figref{fig:PD_Q4_Importance_of_self-recovery}), participants provide feedback on robustness and trust.
\SI{50.0}\% feel \sysname's navigation behavior matched their expectations (\figref{fig:PD_Q1_Navigation_vs_expectations}), and \SI{62.5}\% report improved perceived reliability due to staged parsing and verification (\figref{fig:PD_Q2_Perception_of_reliability}).
Participants further emphasize dynamic robustness: \SI{58.3}\% state that pop-up adaptation increased perceived intelligence (\figref{fig:PD_Q3_Adapting_to_pop}), and \SI{79.2}\% consider self-recovery critical for trusting automation tools (\figref{fig:PD_Q4_Importance_of_self-recovery}).

\fakeparagraph{Summary}
The study suggests three practical demands for mobile GUI automation:
1) robustness to UI changes and disruptions is a dominant unmet need;
2) privacy is a primary barrier, motivating on-device solutions;
3) trust hinges on closed-loop execution, especially self-recovery.
These findings support \sysname's design choice of GUI agent and meta-controller-based verification and recovery for reliable real-world GUI automation.

\begin{figure}[t]
  \centering
  \captionsetup[subfloat]{labelformat=empty}
  \subfloat[]{
    \centering
 \includegraphics[height=0.12\textwidth]{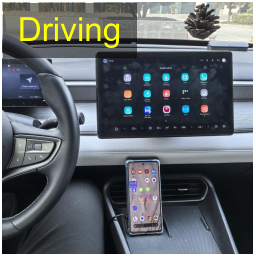}
    \label{fig:driving}
  }
  \subfloat[]{
\includegraphics[height=0.12\textwidth]{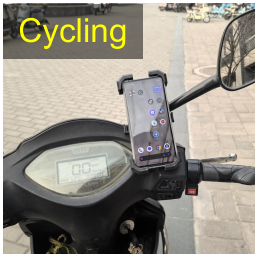}
    \label{fig:cycling}
  }
  \subfloat[]{
\includegraphics[height=0.12\textwidth]{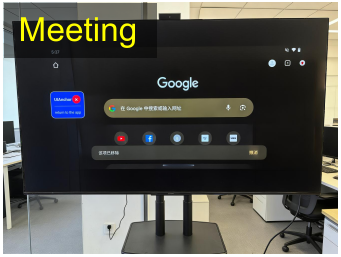}
    \label{fig:meeting}
  }
  \subfloat[]{
\includegraphics[height=0.12\textwidth]{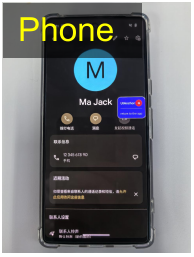}
    \label{fig:phone}
  }
  \subfloat[]{
\includegraphics[height=0.12\textwidth]{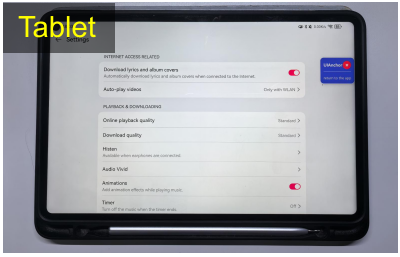}
    \label{fig:tablet}
  }
  \subfloat[]{
\includegraphics[height=0.12\textwidth]{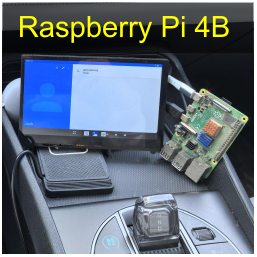}
    \label{fig:raspberry}
  }
  \vspace{-4mm}
  \caption{Representative scenarios and devices used in our real-world case study.}
  \vspace{-3mm}
  \label{fig:case_study_pic}
\end{figure}

\subsection{Real-World Case Study}
We further validate \sysname in the kinds of moments where mobile automation is most valuable, when users are \emph{in motion}, \emph{hands-busy}, or \emph{attention-limited}. 
As shown in \figref{fig:case_study_pic}, we conduct a real-world study spanning \textbf{six} practical conditions along two axes: \emph{scenario} and \emph{device}. 
For \textit{scenarios}, we test \textbf{Driving} (in-vehicle use while driving), \textbf{Cycling} (delivery riders on the move), and \textbf{Meeting} (professional settings with social/attention constraints), each with \textbf{12} scenario-specific tasks reflecting realistic intents under interruptions and UI volatility.
For \textit{devices}, we evaluate a \textbf{Phone} (Pixel~7, 1080$\times$2400), a \textbf{Tablet} (MatePad Pro, 2560$\times$1600), and a \textbf{Raspberry Pi} setup (1080p monitor), running \textbf{12} cross-app common tasks per device to stress UI adaptation across screen sizes and interaction paradigms. 
We compare against a strong baseline, \ie the M3A agent with Qwen2.5-VL-7B~\cite{Qwen2.5-VL}. 
In each scenario, we carry the device and system into real use and continuously evaluate for \textit{2-3 hours}, reporting averaged results across five users.

\begin{figure}[t]
  \centering
  \captionsetup[subfloat]{labelformat=empty}
  \subfloat[]{
    \centering
 \includegraphics[height=0.12\textwidth]{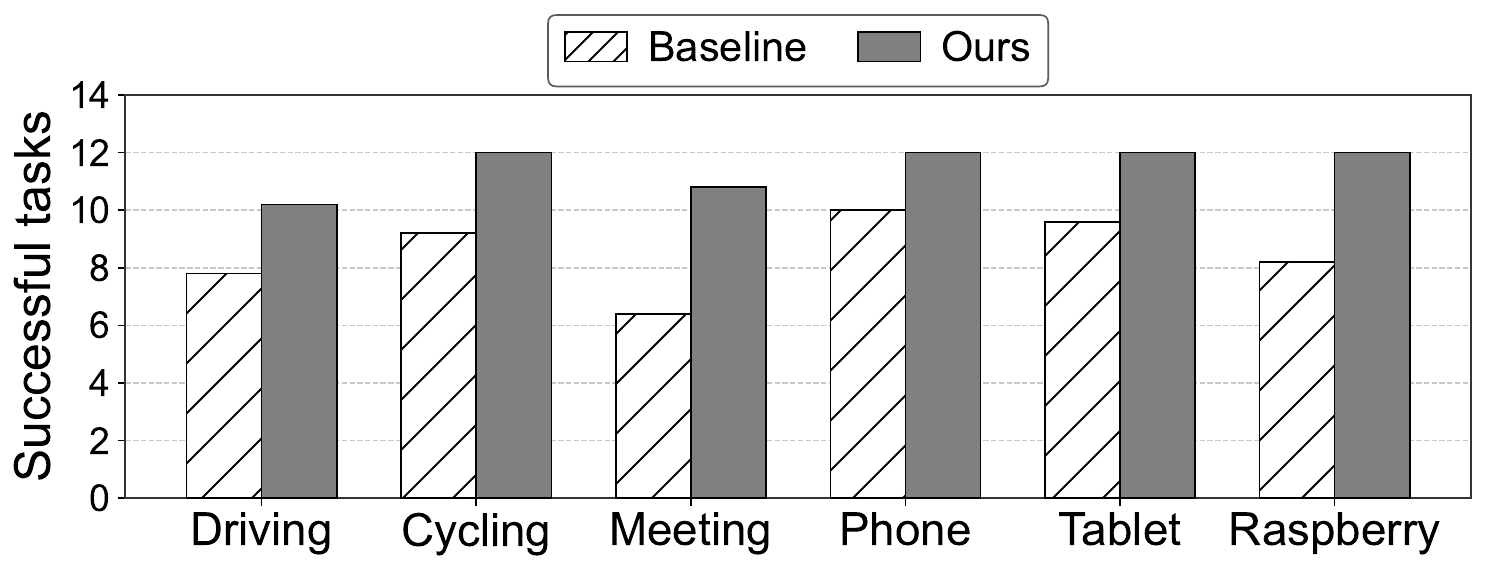}
    \label{fig:case_successful_tasks_comparison}
  }
  \subfloat[]{
\includegraphics[height=0.12\textwidth]{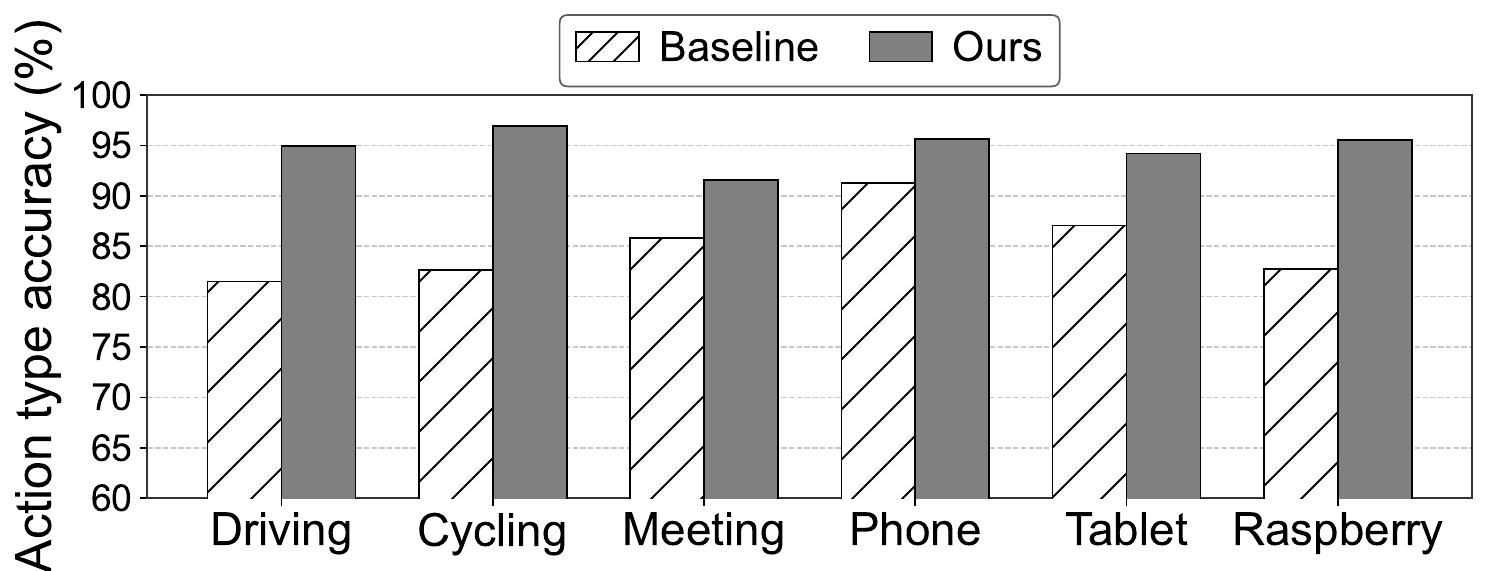}
    \label{fig:case_action_type_accuracy_comparison}
  }
  \subfloat[]{
\includegraphics[height=0.12\textwidth]{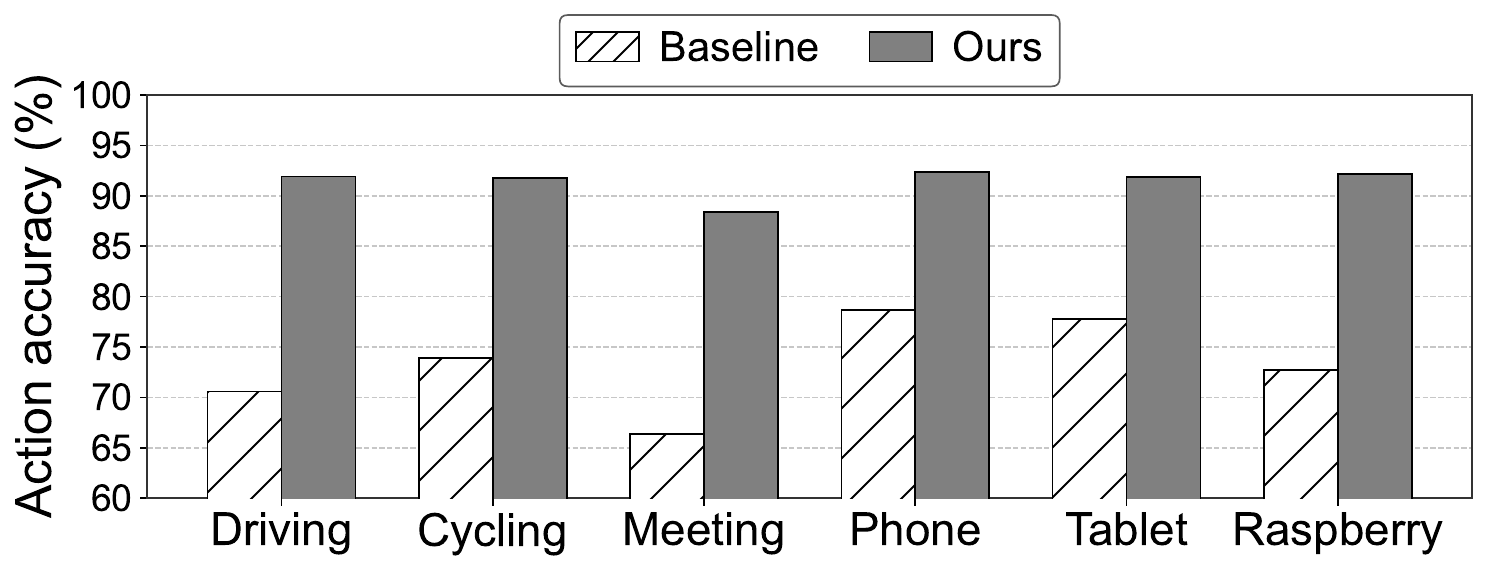}
    \label{fig:case_action_accuracy_comparison}
  }
  \vspace{-4mm}
  \caption{Performance comparison between baseline and \sysname method across six scenarios.}
  \vspace{-3mm}
  \label{fig:case_study}
\end{figure}

Across all conditions, \sysname consistently outperforms the baseline across all metrics (\figref{fig:case_study}). 
\textit{First}, it completes more tasks in every scenario, with the largest gains in the most human-constrained contexts, such as \textbf{Cycling} (12 vs.\ 9.2) and \textbf{Meeting} (10.8 vs.\ 6.4) successful tasks (\figref{fig:case_successful_tasks_comparison}). 
\textit{Second}, it makes substantially better interaction decisions under real-world disturbances, improving action-type accuracy by over 13\% in \textbf{Driving} and \textbf{Cycling}(\figref{fig:case_action_type_accuracy_comparison}). 
\textit{Third}, end-to-end action accuracy sees similarly large gains, notably in \textbf{Driving} (91.92\% vs.\ 70.59\%) (\figref{fig:case_action_accuracy_comparison}). 
\textit{Fourth}, device-wise, \sysname generalizes well across form factors, with consistent improvements on \textbf{Tablet} and \textbf{Raspberry Pi}, indicating robust cross-resolution UI understanding. 
\textbf{Summary}. This real-world case study shows that \sysname translates benchmark gains into practical robustness, with substantially fewer perception misses, less state drift, and more stable recovery when interaction is inconvenient, unsafe, or inaccessible. These results support practical promise in such scenarios, while stronger claims about driving, accessibility, or safety-critical deployment require dedicated task-based user studies and broader validation.

\section{RELATED WORK}
\subsection{Mobile Task Automation}
Mobile task automation translates high-level intents (often natural language) into executable multi-step interactions on smartphones. 
Prior work falls into three paradigms:
\textbf{API-based} methods map user intents to privileged APIs~\cite{erdogan2024tinyagent,xie2024droidcall}. They are efficient and reliable, but fundamentally limited by API availability and permission scope.
\textbf{GUI-to-API} hybrids first mine GUIs offline to infer reusable functions/scripts, then invoke them online (e.g., AXIS~\cite{lu2024axis}, AutoDroid-v2~\cite{wen2024autodroidv2}). They improve coverage but require costly offline exploration and remain brittle to UI/version changes.
\textbf{GUI-based} agents act directly through the visual interface via taps/typing/swipes, offering maximal generality across system and third-party apps. 
However, open-loop execution and imperfect UI grounding often cause compounding errors under dynamic mobile UIs.
\sysname follows the GUI-based paradigm, but strengthens reliability via (i) structured, high-recall UI parsing to anchor \textit{what} to act on, and (ii) a meta-controller for state verification and targeted recovery to anchor \textit{whether} actions succeed.

\subsection{GUI-Based Mobile Task Automation}
\label{subsec:related_gui_agents}
GUI agents automate tasks by operating directly on app interfaces. 
Prior work falls into two lines:
\textbf{UI-tree–based LLM agents} reason over accessibility/UI trees to select actions (e.g., AutoDroid~\cite{wen2024autodroid}, MobileGPT~\cite{lee2024mobilegpt}, V-Droid~\cite{dai2025advancing}, AndroidGen~\cite{lai2025androidgen}). 
They are token-efficient and interpretable, and can be accurate when metadata is complete, but often break in real apps with \emph{incomplete, noisy, or restricted} UI trees (\eg custom rendering, dynamic feeds, obfuscation), leading to missed targets, state loss, and brittle execution.
\textbf{Screenshot-grounded VLM agents} operate on screenshots and generalize better across system and third-party apps~\cite{cheng2024seeclick,hong2024cogagent,ye2025mobileagentv3,zhang2025appagent}. 
However, most are \emph{monolithic}, entangling perception, planning, and execution, which makes grounding fragile and errors easy to compound under mobile constraints.
Overall, both paradigms suffer from \emph{error amplification} in dynamic UIs. 
\sysname adopts screenshot grounding for generality, but improves reliability by decomposing parsing/planning/execution and coordinating them with a meta-controller for step-wise verification and targeted recovery.

\subsection{Multi-Agent Systems for Mobile Workflows}
\label{subsec:related_multi_agent}
Multi-agent designs improve long-horizon mobile automation by assigning specialized roles. Prior work broadly falls into three lines:
\textbf{Orchestrator-centric architectures} use a central planner to decompose tasks and dispatch subtasks to tool/GUI agents~\cite{lee2024mobilegpt}. They improve modularity, but the orchestrator can become a reasoning bottleneck and often fails to track step-level state under dynamic UIs.
\textbf{Reflective frameworks} add a monitor/reflector for failure detection and recovery~\cite{li2025mobileuse,ye2025mobileagentv3,wang2025mobile_e}. They help correct major mistakes, but are usually coarse-grained, can be unreliable (hallucinated judgments), and add latency via repeated re-parsing/re-planning.
\textbf{Experience-driven systems} introduce memory or self-evolution to reuse trajectories~\cite{wang2025mobile_e,zhang2025appagent, wen2024autodroid}, yet stored experience is often rigid (coarse granularity, weak state abstraction), limiting reuse when UI layouts or contexts shift.
In contrast, \sysname treats parsing, planning, and execution as peer agents coordinated by a \textit{meta-controller} with \textit{fine-grained}, \textit{per-step} verification and targeted recovery, improving grounding and long-horizon robustness in dynamic mobile UIs.
\section{CONCLUSION}
This paper addresses the reliability bottleneck of mobile GUI automation under today’s \emph{service-composed} usage trends (\eg app switching, forms, pop-ups/permissions, copy-paste), which demand fine-grained actions on cross-app, mm-scale, dynamic, and heterogeneous UIs.
While screenshot-grounded VLM agents enable API-free control, real apps expose parsing errors plus open-loop execution cascade into long failure chains.
We present \sysname, which jointly anchors \emph{what to act on} and \emph{whether actions succeed} by integrating structured UI grounding with runtime execution verification. 
A two-stage parser produces confidence-aware, context-grounded UI states, while a lightweight meta-controller performs per-action verification, maintains runtime meta-state memory (\eg toasts and activity transitions), and enables targeted recovery.
Despite a compact 4B backbone with lightweight LoRA adapters, UIAnchor substantially improves end-to-end success on hard L4 tasks, significantly outperforming GPT-4o and a 7B GUI agent; with edge/cloud assistance, it runs at $\sim$2\,s per step (human-like), reduces latency/energy, and generalizes to unseen apps and evolving cross-platform GUIs.
Future work will incorporate richer system signals beyond screenshots, improve UI-shift adaptation via continual learning, and further reduce latency while preserving closed-loop reliability.

\begin{acks}
This work was partially supported by the National Natural Science Foundation of China (62522215, 62532009, 62472354, 62572412). 
Zimu Zhou's research is supported by City University of Hong Kong Shenzhen Research Institute (CityUHKSRI). 
\end{acks}

\bibliography{acmart}
\bibliographystyle{ACM-Reference-Format}
\end{document}